\documentclass[aps,prl,twocolumn,showpacs,groupedaddress,floatfix,footinbib]{revtex4-1}
\usepackage{graphicx}
\usepackage{dcolumn}
\usepackage{bm}				
\usepackage{bbm}
\usepackage{amsfonts}
\usepackage{amssymb}
\usepackage{amsmath}
\usepackage{ulem} 
\usepackage{graphicx}		
\usepackage{xcolor, soul}	
\usepackage{verbatim}
\usepackage[T1]{fontenc}
\usepackage{pifont}
\usepackage{multirow}
\usepackage{notes2bib}
\usepackage{bibunits}
\usepackage{appendix}

\DeclareMathOperator{\sgn}{sgn}
\DeclareMathOperator{\pf}{Pf}

\begin{document}
\title{Odd--Parity Spin--Triplet Superconductivity in Centrosymmetric Antiferromagnetic Metals}

\author{Seung Hun Lee$^{1,2,3}$}
\author{Bohm-Jung Yang$^{1,2,3}$}
\email{bjyang@snu.ac.kr}
\affiliation{$^1$Center for Correlated Electron Systems, Institute for Basic Science (IBS), Seoul 08826, Korea\\$^2$Department of Physics and Astronomy, Seoul National University, Seoul 08826, Korea\\$^{3}$Center for Theoretical Physics (CTP), Seoul National University, Seoul 08826, Korea}
\date{\today}

\begin{abstract}
We propose a route to achieve odd-parity spin-triplet superconductivity in metallic collinear antiferromagnets with inversion symmetry. Owing to the existence of hidden antiunitary symmetry, which we call the effective time-reversal symmetry (eTRS), the Fermi surfaces of ordinary antiferromagnetic metals are generally spin-degenerate, and spin-singlet pairing is favored. However, by introducing a local inversion symmetry breaking perturbation that also breaks the eTRS, we can lift the degeneracy to obtain spin-polarized Fermi surfaces. In the weak-coupling limit, the spin-polarized Fermi surfaces constrain the electrons to form spin-triplet Cooper pairs with odd-parity. Interestingly, all the odd-parity superconducting ground states we obtained host nontrivial band topologies manifested as chiral topological superconductors, second-order topological superconductors, and nodal superconductors. We propose that layered double-perovskites with collinear antiferromagnetism, sandwiched by conventional superconductors, are promising candidate systems where our theoretical ideas can be applied to.
\end{abstract}

\maketitle
\begin{bibunit}

\indent
\textit{Introduction.}---Magnetism and superconductivity are two representative quantum mechanical phenomena arising from spontaneous symmetry breaking. For decades, not only the individual phenomenon but also the interplay between them has been a central topic in condensed matter physics. 
Especially, motivated by the observation that the superconducting region usually appears near the magnetic quantum critical point~\cite{nagaosa1997superconductivity,taillefer2010scattering,sachdev2012entangling}, 
the pairing instability mediated by critical spin flucutations has been extensively studied~\cite{nagaosa1997superconductivity,taillefer2010scattering,sachdev2012entangling,kuwabara2000spin,meng2014odd,sumita2017multipole,almeida2017induced,setty2019ultranodal,fay1980coexistence,tada2013spin,ishizuka2018odd}. On the other hand, compared to the critical fluctuation driven superconductivity, the nature of the superconducting phase coexisting with stable magnetism has received relatively less attention~\cite{machida1980theory,fujimoto2006emergent,milovanovic2012coexistence,qi2017coexistence,powell2003gap,cheung2016topological,kkadzielawa2018spin}.
However, various materials that exhibit magnetism and superconductivity simultaneously have been reported such as heavy fermion superconductors~\cite{feyerherm1994coexistence,pagliuso2001coexistence,knebel2006coexistence,saxena2000superconductivity,aoki2001coexistence,huy2007superconductivity,linder2008coexistence,gasparini2010superconducting,wu2017pairing,ran2019nearly,sundar2019coexistence}, iron-based superconductors~\cite{ikeda2018new,lu2015coexistence,pratt2009coexistence}, twisted double bilayer graphene~\cite{liu2019spin,lee2019theory,wu2019ferromagnetism}, etc. 
Considering that magnetism strongly modifies the symmetry of the ground state, which in turn constrains possible pairing channels, coexisting magnetism and superconductivity has a great potential to realize unconventional superconductivity.

In fact, the structure of Cooper pairs can be significantly affected by magnetic ordering. For example, in ferromagnets, there is no Kramers degeneracy at general $k$-points due to spin-splitting, and the Fermi surface is spin-polarized. Therefore, in the weak-pairing limit, Cooper pairs must be formed by equal-spin electrons, and the spin part of their wave function must be a triplet~\cite{sigrist2009introduction}.
\begin{figure}[t!]
	\begin{center}
		\includegraphics[scale=0.47]{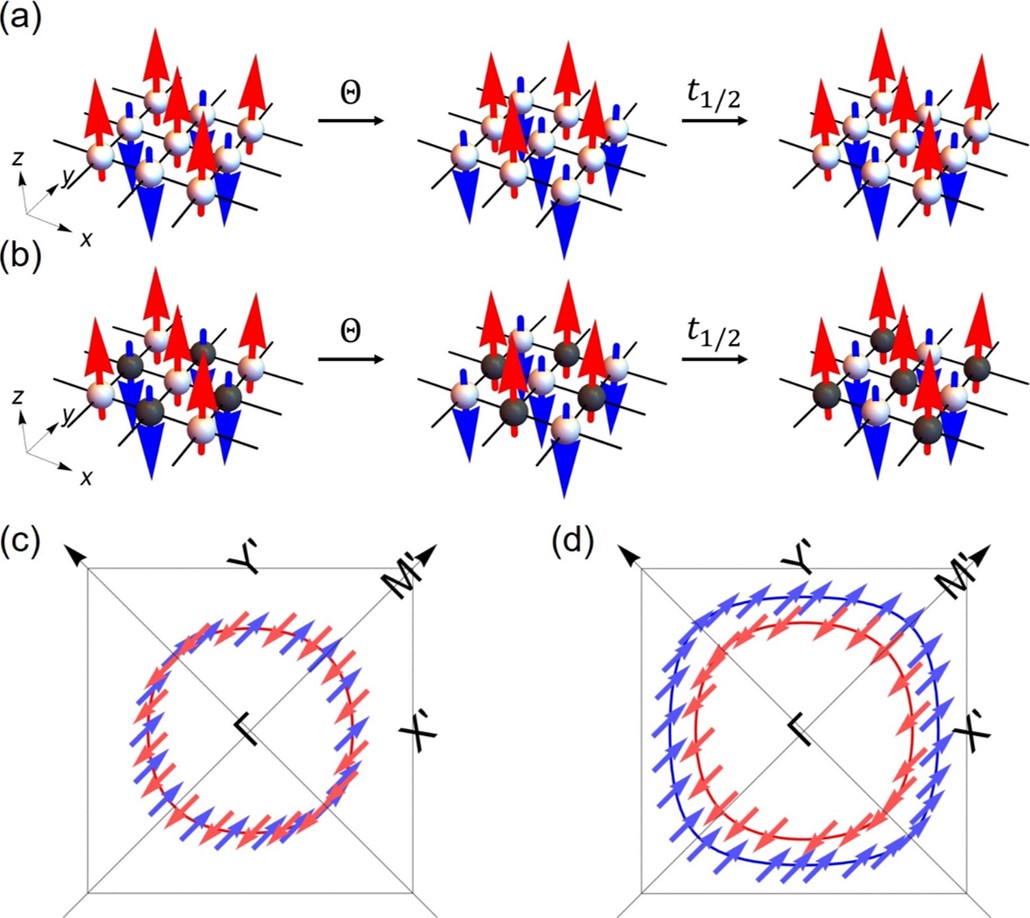}
		\caption{(a) A two-dimensional (2D) collinear antiferromagnet invariant under the eTRS $\tilde{\mathit{\Theta}}=t_{1/2}\mathit{\Theta}$. 
			(b) A 2D collinear antiferromagnet with staggered sublattice potential $\epsilon_{\textrm{sp}}$, which breaks $\tilde{\mathit{\Theta}}$. The atoms with different on-site potential energies ($\epsilon_{\textrm{sp}}\neq0$) are distinguished by white and grey colors. 
			(c) The spin texture on the Fermi surface when $\epsilon_{\textrm{sp}}=0$. The Fermi surface is spin-degenerate.
			(d) Similar figure as (c) when $\epsilon_{\textrm{sp}}\neq0$, where the Fermi surfaces are spin-polarized.
		}\label{afmlattice}
	\end{center}
\end{figure}
On the other hand, a collinear antiferromagnetic (AFM) ordering constrains Cooper pairs in a different manner~\cite{ramazashvili2008kramers,ramazashvili2009kramers,vsmejkal2019crystal}. Since a collinear AFM ordering preserves an $\textit{effective}$ time-reversal $\tilde{\mathit{\Theta}}$ symmetry (eTRS), defined as time-reversal operation $\mathit{\Theta}$ followed by a half lattice translation $t_{1/2}$~\cite{ramazashvili2008kramers,ramazashvili2009kramers,vsmejkal2019crystal}, if the system possesses additional inversion $P$ symmetry, the Kramers degeneracy at general $k$-points remains unlifted (see Fig.~1(a)), unlike in ferromagnets~\footnote{The combination of inversion symmetry and the eTRS is an antiunitary symmetry whose square is $-1$. Since the combined symmetry is local in the momentum space, it protects the Kramers degeneracy at any $k$-point.}. Dominant spin-singlet pairing reported for several AFM superconductors~\cite{machida1980theory,fujimoto2006emergent} in earlier studies can be understood in this way. Therefore, to achieve stable spin-triplet pairing in the AFM system as in ferromagnetic systems, it is necessary to break the eTRS while keeping global inversion symmetry.

In this letter, we propose a way to realize odd-parity spin-triplet superconductivity in centrosymmetric collinear antiferromagnets. Here, the central idea is to introduce the perturbations that break local inversion symmetry, such as staggered potential (SP) or antisymmetric spin-orbit coupling (ASOC)~\cite{fischer2011superconductivity,goryo2012possible}.
Since SP or ASOC makes the sublattices inequivalent, eTRS is also broken~\cite{vsmejkal2019crystal}.
Thus, in the presence of SP or ASOC, the Fermi surface of the AFM system becomes spin-split, so that spin-triplet pairing can be predominant as in ferromagnets. 
Furthermore, it  is found that the odd-parity spin-triplet pairing drives the AFM system with SP to be one of the following topological superconductors (TSCs): a chiral (spin-chiral) TSC with a non-zero Chern (spin-Chern) number, and a nodal TSC.
The chiral TSC is robust against the inclusion of spin-orbit coupling (SOC) while the stability of the spin-chiral TSC against SOC requires mirror or spin-reflection symmetry. Interestingly, once the mirror or spin-reflection symmetry is broken, the spin-chiral TSC with SOC turns into a second-order TSC.

\textit{Model.}---We consider a tight-binding model for a N\'eel ordered antiferromagnet on a square lattice. For simplicity, we include up to the nearest-neighbor (NN) hoppings in our model Hamiltonian $\hat{H}=\sum_{k}\textbf{c}_{\textbf{k}}^{\dagger}\mathcal{H}_{\textrm{nn}}(\textbf{k})\textbf{c}_{\textbf{k}}$
where $\textbf{c}_{\textbf{k}}^{\dagger}=(c_{\textbf{k}A\uparrow}^{\dagger},c_{\textbf{k}B\uparrow}^{\dagger},c_{\textbf{k}A\downarrow}^{\dagger},c_{\textbf{k}B\downarrow}^{\dagger})$, and $\mathcal{H}_{\textrm{nn}}(\textbf{k})=2t(\cos{k_{x}}+\cos{k_{y}})\sigma_{0}\tau_{x}\equiv\epsilon_{\textrm{nn}}(\textbf{k})\sigma_{0}\tau_{x}$. $\sigma$ and $\tau$ are Pauli matrices which represent the spin ($\uparrow$, $\downarrow$) and sublattice ($A$, $B$) degrees of freedom, respectively. 
Then, to describe the effect of the collinear AFM ordering, we add a mean-field approximated exchange coupling term $-\textbf{m}\cdot\bm{\sigma}\tau_{z}$ to the Hamiltonian. Now the Hamiltonian becomes
\begin{equation}
\mathcal{H}_{0}(\textbf{k})=\epsilon_{\textrm{nn}}(\textbf{k})\sigma_{0}\tau_{x}-\textbf{m}\cdot\bm{\sigma}\tau_{z},
\end{equation}
whose energy spectrum is doubly degenerate at every k-point in the Brillouin zone due to the inversion $P=\sigma_0\tau_0$ and eTRS $\tilde{\mathit{\Theta}}=i\sigma_{y}\tau_{x}K$~\footnote{More rigorously, $\tilde{\mathit{\Theta}}_{\textbf{k}}=t_{1/2}\mathit{\Theta}=ie^{i\textbf{k}\cdot(\textbf{a}/2)}\sigma_{y}\tau_{x}K$ and $\tilde{\mathit{\Theta}}_{\textbf{k}}\mathcal{H}(\textbf{k})\tilde{\mathit{\Theta}}_{\textbf{k}}^{-1}=\mathcal{H}(-\textbf{k})$, where $t_{1/2}$ and $\textbf{a}$ denote the half translation operator and the lattice vector, respectively. But the extra phase factor comming from the translation by half the lattice vector is not important in this context, so we omit it and write $\tilde{\mathit{\Theta}}$ as $i\sigma_{y}\tau_{x}K$ from now on.}.

However, if we take into account SP in the form of $\epsilon_{\textrm{sp}}\sigma_{0}\tau_{z}$ or ASOC in the form of $\sum_{i=x,y,z}2v_{i}(\cos{k_{x}}+\cos{k_{y}})\sigma_{i}\tau_{y}\equiv\bm{\epsilon}_{\textrm{asoc}}(\textbf{k})\cdot\bm{\sigma}\tau_{y}$ that breaks inversion symmetry locally while respecting global inversion symmetry, we immediately find that eTRS is broken and spin-degeneracy is lifted.
For the rest of this work, we study the universal properties of the AFM superconductivity considering SP only. In the case of ASOC, since the details of $\textbf{v}$ significantly affect the symmetry of the system, more specific information from materials is necessary to examine its influence. The final form of the normal state Hamiltonian is then given by $\mathcal{H}(\textbf{k})=\mathcal{H}_{0}(\textbf{k})+\epsilon_{\textrm{sp}}\sigma_{0}\tau_{z}$. Fig. \ref{afmlattice} shows spin expectation values of the eigenstates of $\mathcal{H}(\textbf{k})$ on the Fermi surfaces for $\textbf{m}=(m,0,0)$ without $\epsilon_{\textrm{sp}}$ (Fig. \ref{afmlattice} (c)) and with a finite $\epsilon_{\textrm{sp}}$ (Fig. \ref{afmlattice} (d)). In contrast to the spin-degenerate Fermi surface in Fig. \ref{afmlattice} (c), each Fermi surface in Fig. \ref{afmlattice} (d) is indeed spin-polarized.

\textit{Group theoretical classification of pairing functions.---}
To describe superconductivity, we consider short-ranged density-density interactions
\begin{align}\label{Hint}
H_{\textrm{int}}&=-U\int d\textbf{r}\sum_{l=A,B}n_{l\uparrow}(\textbf{r})n_{l\downarrow}(\textbf{r})\nonumber\\
&\ \qquad-V\int d\textbf{r}\sum_{l\neq l'}\sum_{\sigma\sigma'}\sum_{i}n_{l\sigma}(\textbf{r})n_{l'\sigma'}(\textbf{r}+\bm{\delta}_{i}),
\end{align}
where $U$ and $V$ denote the on-site interaction and the NN interaction strength, respectively.
We take the Nambu basis in the form of $\Psi_{\textbf{k}}^{\dagger}=(\textbf{c}_{\textbf{k}}^{\dagger},\textbf{c}_{-\textbf{k}})$, and write the Bogoliubov-de Gennes (BdG) Hamiltonian as $\sum_{\textbf{k}}\Psi_{\textbf{k}}^{\dagger}\mathcal{H}_{\textrm{BdG}}(\textbf{k})\Psi_{\textbf{k}}$ with
\begin{equation}
\mathcal{H}_{\textrm{BdG}}(\textbf{k})=
\begin{pmatrix}
\mathcal{H}(\textbf{k})-\mu	& \Delta(\textbf{k})\\
\Delta^{\dagger}(\textbf{k})	& -\mathcal{H}^{T}(-\textbf{k})+\mu
\end{pmatrix},
\end{equation}
where $\mu$ and $\Delta(\textbf{k})$ denote the chemical potential and mean-field pairing interaction, respectively.
In general, it is able to express the pairing function in the basis of $\tilde{\sigma}_{i}\tau_{j}$'s as $\Delta(\textbf{k})=\sum_{ij}f_{ij}(\textbf{k})\tilde{\sigma}_{i}\tau_{j}$ where $\tilde{\sigma}_{i}\equiv\sigma_{i}(i\sigma_{y})$. Due to the fermionic statistics, $\Delta(\textbf{k})$ must satisfy $\Delta(\textbf{k})=-\Delta^{T}(-\textbf{k})$. Thus, $f_{ij}(\textbf{k})$ has to be an even function of $\textbf{k}$ for $(i,j)=(0,0)$, $(0,x)$, $(0,z)$, $(x,y)$, $(y,y)$, and $(z,y)$, while it has to be an odd function of $\textbf{k}$ otherwise. Next, following the Sigrist-Ueda method~\cite{sigrist1991phenomenological}, we classify possible pairing functions that can arise from $H_{\textrm{int}}$ by the irreducible representations (IRs) of the point group for two representative collinear AFM structures with high symmetry: out-of-plane AFM (O-AFM) ordering along $[001]$ direction, and in-plane AFM (I-AFM) ordering along $[100]$ direction. 
In the presence of the O-AFM (I-AFM) ordering $\textbf{m}=(0,0,m)$ ($(m,0,0)$) together with SP, the system belongs to $C_{4h}^{z}$ ($C_{2h}^{x}$) point group, whose principal axis is the $z$-axis ($x$-axis). 
Hearafter, we denote the $\gamma$-th gap function that belongs to the $\Gamma$ IR by $[f(\textbf{k})\sigma\tau]_{\gamma}^{\Gamma}$.
As each gap function represents an independent pairing channel, we define a corresponding order parameter as $\Delta_{\gamma}^{\Gamma}\equiv-V_{\gamma}^{\Gamma}\langle \textbf{c}_{\textbf{k}}^{T}[f(\textbf{k})\sigma\tau]_{\gamma}^{\Gamma*}\textbf{c}_{\textbf{k}}\rangle$ where $V_{\gamma}^{\Gamma}=U~(V)$ for intra-sublattice (inter-sublattice) channels. Then a general expression of a pairing potential that belongs to the $\Gamma$ representation reads $\Delta^{\Gamma}(\textbf{k})=\sum_{\gamma}\Delta_{\gamma}^{\Gamma}[f(\textbf{k})\sigma\tau]_{\gamma}^{\Gamma}$ and the corresponding BdG Hamiltonian is denoted by $\mathcal{H}_{\textrm{BdG}}^{\Gamma}(\textbf{k})$. 
In the weak-pairing limit, the magnitude of the superconducting gap is approximately given by diagonal elements of $\Delta^{\Gamma}(\textbf{k})$, projected onto the band basis of $\mathcal{H}(\textbf{k})$ on the Fermi surface of the normal state. In Table \ref{1}, we summarize the result of the group theoretical classification and the gap structure analysis for various $\Gamma$s and $\gamma$s.

{
	\renewcommand{\arraystretch}{1.5}
	\begin{table}[h!]
		\caption{Superconducting gap structures (GS) of various pairing channels. FG indicates that the bulk is fully gapped. NP indicates that pairs of nodal points appear. NG means there is no superconducting gap.} \label{1}
		\begin{tabular}{ccccccccc}
			\hline\hline
			& AFM	& 	& $\Delta_{\gamma}^{\Gamma}$	& 	& $[f(\textbf{k})\sigma\tau]_{\gamma}^{\Gamma}$	& 	& GS	& \\ \hline
			& \multirow{5}{*}{O-AFM}	& 	& $\Delta_{1}^{A_{u}}$	& 	& $\sin{k_{x}}\tilde{\sigma}_{x}\tau_{x}+\sin{k_{y}}\tilde{\sigma}_{y}\tau_{x}$	& 	& FG \\
			& 	& 	& $\Delta_{2}^{A_{u}}$	& 	& $\sin{k_{x}}\tilde{\sigma}_{y}\tau_{x}-\sin{k_{y}}\tilde{\sigma}_{x}\tau_{x}$	& 	& FG	& \\
			& 	& 	& $\Delta_{1}^{B_{u}}$	& 	& $\sin{k_{x}}\tilde{\sigma}_{x}\tau_{x}-\sin{k_{y}}\tilde{\sigma}_{y}\tau_{x}$	& 	& FG	& \\
			& 	& 	& $\Delta_{2}^{B_{u}}$	& 	& $\sin{k_{x}}\tilde{\sigma}_{y}\tau_{x}+\sin{k_{y}}\tilde{\sigma}_{x}\tau_{x}$	& 	& FG	& \\
			& 	& 	& Others	& 	& 	& 	& NG	& \\ \hline
			& \multirow{5}{*}{I-AFM}	& 	& $\Delta_{1}^{A_{u}}$	& 	& $\sin{k_{y}}\tilde{\sigma}_{y}\tau_{x}$	& 	& NP	& \\
			& & 	& $\Delta_{2}^{A_{u}}$	& 	& $\sin{k_{y}}\tilde{\sigma}_{z}\tau_{x}$	& 	& NP	& \\
			& & 	& $\Delta_{1}^{B_{u}}$	& 	& $\sin{k_{x}}\tilde{\sigma}_{y}\tau_{x}$	& 	& NP	& \\
			& & 	& $\Delta_{2}^{B_{u}}$	& 	& $\sin{k_{x}}\tilde{\sigma}_{z}\tau_{x}$	& 	& NP	& \\
			& & 	& Others	& 	& 	& 	& NG \\ \hline\hline
		\end{tabular}
	\end{table}
}

Referring to Table \ref{1}, only the gap functions in the $A_{u}$ and $B_{u}$ IRs can open superconducting gap on the Fermi surfaces in the weak-pairing limit, while others cannot. It means that, only the odd-parity spin-triplet channels in the $A_{u}$ and $B_{u}$ IRs can contribute to the superconducting instability, for both the O-AFM and the I-AFM cases. The gap structures of $\Delta^{\Gamma}(\textbf{k})$s when there are two Fermi surfaces around the $\Gamma$ point are visualized in Fig. \ref{gapstructure}.
\begin{figure}[h!]
	\begin{center}
		\includegraphics[scale=0.38]{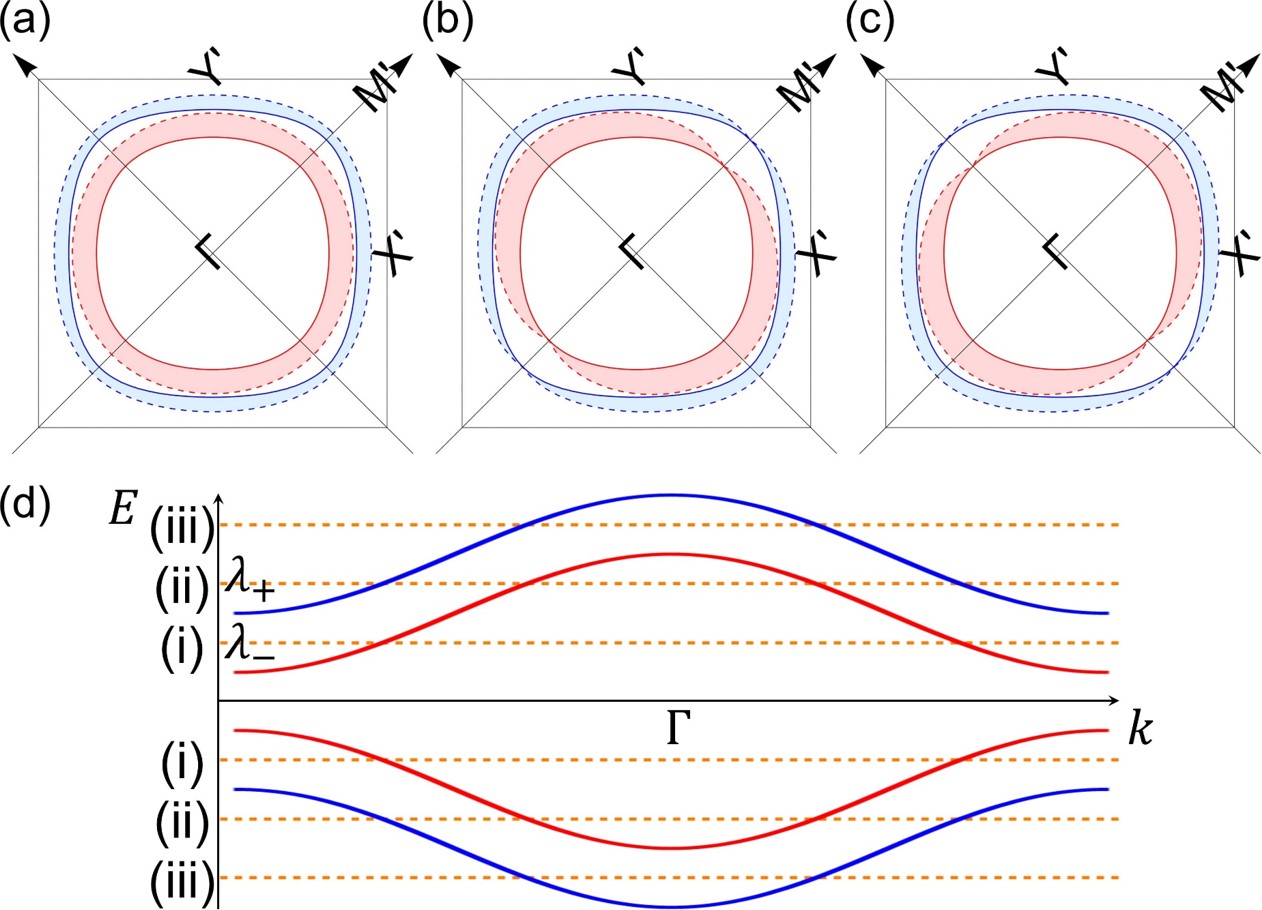}
		\caption{The superconducting gap structures for (a) $\Delta^{A_{u}}(\textbf{k})$ and $\Delta^{B_{u}}(\textbf{k})$ of the O-AFM, (b) $\Delta^{A_{u}}(\textbf{k})$ of the I-AFM, (c) $\Delta^{B_{u}}(\textbf{k})$ of the I-AFM. The red and blue solid lines represent the Fermi surfaces. Distance between a solid line and a dashed line with the same color represents the size of the relevant superconducting gap on the Fermi surface.
			(d) Schematic figure for Case (i), (ii), and (iii) with different Fermi levels shown together with the energy spectrum of the normal state, $\pm\lambda_{\pm}(\textbf{k})$.}\label{gapstructure}
	\end{center}
\end{figure}

\textit{Mean field theory and Ginzburg-Landau free energy.}---To determine the leading instability and find the exact forms of $\Delta^{\Gamma}(\textbf{k})$s in the superconducting states, we proceed to solve the linearized gap equation. The free energy of the system is given by
\begin{align}
F&=\frac{1}{V}\sum_{\gamma}|\Delta_{\gamma}^{\Gamma}|^{2}-\frac{1}{\beta}\sum_{N}\sum_{\textbf{k}n}\ln{\left[\omega_{N}^{2}+\xi_{n}^{2}(\textbf{k})\right]},
\end{align}
where $\omega_{N}$ is the $N$-th fermionic Matsubara frequency, and $\xi_{n}(\textbf{k})$ is the negative eigenvalue of $\mathcal{H}_{\textrm{BdG}}^{\Gamma}(\textbf{k})$ with $n$-th smallest absolute value. The range of the summation over $n$ changes depending on $\mu$, because we are interested only in the energy bands which cross the Fermi level. Thus, we consider three different cases: Case (i) $\min\lambda_{-}(\textbf{k})<|\mu|<\min\lambda_{+}(\textbf{k})$, Case (ii) $\min\lambda_{+}(\textbf{k})<|\mu|<\max\lambda_{-}(\textbf{k})$, and Case (iii) $\max\lambda_{-}(\textbf{k})<|\mu|<\max\lambda_{+}(\textbf{k})$, where $\lambda_{\pm}=\sqrt{(m\pm\epsilon_{\textrm{sp}})^{2}+\epsilon_{\textrm{nn}}^{2}(\textbf{k})}$. Here $\min\lambda$ ($\max\lambda$) denotes the mininum (maximum) value of $\lambda$. 
In Case (i) (Case (iii)), only $\xi_{1}$ ($\xi_2$) is included in the summation, while in Case (ii), $\xi_{1}$ and $\xi_{2}$ are included (see Fig. \ref{gapstructure} (d)).

Using polar forms of the complex numbers $\Delta_{\gamma}^{\Gamma}=|\Delta_{\gamma}^{\Gamma}|e^{i\theta_{\gamma}^{\Gamma}}$, the equilibrium conditions are given by
\begin{align}\label{angleminimum}
\frac{\partial F}{\partial \theta_{\gamma}^{\Gamma}}&=-\frac{1}{\beta}\sum_{N}\sum_{\textbf{k}n}\frac{\partial\xi_{n}^{2}(\textbf{k})/\partial\theta_{\gamma}^{\Gamma}}{\omega_{N}^{2}+\xi_{n}^{2}(\textbf{k})}=0,\\
\frac{\partial F}{\partial |\Delta_{\gamma}^{\Gamma}|}&=\frac{2}{V}|\Delta_{\gamma}^{\Gamma}|-\frac{1}{\beta}\sum_{N}\sum_{\textbf{k}n}\frac{\partial\xi_{n}^{2}(\textbf{k})/\partial|\Delta_{\gamma}^{\Gamma}|}{\omega_{N}^{2}+\xi_{n}^{2}(\textbf{k})}=0,\label{deltaminimum}
\end{align}
where Eq. (\ref{deltaminimum}) is the so-called linearized gap equation.
For $\Gamma=A_{u}$ and $B_{u}$, our model gives that both $\partial F/\partial \theta_{1}^{\Gamma}$ and $\partial F/\partial \theta_{2}^{\Gamma}$ are proportional to $|\Delta_{1}^{\Gamma}||\Delta_{2}^{\Gamma}|\cos{(\theta_{1}^{\Gamma}-\theta_{2}^{\Gamma})}$, implying that the free energy is minimized when either one of the two order parameters vanishes, or $\cos{(\theta_{1}^{\Gamma}-\theta_{2}^{\Gamma})}=0$ (i.e. $|\Delta_{1}^{\Gamma}|=\pm i|\Delta_{2}^{\Gamma}|$). 
However, when $|\Delta_{1}^{\Gamma}|=\pm i|\Delta_{2}^{\Gamma}|$, the pairing interaction can induce the gap only on one of the two possible Fermi surfaces. Thus, $(\Delta_{1}^{\Gamma},\Delta_{2}^{\Gamma})=(\Delta,0)$ or $(0,\Delta)$ is favored for Case (ii), while one of the two solutions $(\Delta_{1}^{\Gamma},\Delta_{2}^{\Gamma})=\Delta(i,1)$ and $\Delta(1,i)$ is favored for Case (i) and (iii). The transition temperature for each case can be calculated by solving Eq. (\ref{deltaminimum}). However, we note that our model does not have enough anisotropy to differentiate the transition temperatures of the superconducting states in the $A_{u}$ and $B_{u}$ IRs, unless extra perturbations allowed by the symmetry enter the Hamiltonian. For example, in the I-AFM case, the two representations can be distinguished if the hopping constants along the $x$ and $y$ directions are different.

\textit{Topological Superconductivity (TSC).}---Odd-parity pairings play a key role in TSC in centrosymmetric systems~\cite{fu2010odd,sato2010topological,nakosai2012topological}. Here we study the topological properties of the odd-parity superconducting states in the $A_{u}$ and $B_{u}$ IRs, obtained above. Both the O-AFM and the I-AFM superconductors belong to the $D$ symmetry class in the Altland-Zirnbauer (AZ) classification table~\cite{altland1997nonstandard,kitaev2009periodic,chiu2016classification,schnyder2008classification}. However, the quasiparticle spectrum of the O-AFM superconductor is fully gapped, while that of the I-AFM superconductor has gapless nodes. Thus, we treat the two cases separately.

We first consider the O-AFM case.
To check the topological properties of the O-AFM superconductors, we calculate the Wilson loop eigenvalue spectrum of the occupied bands of $\mathcal{H}_{\textrm{BdG}}^{\Gamma}(\textbf{k})$~\cite{fidkowski2011model,alexandradinata2014wilson}. Since the spin-up and spin-down sectors of $\mathcal{H}_{\textrm{BdG}}^{\Gamma}(\textbf{k})$ are totally decoupled, $\mathcal{H}_{\textrm{BdG}}^{\Gamma}(\textbf{k})$ can be reduced into two blocks as
\begin{align}\label{blocks}
\mathcal{H}_{\textrm{BdG}}^{\Gamma}(\textbf{k})=
\begin{pmatrix}
\mathcal{H}_{\textrm{BdG}}^{\Gamma,\uparrow\uparrow}(\textbf{k})	& 0 \\
0	& \mathcal{H}_{\textrm{BdG}}^{\Gamma,\downarrow\downarrow}(\textbf{k})
\end{pmatrix}.
\end{align}
We find that $\mathcal{H}_{\textrm{BdG}}^{\Gamma,\downarrow\downarrow}(\textbf{k})$ ($\mathcal{H}_{\textrm{BdG}}^{\Gamma,\uparrow\uparrow}(\textbf{k})$) has a non-trivial winding in its Wilson loop spectrum in Case (i) (Case (iii)), while the other block does not. It indicates that the occupied bands of $\mathcal{H}_{\textrm{BdG}}^{\Gamma,\downarrow\downarrow}(\textbf{k})$ ($\mathcal{H}_{\textrm{BdG}}^{\Gamma,\uparrow\uparrow}(\textbf{k})$) has a non-zero Chern number. 
On the other hand, both blocks carry finite Chern numbers but with opposite signs in Case (ii), making the total Chern number of the system zero. This result agrees well with the Fu-Berg-Sato criteria for diagnosing band topology in centrosymzmetric systems~\cite{fu2010odd,sato2010topological}. We note that $\mathcal{C}_{\textrm{BdG}}^{A_{u},\sigma\sigma}=-\mathcal{C}_{\textrm{BdG}}^{B_{u},\sigma\sigma}$ where $\mathcal{C}_{\textrm{BdG}}^{\Gamma,\sigma\sigma}$ denotes the Chern number carried by the occupied bands of $\mathcal{H}_{\textrm{BdG}}^{\Gamma,\sigma\sigma}$ ($\sigma\sigma=\uparrow\uparrow$ or $\downarrow\downarrow$). 
Corresponding to the nontrivial bulk topology, gapless modes appear on the edges of the system~\cite{laughlin1981quantized,halperin1982quantized,hatsugai1993chern,hatsugai1993edge,qi2006general,fukui2012bulk}. 
To confirm this, we have performed the finite-size tight-binding model calculation for the system in a ribbon geometry. Fig. \ref{Fig4} (a) displays the result for $\Gamma=A_{u}$ in Case (ii).
In fact, the two blocks in Eq. (\ref{blocks}) are nothing but the mirror $M_z$ invariant sectors with the eigenvalues $\pm1$, where $M_z:(x,y,z)\rightarrow(x,y,-z)$. 
Thus, the Case (ii) O-AFM superconducting state can be interpreted as a mirror Chern superconductor, whose $M_z=\pm1$ eigensectors are analogous to chiral $p$-wave superconductors with the Chern number $\pm1$~\cite{sato2017topological}. 
In Case (i) or (iii) with a nonzero Chern number, the thermal Hall conductance and the spin Nernst conductance are quantized due to the spin-polarized edge channels~\cite{imai2017thermal}. 
In Case (ii), although the net thermal Hall conductivity vanishes, we expect a finite spin Nernst conductance, since the edge quasiparticles with opposite spins propagate in opposite directions. 

If the system preserves the $M_{z}$ symmetry, the above discussion is still valid even in the presence of SOC. 
When $M_{z}$ is broken, the two edge channels of Case (ii) superconductor mix and open a gap (Fig. \ref{Fig4} (b)), leading to a second-order TSC protected by inversion symmetry~\cite{khalaf2018higher,ahn2020higher,hwang2019fragile} (see Supplemental Materials (SM)).
\begin{figure}[h!]
	\begin{center}
		\includegraphics[scale=0.42]{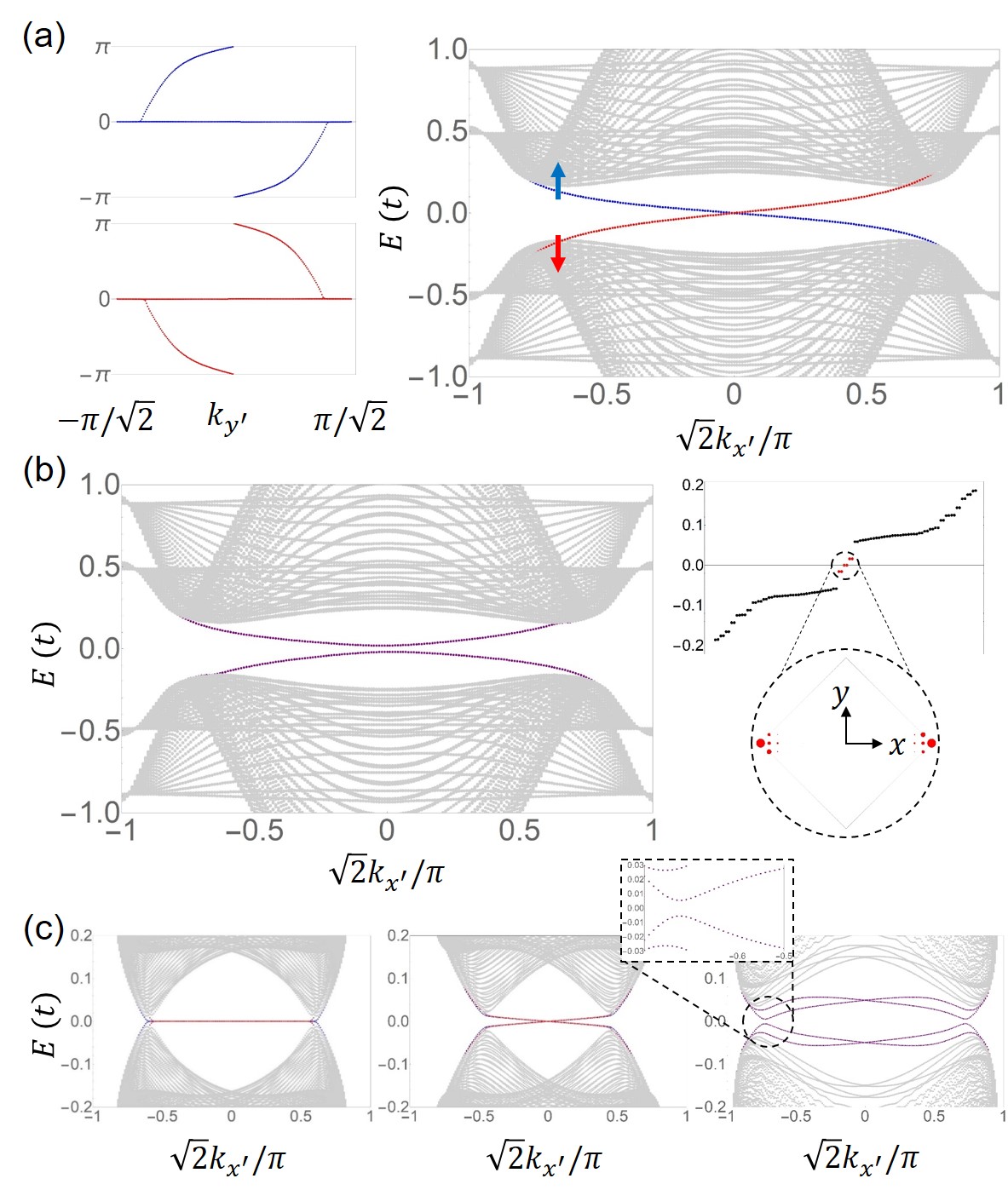}
		\caption{
			Edge spectra for Case (ii) O-AFM superconductor (a,b) and I-AFM superconductor (c).
			(a) Left panel: The Wilson loop eigenvalue spectra of $\mathcal{H}_{\textrm{BdG}}^{A_{u},\uparrow\uparrow}(\textbf{k})$ (upper) and $\mathcal{H}_{\textrm{BdG}}^{A_{u},\downarrow\downarrow}(\textbf{k})$ (lower) in the Case (ii) O-AFM superconductor. The two block Hamiltonians have the opposite Chern numbers. 
			Right panel: Energy spectrum of a finite-size system with a ribbon geometry extended along $x'(1,1,0)$ direction while having a finite length along $y'(-1,1,0)$ direction. The chiral edge modes originating from spin-up (blue) and spin-down (red) bands. Eigenstates localized on only one side of the two edges are shown. 
			(b) The edge gap opened by the $M_{z}$ breaking SOC (left) and the localized charge distribution of the in-gap states near the zero energy (right). 
			(c) The two Fermi arc-like states connecting the two pairs of nodes projected onto the edge momentum space of the Case (ii) I-AFM superconductor (left). When additional pairing terms respecting symmetries are included, the bulk spectrum becomes fully gapped, but there remain the zero-energy edge states at $k_{x'}=0$ (middle). In the presence of SOC that breaks the spin-reflection symmetry, the edge states are also gapped and the system becomes a second-order TSC (right).}\label{Fig4}
	\end{center}
\end{figure}

In the I-AFM case, as $\Delta^{A_{u}}(\textbf{k})\propto\sin{k_{x}}$ ($\Delta^{B_{u}}(\textbf{k})\propto\sin{k_{y}}$), there are nodes at the points where the $k_{y}$ ($k_{x}$) axis intersects the Fermi surfaces. However, these nodes are not topologically protected, but rather accidental. i.e. a random perturbation that respects the symmetries can open the bulk gap (Fig. 4 (c)). In the present case, the spin-up and spin-down subspaces are well-separated due to the spin-reflection $S_{x}$ symmetry where $S_{x}=i\sigma_{x}$. 
Thus, when the bulk becomes fully gapped, the I-AFM superconductor becomes a spin-chiral (chiral) TSC for Case (ii) (Case (i) and (iii)), similar to the O-AFM cases.
When SOC exists, the edge states of the spin-chiral TSC are gapped due to the $S_x$ symmetry breaking. Interestingly, the resulting gapped phase turns out to be a second-order TSC protected by inversion symmetry~\cite{khalaf2018higher,ahn2020higher}.
The edge states of various TSCs for I-AFM superconductors are shown in Fig. \ref{Fig4} where we solve the Hamiltonian of the finite-size system in the same geometry as for the O-AFM case.

\textit{Conclusions.}--We have explored the possible superconducting states in two-dimensional AFM metallic systems with SP. 
Despite the simplicity of the model considered,
the resulting superconducting states exhibit rich intriguing physical properties.
Depending on the patterns of background magnetic orderings, which determine the magnetic space group symmetry of the system, the itinerant electrons experience different pairing instability, leading to various odd-parity antiferromagnetic superconductors with distinct topological properties.
We believe that the superconductivity of AFM metallic systems provides a promising platform for searching new types of TSCs.
It is worth noting that the Cooper pairs in our model have a purely odd parity, since the system keeps the global inversion symmetry, distinct from parity-mixed spin-triplet superconductors featured in previous studies focused on noncentrosymmetric antiferromagnets~\cite{qi2017coexistence,fujimoto2006emergent,sumita2018unconventional}.

Finally, we propose that a class of materials called double perovskites with a formula $\textrm{A}_{2}\textrm{BB}'\textrm{O}_{6}$~\cite{Saha_Dasgupta_2020} is a promising candidate where our theoretical idea can be tested. 
The double perovskites share the same lattice structure as conventional centrosymmetric perovskites. However, composed of two different species of transition metals B and $\textrm{B}'$, they can be seen as lattice systems with SP.
Moreover, a few double perovskites such as $\textrm{Sr}_{2}\textrm{FeMo}\textrm{O}_{6}$ exhibit antiferromagnetic metallicity~\cite{kobayashi1998room,sanyal2009evidence}.
Since the layered structure of double perovskites is also available by partitioning the bulk crystal with organic cations~\cite{connor2018layered}, we anticipate that our model, the square lattice antiferromagnet with SP on its sublattices, can be realized. More specifically, if a layer of double perovskite, sandwiched by two layers of superconducting materials, acquires pairing interaction via the proximity effect, it may be a feasible playground where our theory can be applied to.

\begin{acknowledgments}
	We thank Junyeong Ahn, Sungjoon Park, Yoonseok Hwang, and Se Young Park for helpful discussions. S.L. was supported by IBSR009-D1. 
	B.-J.Y. was supported by the Institute for Basic Science in Korea (Grant No. IBS-R009-D1) and Basic Science Research Program through the National Research Foundation of Korea (NRF) (Grant No. 0426-20200003). This work was supported in part by the U.S. Army Research Office under Grant Number W911NF-18-1-0137.
\end{acknowledgments}

\end{bibunit}

\newpage

\onecolumngrid

\appendix
\begin{bibunit}                         
\renewcommand{\appendixpagename}{\center\large Supplementary Information: Odd-Parity Spin-Triplet Superconductivity in Antiferromagnetic Metals}

\appendixpage

\setcounter{page}{1}
\setcounter{equation}{0}
\setcounter{figure}{0}
\setcounter{table}{0}
\setcounter{equation}{0}

\renewcommand{\theequation}{S\arabic{equation}}
\renewcommand{\thetable}{S\arabic{table}}
\renewcommand{\thefigure}{S\arabic{figure}}
	
	\author{Seung Hun Lee$^{1,2}$}
	\email{sh2lee@snu.ac.kr}
	\author{Bohm Jung Yang$^{1,2,3}$}
	\email{bjyang@snu.ac.kr}
	\affiliation{$^1$Center for Correlated Electron Systems, Institute for Basic Science (IBS), Seoul 08826, Korea\\$^2$Department of Physics and Astronomy, Seoul National University, Seoul 08826, Korea\\$^{3}$Center for Theoretical Physics (CTP), Seoul National University, Seoul 08826, Korea}
	
	\date{\today}
	
	\maketitle
	
	\noindent
	\textbf{Outline}\\
	\indent In this supplementary material, we provide details of discussions made in the main text.
	\\
	\\
	\noindent
	\textbf{Group theoretical classification of pairing functions}\\
	\indent In this section, we list the results of the group theoretical classification of the possible pairing potential terms. In general, the pairing interaction $\Delta(\textbf{k})$ in the BdG Hamiltonian
	\begin{equation}
	\mathcal{H}_{\textrm{BdG}}(\textbf{k})=
	\begin{pmatrix}
	\mathcal{H}(\textbf{k})-\mu	& \Delta(\textbf{k})\\
	\Delta^{\dagger}(\textbf{k})	& -\mathcal{H}^{T}(-\textbf{k})+\mu
	\end{pmatrix}
	\end{equation}
	can be written in the form of $\Delta(\textbf{k})=\sum_{ij}f_{ij}(\textbf{k})\tilde{\sigma}_{i}\tau_{j}\equiv\sum_{ij}f_{ij}(\textbf{k})\sigma_{i}(i\sigma_{y})\tau_{j}$, where $\sigma_{i}$ and $\tau_{j}$ are the pauli matrices acting on the spin space and the sublattice space. Due to the fermionic statistics, $\Delta(\textbf{k})$ must satisfy $\Delta(\textbf{k})=-\Delta^{T}(-\textbf{k})$. Possible pairing functions that can arise from the on-site and NN electron-electron interaction are listed below.
	\begin{align}
	\textrm{Even functions in}\,\textbf{k}:\,&\Delta_{00}\tilde{\sigma}_{0}\tau_{0}, \Delta_{0x}\tilde{\sigma}_{0}\tau_{x}, \Delta_{0z}\tilde{\sigma}_{0}\tau_{z},\nonumber\\
	&\Delta_{xy}^{1}\cos{k_{x}}\tilde{\sigma}_{x}\tau_{y}, \Delta_{yy}^{1}\cos{k_{x}}\tilde{\sigma}_{y}\tau_{y}, \Delta_{zy}^{1}\cos{k_{x}}\tilde{\sigma}_{z}\tau_{y},\nonumber\\ &\Delta_{xy}^{2}\cos{k_{y}}\tilde{\sigma}_{x}\tau_{y},  \Delta_{yy}^{2}\cos{k_{y}}\tilde{\sigma}_{y}\tau_{y},  \Delta_{zy}^{2}\cos{k_{y}}\tilde{\sigma}_{z}\tau_{y}.
	\end{align}
	\begin{align}
	\textrm{Odd functions in}\,\textbf{k}:\,&\Delta_{0y}^{1}\sin{k_{x}}\tilde{\sigma}_{0}\tau_{y},gyd \Delta_{0y}^{2}\sin{k_{y}}\tilde{\sigma}_{0}\tau_{y}\nonumber\\
	&\Delta_{x0}^{1}\sin{k_{x}}\tilde{\sigma}_{x}\tau_{0}, \Delta_{y0}^{1}\sin{k_{x}}\tilde{\sigma}_{y}\tau_{0}, \Delta_{z0}^{1}\sin{k_{x}}\tilde{\sigma}_{z}\tau_{0},\nonumber\\ &\Delta_{x0}^{2}\sin{k_{y}}\tilde{\sigma}_{x}\tau_{0}, \Delta_{y0}^{2}\sin{k_{y}}\tilde{\sigma}_{y}\tau_{0}, \Delta_{z0}^{2}\sin{k_{y}}\tilde{\sigma}_{z}\tau_{0},\nonumber\\
	&\Delta_{xx}^{1}\sin{k_{x}}\tilde{\sigma}_{x}\tau_{x}, \Delta_{yx}^{1}\sin{k_{x}}\tilde{\sigma}_{y}\tau_{x}, \Delta_{zx}^{1}\sin{k_{x}}\tilde{\sigma}_{z}\tau_{x},\nonumber\\
	&\Delta_{xx}^{2}\sin{k_{y}}\tilde{\sigma}_{x}\tau_{x}, \Delta_{yx}^{2}\sin{k_{y}}\tilde{\sigma}_{y}\tau_{x}, \Delta_{zx}^{2}\sin{k_{y}}\tilde{\sigma}_{z}\tau_{x},\nonumber\\
	&\Delta_{xz}^{1}\sin{k_{x}}\tilde{\sigma}_{x}\tau_{z}, \Delta_{yz}^{1}\sin{k_{x}}\tilde{\sigma}_{y}\tau_{z}, \Delta_{zz}^{1}\sin{k_{x}}\tilde{\sigma}_{z}\tau_{z},\nonumber\\
	&\Delta_{xz}^{2}\sin{k_{y}}\tilde{\sigma}_{x}\tau_{z}, \Delta_{yz}^{2}\sin{k_{y}}\tilde{\sigma}_{y}\tau_{z}, \Delta_{zz}^{2}\sin{k_{y}}\tilde{\sigma}_{z}\tau_{z}.
	\end{align}
	\indent Next, we classify the pairing functions in Eq. (4) and (5) by the IRs of the point group for two different cases: Out-of-plane antiferromagnetic (OAFM) ordering with SP, and in-plane antiferromagnetic (IAFM) ordering along $[100]$ direction with SP~\cite{sigrist1991phenomenological}.
	\\
	\\
	OAFM\\
	\indent In the OAFM case, the system belongs to $C_{4h}^{z}$ point group. The elements of the $C_{4h}^{z}$ point group such as three rotations around the $z$-axis $C_{4z}$, $C_{2z}$, $C_{4z}^{-1}$, inversion symmetry $P$, and a mirror reflection across the $xy$-plane $\sigma_{xy}$. We can classify the matrix part and the momentum dependent function part as TABLE \ref{1}. It follows from TABLE \ref{1} that every combinations of $\tilde{\sigma}_{i}\tau_{j}$'s and $f_{ij}(\textbf{k})$'s can also be classified by the IRs of the point group (TABLE \ref{2}).
	{
		\renewcommand{\arraystretch}{1.5}
		\begin{table}[h!]
			\parbox{.45\linewidth}{
				\centering
				\caption{} \label{1}
				\begin{tabular}{ccccccccccccccc}
					\hline\hline
					$\tilde{\sigma}_{i}\tau_{j}$ part&  & $E$  &  & $C_{4z}$                      &  & $C_{2z}$ &  & $C_{4z}^{-1}$                 &  & $i$  &  & $\sigma_{xy}$ &        \\ \hline
					$\tilde{\sigma}_{0,z}\tau_{j}$ &  & $+1$ &  & $+1$                          &  & $+1$     &  & $+1$                          &  & $+1$ &  & $+1$          &   \\
					$\tilde{\sigma}_{x}\tau_{j}$   &  & $+1$ &  & $\tilde{\sigma}_{y}\tau_{i}$  &  & $-1$     &  & $-\tilde{\sigma}_{y}\tau_{j}$ &  & $+1$ &  & $-1$          &   \\
					$\tilde{\sigma}_{y}\tau_{j}$   &  & $+1$ &  & $-\tilde{\sigma}_{x}\tau_{i}$ &  & $-1$     &  & $\tilde{\sigma}_{x}\tau_{j}$  &  & $+1$ &  & $-1$          &   \\
					\hline
					$f_{ij}(\textbf{k})$ part &  & $E$  &  & $C_{4z}$       &  & $C_{2z}$ &  & $C_{4z}^{-1}$  &  & $i$  &  & $\sigma_{xy}$ &                                                \\ \hline
					$\cos{k_{x}}+\cos{k_{y}}$ &  & $+1$ &  & $+1$           &  & $+1$     &  & $+1$           &  & $+1$ &  & $+1$          &                                           \\
					$\cos{k_{x}}-\cos{k_{y}}$ &  & $+1$ &  & $-1$           &  & $+1$     &  & $-1$           &  & $+1$ &  & $+1$          &                                           \\
					$\sin{k_{x}}$             &  & $+1$ &  & $\sin{k_{y}}$  &  & $-1$     &  & $-\sin{k_{y}}$ &  & $-1$ &  & $+1$          &   \\
					$\sin{k_{y}}$             &  & $+1$ &  & $-\sin{k_{x}}$ &  & $-1$     &  & $\sin{k_{x}}$  &  & $-1$ &  & $+1$          &                                           \\ \hline\hline
				\end{tabular}
			}
			\hfill
			\parbox{.45\linewidth}{
				\centering
				\caption{} \label{2}
				\begin{tabular}{ccccc}
					\hline\hline
					& & singlet                                               &  & triplet                                                                                                                                                                                                                                                                                   \\ \hline
					$A_{g}$ & \vline & $\tilde{\sigma}_{0}\tau_{0,x,z}$                      & \vline & $(\cos{k_{x}}+\cos{k_{y}})\tilde{\sigma}_{z}\tau_{y}$                                                                                                                                                                                                                                     \\ \hline
					$B_{g}$ & \vline &                                                       & \vline & $(\cos{k_{x}}-\cos{k_{y}})\tilde{\sigma}_{z}\tau_{y}$                                                                                                                                                                                                                                     \\ \hline
					$E_{g}$ & \vline &                                                       & \vline & $(\cos{k_{x}}\pm\cos{k_{y}})\lbrace\tilde{\sigma}_{x}\tau_{y},\tilde{\sigma}_{y}\tau_{y}\rbrace$
					\\ \hline
					$A_{u}$ & \vline &  								                       & \vline &
					\begin{tabular}[c]{@{}c@{}}
						$\sin{k_{x}}\tilde{\sigma}_{x}\tau_{0,x,z}+\sin{k_{y}}\tilde{\sigma}_{y}\tau_{0,x,z}$	\\
						$\sin{k_{x}}\tilde{\sigma}_{y}\tau_{0,x,z}-\sin{k_{y}}\tilde{\sigma}_{x}\tau_{0,x,z}$
					\end{tabular}                                                                                                                                                                                                                                     \\ \hline
					$B_{u}$ & \vline &                                                       & \vline & 
					\begin{tabular}[c]{@{}c@{}}
						$\sin{k_{x}}\tilde{\sigma}_{x}\tau_{0,x,z}-\sin{k_{y}}\tilde{\sigma}_{y}\tau_{0,x,z}$	\\
						$\sin{k_{x}}\tilde{\sigma}_{y}\tau_{0,x,z}+\sin{k_{y}}\tilde{\sigma}_{x}\tau_{0,x,z}$
					\end{tabular}                                                                                                                                                                                            \\ \hline
					$E_{u}$ & \vline & $\lbrace\sin{k_{x}},\sin{k_{y}}\rbrace\tilde{\sigma}_{0}\tau_{y}$ & \vline & $\lbrace\sin{k_{x}},\sin{k_{y}}\rbrace\tilde{\sigma}_{z}\tau_{0,x,z}$ \\ \hline\hline
				\end{tabular}
			}
		\end{table}
	}
	\newpage
	\noindent IAFM\\
	\indent In the IAFM case, the system belongs to $C_{2h}^{x}$ point group. Again, all the mirror and rotation symmetries do not interchange the sublattices. Repeating the similar process as for the OAFM case, we obtain TABLE \ref{3} and TABLE \ref{4}.
	{
		\renewcommand{\arraystretch}{1.5}
		\begin{table}[h!]
			\parbox{.45\linewidth}{
				\centering
				\caption{} \label{3}
				\begin{tabular}{ccccccccccc}
					\hline\hline
					$\tilde{\sigma}_{i}\tau_{j}$ part &  & $E$  &  & $C_{2x}$ &  & $i$  &  & $\sigma_{yz}$ &  \\ \hline
					$\tilde{\sigma}_{0,x}\tau_{j}$ &  & $+1$ &  & $+1$     &  & $+1$ &  & $+1$          &  \\
					$\tilde{\sigma}_{y,z}\tau_{j}$     &  & $+1$ &  & $-1$     &  & $+1$ &  & $-1$          &  \\ 
					\hline
					$f_{ij}(\textbf{k})$ part	&  & $E$  &  & $C_{2x}$ &  & $i$  &  & $\sigma_{yz}$ &  \\ \hline
					$\cos{k_{x,y}}$ &  & $+1$ &  & $+1$     &  & $+1$ &  & $+1$          &  \\
					$\sin{k_{x}}$ &  & $+1$ &  & $+1$     &  & $-1$ &  & $-1$          &  \\
					$\sin{k_{y}}$ &  & $+1$ &  & $-1$     &  & $-1$ &  & $+1$          &  \\ \hline\hline
				\end{tabular}
			}
			\hfill
			\parbox{.45\linewidth}{
				\centering
				\caption{} \label{4}
				\begin{tabular}{ccccc}
					\hline\hline
					IR		&  & singlet                                 &  & triplet                                                                                    \\ \hline
					$A_{g}$ & \vline & $\tilde{\sigma}_{0}\tau_{0,x,z}$        & \vline & $\cos{k_{x,y}}\tilde{\sigma}_{x}\tau_{y}$                                                \\ \hline
					$B_{g}$ & \vline &                                         & \vline & $\cos{k_{x,y}}\tilde{\sigma}_{y,z}\tau_{y}$                                                  \\ \hline
					$A_{u}$ & \vline & $\sin{k_{x}}\tilde{\sigma}_{0}\tau_{y}$ & \vline &
					\begin{tabular}[c]{@{}c@{}}
						$\sin{k_{y}}\tilde{\sigma}_{y,z}\tau_{0,x,z}$	\\
						$\sin{k_{x}}\tilde{\sigma}_{x}\tau_{0,x,z}$
					\end{tabular} \\ \hline
					$B_{u}$ & \vline & $\sin{k_{y}}\tilde{\sigma}_{0}\tau_{y}$ & \vline &
					\begin{tabular}[c]{@{}c@{}}
						$\sin{k_{x}}\tilde{\sigma}_{y, z}\tau_{0,x,z}$	\\
						$\sin{k_{y}}\tilde{\sigma}_{x}\tau_{0,x,z}$
					\end{tabular} \\ \hline\hline
				\end{tabular}
			}
		\end{table}
	}
	\\
	\\
	\noindent
	\textbf{Matrix transformation of pairing to the band basis}\\
	\indent In this section, we present the exact forms of the transformation matrices used in the superconducting gap structure analysis.
	\\
	\\
	OAFM\\
	\indent The normal state Hamiltonian for the OAFM case reads
	\begin{equation}
	\mathcal{H}(\textbf{k})=\epsilon_{\textrm{nn}}(\textbf{k})\sigma_{0}\tau_{x}+m\sigma_{z}\tau_{z}+\epsilon_{\textrm{sp}}\sigma_{0}\tau_{z}-\mu\sigma_{0}\tau_{0}.
	\end{equation}
	The transformation matrix $U(\textbf{k})$ is given by
	\begin{align}
	U^{T}(\textbf{k})=\begin{pmatrix}
	-\frac{\epsilon_{\textrm{nn}}(\textbf{k})}{2\lambda_{+}(\textbf{k})}	& \frac{\epsilon_{\textrm{nn}}^{2}(\textbf{k})}{\epsilon_{\textrm{nn}}^{2}(\textbf{k})+\left(m+\epsilon_{\textrm{sp}}-\lambda_{+}(\textbf{k})\right)^{2}}	& 0	& 0\\
	0	& 0	&  -\frac{\epsilon_{\textrm{nn}}(\textbf{k})}{2\lambda_{-}(\textbf{k})}	& \frac{\epsilon_{\textrm{nn}}^{2}(\textbf{k})}{\epsilon_{\textrm{nn}}^{2}(\textbf{k})+\left(m-\epsilon_{\textrm{sp}}+\lambda_{-}(\textbf{k})\right)^{2}}\\
	0	& 0	& \frac{\epsilon_{\textrm{nn}}(\textbf{k})}{2\lambda_{-}(\textbf{k})}	& \frac{\epsilon_{\textrm{nn}}^{2}(\textbf{k})}{\epsilon_{\textrm{nn}}^{2}(\textbf{k})+\left(m-\epsilon_{\textrm{sp}}-\lambda_{-}(\textbf{k})\right)^{2}}\\
	\frac{\epsilon_{\textrm{nn}}(\textbf{k})}{2\lambda_{+}(\textbf{k})}	& \frac{\epsilon_{\textrm{nn}}^{2}(\textbf{k})}{\epsilon_{\textrm{nn}}^{2}(\textbf{k})+\left(m+\epsilon_{\textrm{sp}}+\lambda_{+}(\textbf{k})\right)^{2}}	& 0	& 0
	\end{pmatrix},
	\end{align}
	where $\lambda_{\pm}(\textbf{k})=\sqrt{(m\pm\epsilon_{\textrm{sp}})^{2}+\epsilon_{\textrm{nn}}^{2}(\textbf{k})}$.
	\\
	\\
	IAFM\\
	\indent The normal state Hamiltonian for the IAFM case reads
	\begin{equation}
	\mathcal{H}(\textbf{k})=\epsilon_{\textrm{nn}}(\textbf{k})\sigma_{0}\tau_{x}+m\sigma_{x}\tau_{z}+\epsilon_{\textrm{sp}}\sigma_{0}\tau_{z}-\mu\sigma_{0}\tau_{0}.
	\end{equation}
	The transformation matrix $U(\textbf{k})$ is given by
	\begin{align}
	&U^{T}(\textbf{k})=\nonumber\\
	&\quad\begin{pmatrix}
	\frac{\epsilon_{\textrm{nn}}(\textbf{k})(m+\epsilon_{\textrm{sp}}-\lambda_{+}(\textbf{k}))}{2\left(\epsilon_{\textrm{nn}}^{2}(\textbf{k})+(m+\epsilon_{\textrm{sp}}-\lambda_{+}(\textbf{k}))^{2}\right)}	& \frac{\epsilon_{\textrm{nn}}^{2}(\textbf{k})}{2\left(\epsilon_{\textrm{nn}}^{2}(\textbf{k})+(m+\epsilon_{\textrm{sp}}-\lambda_{+}(\textbf{k}))^{2}\right)}	& \frac{\epsilon_{\textrm{nn}}(\textbf{k})(m+\epsilon_{\textrm{sp}}-\lambda_{+}(\textbf{k}))}{2\left(\epsilon_{\textrm{nn}}^{2}(\textbf{k})+(m+\epsilon_{\textrm{sp}}-\lambda_{+}(\textbf{k}))^{2}\right)}	& \frac{\epsilon_{\textrm{nn}}^{2}(\textbf{k})}{2\left(\epsilon_{\textrm{nn}}^{2}(\textbf{k})+(m+\epsilon_{\textrm{sp}}-\lambda_{+}(\textbf{k}))^{2}\right)}\\
	\frac{\epsilon_{\textrm{nn}}(\textbf{k})(m-\epsilon_{\textrm{sp}}+\lambda_{-}(\textbf{k}))}{2\left(\epsilon_{\textrm{nn}}^{2}(\textbf{k})+(m-\epsilon_{\textrm{sp}}+\lambda_{-}(\textbf{k}))^{2}\right)}	& -\frac{\epsilon_{\textrm{nn}}^{2}(\textbf{k})}{2\left(\epsilon_{\textrm{nn}}^{2}(\textbf{k})+(m-\epsilon_{\textrm{sp}}+\lambda_{-}(\textbf{k}))^{2}\right)}	& -\frac{\epsilon_{\textrm{nn}}(\textbf{k})(m-\epsilon_{\textrm{sp}}+\lambda_{-}(\textbf{k}))}{2\left(\epsilon_{\textrm{nn}}^{2}(\textbf{k})+(m-\epsilon_{\textrm{sp}}+\lambda_{-}(\textbf{k}))^{2}\right)}	& \frac{\epsilon_{\textrm{nn}}^{2}(\textbf{k})}{2\left(\epsilon_{\textrm{nn}}^{2}(\textbf{k})+(m-\epsilon_{\textrm{sp}}+\lambda_{-}(\textbf{k}))^{2}\right)}\\
	\frac{\epsilon_{\textrm{nn}}(\textbf{k})(m-\epsilon_{\textrm{sp}}-\lambda_{-}(\textbf{k}))}{2\left(\epsilon_{\textrm{nn}}^{2}(\textbf{k})+(m-\epsilon_{\textrm{sp}}-\lambda_{-}(\textbf{k}))^{2}\right)}	& -\frac{\epsilon_{\textrm{nn}}^{2}(\textbf{k})}{2\left(\epsilon_{\textrm{nn}}^{2}(\textbf{k})+(m-\epsilon_{\textrm{sp}}-\lambda_{-}(\textbf{k}))^{2}\right)}	& -\frac{\epsilon_{\textrm{nn}}(\textbf{k})(m-\epsilon_{\textrm{sp}}-\lambda_{-}(\textbf{k}))}{2\left(\epsilon_{\textrm{nn}}^{2}(\textbf{k})+(m-\epsilon_{\textrm{sp}}-\lambda_{-}(\textbf{k}))^{2}\right)}	& \frac{\epsilon_{\textrm{nn}}^{2}(\textbf{k})}{2\left(\epsilon_{\textrm{nn}}^{2}(\textbf{k})+(m-\epsilon_{\textrm{sp}}-\lambda_{-}(\textbf{k}))^{2}\right)}\\
	\frac{\epsilon_{\textrm{nn}}(\textbf{k})(m+\epsilon_{\textrm{sp}}+\lambda_{+}(\textbf{k}))}{2\left(\epsilon_{\textrm{nn}}^{2}(\textbf{k})+(m+\epsilon_{\textrm{sp}}+\lambda_{+}(\textbf{k}))^{2}\right)}	& \frac{\epsilon_{\textrm{nn}}^{2}(\textbf{k})}{2\left(\epsilon_{\textrm{nn}}^{2}(\textbf{k})+(m+\epsilon_{\textrm{sp}}+\lambda_{+}(\textbf{k}))^{2}\right)}	& \frac{\epsilon_{\textrm{nn}}(\textbf{k})(m+\epsilon_{\textrm{sp}}+\lambda_{+}(\textbf{k}))}{2\left(\epsilon_{\textrm{nn}}^{2}(\textbf{k})+(m+\epsilon_{\textrm{sp}}+\lambda_{+}(\textbf{k}))^{2}\right)}	& \frac{\epsilon_{\textrm{nn}}^{2}(\textbf{k})}{2\left(\epsilon_{\textrm{nn}}^{2}(\textbf{k})+(m+\epsilon_{\textrm{sp}}+\lambda_{+}(\textbf{k}))^{2}\right)}
	\end{pmatrix}.
	\end{align}
	\\
	\noindent
	\textbf{Details on the mean-field theory}\\
	\indent In the two-dimensional square lattice, a short ranged density-density interaction Hamiltonian that involves the on-site and NN interaction is given in the form of
	\begin{align}
	H_{\textrm{int}}&=\int d\textbf{r} d\textbf{r}'\sum_{\alpha\beta\mu\nu}\Gamma_{\mu\nu,\beta\alpha}(\textbf{r},\textbf{r}')c_{\textbf{r},\mu}^{\dagger}c_{\textbf{r}',\nu}^{\dagger}c_{\textbf{\textbf{r}}',\beta}c_{\textbf{r},\alpha}\nonumber\\
	&=U\int d\textbf{r}\sum_{l=A,B}n_{l\uparrow}(\textbf{r})n_{l\downarrow}(\textbf{r})+V\int d\textbf{r}\sum_{l\neq l'}\sum_{\sigma\sigma'}\sum_{i}n_{l\sigma}(\textbf{r})n_{l'\sigma'}(\textbf{r}+\bm{\delta}_{i})\nonumber\\
	&=U\int d\textbf{r}\sum_{l=A,B}c_{\textbf{r},l\uparrow}^{\dagger}c_{\textbf{r},l\uparrow}c_{\textbf{r},l\downarrow}^{\dagger}c_{\textbf{r},l\downarrow}\nonumber\\
	&\qquad+V\int d\textbf{r}\sum_{l\neq l'}\sum_{\sigma\sigma'}\sum_{i}c_{\textbf{r},l\sigma}^{\dagger}c_{\textbf{r},l'\sigma'}c_{\textbf{r}+\bm{\delta}_{i},l'\sigma'}^{\dagger}c_{\textbf{r}+\bm{\delta}_{i},l\sigma}\nonumber\\
	&=U\int d\textbf{r}\int d\textbf{r}'\delta(\textbf{r}-\textbf{r}')\sum_{l=A,B}c_{\textbf{r},l\uparrow}^{\dagger}c_{\textbf{r}',l\downarrow}^{\dagger}c_{\textbf{r}',l\downarrow}c_{\textbf{r},l\uparrow}\nonumber\\
	&\qquad+V\int d\textbf{r}\int d\textbf{r}'\sum_{i}\delta(\textbf{r}+\bm{\delta}_{i}-\textbf{r}')\sum_{l\neq l'}\sum_{\sigma\sigma'}c_{\textbf{r},l\sigma}^{\dagger}c_{\textbf{r}',l'\sigma'}^{\dagger}c_{\textbf{r}',l'\sigma'}c_{\textbf{r},l\sigma}\nonumber\\
	&=\frac{U}{2}\int d\textbf{r} d\textbf{r}'\delta(\textbf{r}-\textbf{r}')\sum_{ll'}\delta_{ll'}\sum_{\sigma\sigma'}(1-\delta_{\sigma\sigma'})c_{\textbf{r},l\sigma}^{\dagger}c_{\textbf{r}',l'\sigma'}^{\dagger}c_{\textbf{r}',l'\sigma'}c_{\textbf{r},l\sigma}\nonumber\\
	&\qquad+V\int d\textbf{r} d\textbf{r}'\sum_{i}\delta(\textbf{r}+\bm{\delta}_{i}-\textbf{r}')\sum_{ll'}(1-\delta_{ll'})\sum_{\sigma\sigma'}c_{\textbf{r},l\sigma}^{\dagger}c_{\textbf{r}',l'\sigma'}^{\dagger}c_{\textbf{r}',l'\sigma'}c_{\textbf{r},l\sigma}\nonumber\\
	&=\int d\textbf{r} d\textbf{r}'\sum_{ll'}\sum_{\sigma\sigma'}\bigg[\frac{U}{2}\delta(\textbf{r}-\textbf{r}')\delta_{ll'}(1-\delta_{\sigma\sigma'})+V(1-\delta_{ll'})\sum_{i}\delta(\textbf{r}+\bm{\delta}_{i}-\textbf{r}')\bigg]c_{\textbf{r},l\sigma}^{\dagger}c_{\textbf{r}',l'\sigma'}^{\dagger}c_{\textbf{r}',l'\sigma'}c_{\textbf{r},l\sigma}.
	\end{align}
	Identifying $\alpha=l\sigma$ and $\beta=l'\sigma'$, we find that
	\begin{equation}\label{Gammarr'}
	\Gamma_{\mu\nu,\beta\alpha}(\textbf{r},\textbf{r}')=\delta_{\alpha\mu}\delta_{\beta\nu}\left[\frac{U}{2}\delta(\textbf{r}-\textbf{r}')\delta_{ll'}(1-\delta_{\sigma\sigma'})+V(1-\delta_{ll'})\sum_{i}\delta(\textbf{r}+\bm{\delta}_{i}-\textbf{r}')\right].
	\end{equation}
	$U$ and $V$ denote the on-site interaction and NN interaction strength, respectively.
	In the momentum space, Eq. (\ref{Gammarr'}) yields
	\begin{align}
	H_{\textrm{int}}&=\int d\textbf{k}d\textbf{k}' d\textbf{q}\delta(\textbf{k}+\textbf{k}')\sum_{\alpha\beta\mu\nu}\Gamma_{\mu\nu,\beta\alpha}(\textbf{q})c_{\textbf{k},\mu}^{\dagger}c_{\textbf{k}',\nu}^{\dagger}c_{\textbf{k}'+\textbf{q},\beta}c_{\textbf{k}-\textbf{q},\alpha}\nonumber\\
	&=\int d\textbf{k}d\textbf{k}''\sum_{\alpha\beta\mu\nu}\Gamma_{\mu\nu,\beta\alpha}(\textbf{k}-\textbf{k}'')c_{\textbf{k},\mu}^{\dagger}c_{-\textbf{k},\nu}^{\dagger}c_{-\textbf{k}'',\beta}c_{\textbf{k}'',\alpha}
	\end{align}
	where we put a constraint $\textbf{k}'=-\textbf{k}$ to restrict our scope to scattering between electron pairs with zero total momentum, and substitute $\textbf{q}$ by $(\textbf{k}-\textbf{k}'')$. $\Gamma_{\mu\nu,\beta\alpha}(\textbf{q})$, the Fourier transform of $\Gamma_{\mu\nu,\beta\alpha}(\textbf{r},\textbf{r}')=\Gamma_{\mu\nu,\beta\alpha}(\textbf{r}-\textbf{r}')\equiv\Gamma_{\mu\nu,\beta\alpha}(\tilde{\textbf{r}})$, is given by
	\begin{align}\label{}
	\Gamma_{\mu\nu,\beta\alpha}(\textbf{q})&=\int d\tilde{\textbf{r}} e^{-i\textbf{q}\cdot\tilde{\textbf{r}}}\Gamma_{\mu\nu,\beta\alpha}(\tilde{\textbf{r}})\nonumber\\
	&=\delta_{\alpha\mu}\delta_{\beta\nu}\frac{U}{2}\delta_{ll'}(1-\delta_{\sigma\sigma'})\int d\tilde{\textbf{r}}e^{-i\textbf{q}\cdot\tilde{\textbf{r}}}\delta(\tilde{\textbf{r}})+\delta_{\alpha\mu}\delta_{\beta\nu}V(1-\delta_{ll'})\int d\tilde{\textbf{r}}e^{-i\textbf{q}\cdot\tilde{\textbf{r}}}\sum_{i}\delta(\tilde{\textbf{r}}+\bm{\delta}_{i})\nonumber\\
	&=\delta_{\alpha\mu}\delta_{\beta\nu}\frac{U}{2}\delta_{ll'}(1-\delta_{\sigma\sigma'})+\delta_{\alpha\mu}\delta_{\beta\nu}V(1-\delta_{ll'})\sum_{i}e^{i\textbf{q}\cdot\bm{\delta}_{i}}.\nonumber\\
	&=\delta_{\alpha\mu}\delta_{\beta\nu}\left[\frac{U}{2}\delta_{ll'}(1-\delta_{\sigma\sigma'})+V(1-\delta_{ll'})(\cos{q_{x}}+\cos{q_{y}})\right].
	\end{align}
	\\
	OAFM -- (i) -- $A_{u}$\\
	\indent For the pairing channels that belong to the $A_{u}$ representation, the relevant interaction is given by
	\begin{align}\label{operators}
	H_{\textrm{int}}^{A_{u}}&=-V\sum_{\textbf{k}\textbf{k}''}\sum_{\alpha\beta\mu\nu}\delta_{\alpha\mu}\delta_{\beta\nu}(1-\delta_{ll'})(\sin{k_{x}}\sin{k_{x}''}+\sin{k_{y}}\sin{k_{y}''})c_{\textbf{k},\mu}^{\dagger}c_{-\textbf{k},\nu}^{\dagger}c_{-\textbf{k}'',\beta}c_{\textbf{k}'',\alpha}\nonumber\\
	&=-V\sum_{\textbf{k}\textbf{k}''}(\sin{k_{x}}\sin{k_{x}''}+\sin{k_{y}}\sin{k_{y}''})(c_{\textbf{k},A\uparrow}^{\dagger}c_{-\textbf{k},B\uparrow}^{\dagger}c_{-\textbf{k}'',B\uparrow}c_{\textbf{k}'',A\uparrow}+c_{\textbf{k},B\uparrow}^{\dagger}c_{-\textbf{k},A\uparrow}^{\dagger}c_{-\textbf{k}'',A\uparrow}c_{\textbf{k}'',B\uparrow}\nonumber\\
	&\qquad\qquad\qquad\qquad\qquad\qquad\qquad\qquad+c_{\textbf{k},A\downarrow}^{\dagger}c_{-\textbf{k},B\downarrow}^{\dagger}c_{-\textbf{k}'',B\downarrow}c_{\textbf{k}'',A\downarrow}+c_{\textbf{k},B\downarrow}^{\dagger}c_{-\textbf{k},A\downarrow}^{\dagger}c_{-\textbf{k}'',A\downarrow}c_{\textbf{k}'',B\downarrow})\nonumber\\
	&=-2V\sum_{\textbf{k}\textbf{k}''}(\sin{k_{x}}\sin{k_{x}''}+\sin{k_{y}}\sin{k_{y}''})(c_{\textbf{k},A\uparrow}^{\dagger}c_{-\textbf{k},B\uparrow}^{\dagger}c_{-\textbf{k}'',B\uparrow}c_{\textbf{k}'',A\uparrow}+c_{\textbf{k},A\downarrow}^{\dagger}c_{-\textbf{k},B\downarrow}^{\dagger}c_{-\textbf{k}'',B\downarrow}c_{\textbf{k}'',A\downarrow}).
	\end{align}
	The partition function of the system is given by
	\begin{align}\label{partitionfunction}
	Z&=\int\mathcal{D}[\textbf{c}_{\textbf{k}}^{\dagger},\textbf{c}_{\textbf{k}}]e^{-S},\nonumber\\
	S&=\int_{0}^{\beta}d\tau\sum_{\textbf{k}}\textbf{c}_{\textbf{k}}^{\dagger}\left[\partial_{\tau}+\mathcal{H}(\textbf{k})\right]\textbf{c}_{\textbf{k}}+H_{\textrm{int}}^{A_{u}}.
	\end{align}
	We define
	$\Delta_{1}^{A_{u}}\equiv-V\langle \textbf{c}_{\textbf{k}}^{T}(\sin{k_{x}}\tilde{\sigma}_{x}\tau_{x}-\sin{k_{y}}\tilde{\sigma}_{y}\tau_{x})\textbf{c}_{\textbf{k}}\rangle$ and $\Delta_{2}^{A_{u}}\equiv-V\langle \textbf{c}_{\textbf{k}}^{T}(-\sin{k_{x}}\tilde{\sigma}_{y}\tau_{x}-\sin{k_{y}}\tilde{\sigma}_{x}\tau_{x})\textbf{c}_{\textbf{k}}\rangle$. Then using the Hubbard-Stratonovich transformation~\cite{altland2010condensed}, we can rewrite  Eq (\ref{partitionfunction}) as
	\begin{align}
	Z&=\int\mathcal{D}[\Delta_{1}^{A_{u}*},\Delta_{1}^{A_{u}},\Delta_{2}^{A_{u}*},\Delta_{2}^{A_{u}},\textbf{c}_{\textbf{k}}^{\dagger},\textbf{c}_{\textbf{k}}]e^{-S},\nonumber\\
	S&=\int_{0}^{\beta}d\tau\sum_{\textbf{k}}\Psi_{\textbf{k}}^{\dagger}\left[\partial_{\tau}+\mathcal{H}_{\textrm{BdG}}^{A_{u}}(\textbf{k})\right]\Psi_{\textbf{k}}+\frac{1}{V}|\Delta_{1}^{A_{u}}|^{2}+\frac{1}{V}|\Delta_{2}^{A_{u}}|^{2},
	\end{align}
	where
	\begin{align}\label{HStransformed}
	\mathcal{H}_{\textrm{BdG}}^{A_{u}}(\textbf{k})&=
	\begin{pmatrix}
	\mathcal{H}(\textbf{k})-\mu	& \Delta^{A_{u}}(\textbf{k})\\
	\Delta^{A_{u}\dagger}(\textbf{k})	& -\mathcal{H}^{T}(-\textbf{k})+\mu
	\end{pmatrix},\nonumber\\
	\Delta^{A_{u}}(\textbf{k})&=\Delta_{1}^{A_{u}}(\sin{k_{x}}\tilde{\sigma}_{x}\tau_{x}+\sin{k_{y}}\tilde{\sigma}_{y}\tau_{x})+\Delta_{2}^{A_{u}}(\sin{k_{x}}\tilde{\sigma}_{y}\tau_{x}-\sin{k_{y}}\tilde{\sigma}_{x}\tau_{x}).
	\end{align}
	Carrying out the Gaussian integral over the fermionic field~\cite{altland2010condensed}, we obtain
	\begin{align}
	Z&=\int\mathcal{D}[\Delta_{1}^{A_{u}*},\Delta_{1}^{A_{u}},\Delta_{2}^{A_{u}*},\Delta_{2}^{A_{u}}]e^{-S},\nonumber\\
	S&=\int_{0}^{\beta}d\tau\sum_{\textbf{k}}\textrm{Tr}\ln\det{\left[\partial_{\tau}+\mathcal{H}_{\textrm{BdG}}(\textbf{k})\right]}+\frac{1}{V}|\Delta_{1}^{A_{u}}|^{2}+\frac{1}{V}|\Delta_{2}^{A_{u}}|^{2}.
	\end{align}
	From the relation $Z=e^{-\beta F}$, the free energy of the system is given by
	\begin{align}
	F&=\frac{1}{V}|\Delta_{1}^{A_{u}}|^{2}+\frac{1}{V}|\Delta_{2}^{A_{u}}|^{2}-\frac{1}{\beta}\sum_{N}\sum_{\textbf{k}n}\ln{\left[\omega_{N}^{2}+\xi_{n}^{2}(\textbf{k})\right]},
	\end{align}
	where $\omega_{N}$ is the $N$-th fermionic Matsubara frequency, and $\xi_{n}(\textbf{k})$ is an are the eigenvalues of $\mathcal{H}_{\textrm{BdG}}(\textbf{k})$ given by
	\begin{align}
	\xi_{1}^{2}(\textbf{k})&=\lambda_{-}^{2}(\textbf{k})+\mu^{2}+\eta_{+}(\sin^{2}{k_{x}}+\sin^{2}{k_{y}})-2\sqrt{\lambda_{-}^{2}(\textbf{k})\mu^{2}+\eta_{+}(\sin^{2}{k_{x}}+\sin^{2}{k_{y}})(m-\epsilon_{sp})^{2}},\nonumber\\
	\xi_{2}^{2}(\textbf{k})&=\lambda_{+}^{2}(\textbf{k})+\mu^{2}+\eta_{-}(\sin^{2}{k_{x}}+\sin^{2}{k_{y}})-2\sqrt{\lambda_{+}^{2}(\textbf{k})\mu^{2}+\eta_{-}(\sin^{2}{k_{x}}+\sin^{2}{k_{y}})(m+\epsilon_{sp})^{2}},\nonumber\\
	\xi_{3}^{2}(\textbf{k})&=\lambda_{-}^{2}(\textbf{k})+\mu^{2}+\eta_{+}(\sin^{2}{k_{x}}+\sin^{2}{k_{y}})+2\sqrt{\lambda_{-}^{2}(\textbf{k})\mu^{2}+\eta_{+}(\sin^{2}{k_{x}}+\sin^{2}{k_{y}})(m-\epsilon_{sp})^{2}},\nonumber\\
	\xi_{4}^{2}(\textbf{k})&=\lambda_{+}^{2}(\textbf{k})+\mu^{2}+\eta_{-}(\sin^{2}{k_{x}}+\sin^{2}{k_{y}})+2\sqrt{\lambda_{+}^{2}(\textbf{k})\mu^{2}+\eta_{-}(\sin^{2}{k_{x}}+\sin^{2}{k_{y}})(m+\epsilon_{sp})^{2}}.
	\end{align}
	Here, $\eta_{\pm}\equiv|\Delta_{1}^{A_{u}}|^{2}+|\Delta_{2}^{A_{u}}|^{2}\pm i(\Delta_{1}^{A_{u}}\Delta_{2}^{A_{u}*}-\Delta_{2}^{A_{u}}\Delta_{1}^{A_{u}*}).$ For Case (i) ($|\mu|<\min\lambda_{-}(\textbf{k})$), we need to consider $n=1$ only. By differentiating $F$ with repect to $\theta_{1}^{A_{u}}$ and $\theta_{2}^{A_{u}}$, we obtain
	\begin{align}\label{angleminimumIiAu}
	\frac{\partial F}{\partial \theta_{1}^{A_{u}}}&=-\frac{1}{\beta}\sum_{N}\sum_{\textbf{k}}\frac{\partial\xi_{1}^{2}(\textbf{k})/\partial\theta_{1}^{A_{u}}}{\omega_{N}^{2}+\xi_{1}^{2}(\textbf{k})}\nonumber\\
	&=-\frac{1}{\beta}\sum_{N}\sum_{\textbf{k}}-\left[1-\frac{(m-\epsilon_{sp})^{2}}{|\lambda_{-}(\textbf{k})\mu|}\right]\frac{\sin^{2}{k_{x}}+\sin^{2}{k_{y}}}{|\lambda_{-}(\textbf{k})-\mu|\left[\omega_{N}^{2}+\xi_{1}^{2}(\textbf{k})\right]}|\Delta_{1}^{A_{u}}||\Delta_{2}^{A_{u}}|\cos{(\theta_{1}^{A_{u}}-\theta_{2}^{A_{u}})}=0,\nonumber\\
	\frac{\partial F}{\partial \theta_{2}^{A_{u}}}&=-\frac{1}{\beta}\sum_{N}\sum_{\textbf{k}}\frac{\partial\xi_{1}^{2}(\textbf{k})/\partial\theta_{2}^{A_{u}}}{\omega_{N}^{2}+\xi_{1}^{2}(\textbf{k})}\nonumber\\
	&=-\frac{1}{\beta}\sum_{N}\sum_{\textbf{k}}\left[1-\frac{(m-\epsilon_{sp})^{2}}{|\lambda_{-}(\textbf{k})\mu|}\right]\frac{\sin^{2}{k_{x}}+\sin^{2}{k_{y}}}{|\lambda_{-}(\textbf{k})-\mu|\left[\omega_{N}^{2}+\xi_{1}^{2}(\textbf{k})\right]}|\Delta_{1}^{A_{u}}||\Delta_{2}^{A_{u}}|\cos{(\theta_{1}^{A_{u}}-\theta_{2}^{A_{u}})}=0,
	\end{align}
	From Eq. (\ref{angleminimumIiAu}), we extract three extremum conditions for $F$: $|\Delta_{1}^{A_{u}}|=0$, $|\Delta_{2}^{A_{u}}|=0$, and $\cos{(\theta_{1}^{A_{u}}-\theta_{2}^{A_{u}})}=0$. Then, to calculate the transition temperature, we differentiate $F$ with respect to $|\Delta_{1}^{A_{u}}|$ and $|\Delta_{1}^{A_{u}}|$ for each condition. For the first condition, we can derive
	\begin{align}\label{thecon1}
	\frac{\partial F}{\partial |\Delta_{2}^{A_{u}}|}\Bigg\rvert_{|\Delta_{1}^{A_{u}}|=0}&=\frac{2}{V}|\Delta_{2}^{A_{u}}|-\frac{1}{\Omega}\int_{1\textrm{BZ}} d^{2}k\frac{\tanh{(\beta\xi_{1,\textbf{k}}/2)}}{2\xi_{1,\textbf{k}}}\left[1-\frac{(m-\epsilon_{sp})^{2}}{|\lambda_{-}(\textbf{k})\mu|}\right]\frac{\sin^{2}{k_{x}}+\sin^{2}{k_{y}}}{|\lambda_{-}(\textbf{k})-\mu|}|\Delta_{2}^{A_{u}}|=0,
	\end{align}
	where we have used the fermionic Matsubara frequency summation relation
	\begin{equation}
	\frac{1}{\beta}\sum_{N}\frac{1}{\omega_{N}^{2}+\xi_{n,\textbf{k}}^{2}}=\frac{\tanh{(\beta\xi_{n,\textbf{k}}/2)}}{2\xi_{n,\textbf{k}}},
	\end{equation}
	and applied an approximation that near the transition temperature the superconducting order parameters are very small compared to the other parameters. In Eq. (\ref{thecon1}), $\Omega$ is the volume of the system. Since $\xi_{n}^{-1}\tanh{(\beta\xi_{n}(\textbf{k})/2)}$ diverges around the Fermi level $\xi_{n}=0$, we evaluate other factors in the integrand on the $n$-th Fermi surface on which $\xi_{n}=0$, and take the resulting values as their representative values. Substituting the variables of integration by energy $\xi_{n}$ and a component of momentum $k_{Fn}^{\parallel}$ which is parallel to the $n$-th Fermi surface, we obtain
	\begin{align}\label{thecon1approx}
	\frac{2}{V}|\Delta_{2}^{A_{u}}|-\frac{1}{\Omega}\int_{-\omega_{0}}^{\omega_{0}} d\xi_{1}\frac{\tanh{(\beta\xi_{1}/2)}}{2\xi_{1}}\int_{\textrm{FS}_{1}} dk_{F1}^{\parallel}D_{1}(\textbf{k}_{F1})\left[1-\frac{(m-\epsilon_{sp})^{2}}{|\lambda_{-}(\textbf{k}_{F1})\mu|}\right]\frac{\sin^{2}{k_{F1,x}}+\sin^{2}{k_{F1,y}}}{|\lambda_{-}(\textbf{k}_{F1})-\mu|}|\Delta_{2}^{A_{u}}|=0,
	\end{align}
	where we cut the region of integration with respect to $\xi_{n}$ to a small interval from $-\omega_{0}$ to $\omega_{0}$. $D_{n}(\textbf{k}_{Fn})$ is the density of states on the $n$-th Fermi surface. Then, we can simply write Eq. (\ref{thecon1approx}) as
	\begin{align}\label{thecon1-1}
	\frac{\partial F}{\partial |\Delta_{2}^{A_{u}}|}\Bigg\rvert_{|\Delta_{1}^{A_{u}}|=0}&=\frac{2}{V}|\Delta_{2}^{A_{u}}|-\frac{I(\beta)}{\Omega}K_{1}|\Delta_{2}^{A_{u}}|=0,
	\end{align}
	where
	\begin{align}\label{substitution}
	I(\beta)&\equiv\int_{-\omega_{0}}^{\omega_{0}}d\xi\tanh{(\beta\xi_{n}/2)}/(2\xi_{n})=\ln(2e^{\gamma}\beta\omega_{0}/\pi),\nonumber\\
	K_{1}&=\int_{\textrm{FS}_{1}} dk_{F1}^{\parallel}D_{1}(\textbf{k}_{F1})\left[1-\frac{(m-\epsilon_{sp})^{2}}{|\lambda_{-}(\textbf{k}_{F1})\mu|}\right]\frac{\sin^{2}{k_{F1,x}}+\sin^{2}{k_{F1,y}}}{|\lambda_{-}(\textbf{k}_{F1})-\mu|}.
	\end{align}
	From Eq. (\ref{thecon1-1}), we immediately find that the transition temperature satisfies $I(\beta_{c})=2\Omega/VK_{1}$ ($\therefore T_{c}=(2\omega_{0}/\pi k_{B})\exp{[\gamma-(2\Omega/VK_{1})]}$).
	For the second condition, we have
	\begin{align}\label{}
	\frac{\partial F}{\partial |\Delta_{1}^{A_{u}}|}\Bigg\rvert_{|\Delta_{2}^{A_{u}}|=0}&=\frac{2}{V}|\Delta_{1}^{A_{u}}|-\frac{I(\beta)}{\Omega}K_{1}|\Delta_{1}^{A_{u}}|=0,
	\end{align}
	and the transition temperature is again given by $T_{c}=(2\omega_{0}/\pi k_{B})\exp{[\gamma-(2\Omega/VK_{1})]}$. However, for the third condition, we have a pair of equations,
	\begin{align}\label{}
	\frac{\partial F}{\partial |\Delta_{1}^{A_{u}}|}\Bigg\rvert_{\theta_{2}^{A_{u}}=\theta_{1}^{A_{u}}+\pi/2}&=\frac{2}{V}|\Delta_{1}^{A_{u}}|-\frac{I(\beta)}{\Omega}K_{1}(|\Delta_{1}^{A_{u}}|+|\Delta_{2}^{A_{u}}|)=0,\nonumber\\
	\frac{\partial F}{\partial |\Delta_{2}^{A_{u}}|}\Bigg\rvert_{\theta_{2}^{A_{u}}=\theta_{1}^{A_{u}}+\pi/2}&=\frac{2}{V}|\Delta_{2}^{A_{u}}|-\frac{I(\beta)}{\Omega}K_{1}(|\Delta_{1}^{A_{u}}|+|\Delta_{2}^{A_{u}}|)=0.
	\end{align}
	They give us $|\Delta_{1}^{A_{u}}|=|\Delta_{2}^{A_{u}}|$, and $I(\beta_{c})=\Omega/VK_{1}$ ($T_{c}=(2\omega_{0}/\pi k_{B})\exp{[\gamma-(\Omega/VK_{1})]}$). In the present case ($|\mu|<\min\lambda_{-}(\textbf{k})$), $K_{1}$ has to be a positive real number, so the third condition yields the highest transition temperature. Thus, in the superconducting phase of OAFM -- (i) -- $A_{u}$, the pairing interaction of the leading instability is given by $\Delta^{A_{u}}(\textbf{k})=\Delta_{1}^{A_{u}}(\sin{k_{x}}\tilde{\sigma}_{x}\tau_{x}+\sin{k_{y}}\tilde{\sigma}_{y}\tau_{x})+\Delta_{2}^{A_{u}}(\sin{k_{x}}\tilde{\sigma}_{y}\tau_{x}-\sin{k_{y}}\tilde{\sigma}_{x}\tau_{x})$ where $(\Delta_{1}^{A_{u}},\Delta_{2}^{A_{u}})=(\Delta,i\Delta)$.
	\\
	\\
	OAFM -- (i) -- $B_{u}$\\
	\indent For $B_{u}$ representation, we define $\Delta_{1}^{B_{u}}\equiv-V\langle \textbf{c}_{\textbf{k}}^{T}(\sin{k_{x}}\tilde{\sigma}_{x}\tau_{x}+\sin{k_{y}}\tilde{\sigma}_{y}\tau_{x})\textbf{c}_{\textbf{k}}\rangle$ and $\Delta_{2}^{B_{u}}\equiv-V\langle \textbf{c}_{\textbf{k}}^{T}(-\sin{k_{x}}\tilde{\sigma}_{y}\tau_{x}+\sin{k_{y}}\tilde{\sigma}_{x}\tau_{x})\textbf{c}_{\textbf{k}}\rangle$. Then, the free energy of the system is given by
	\begin{align}
	F&=\frac{1}{V}|\Delta_{1}^{B_{u}}|^{2}+\frac{1}{V}|\Delta_{2}^{B_{u}}|^{2}-\frac{1}{\beta}\sum_{N}\sum_{\textbf{k}n}\ln{\left[\omega_{N}^{2}+\xi_{n}^{2}(\textbf{k})\right]}.
	\end{align}
	The pairing interaction of the leading instability is given by $\Delta^{B_{u}}(\textbf{k})=\Delta_{1}^{B_{u}}(\sin{k_{x}}\tilde{\sigma}_{x}\tau_{x}-\sin{k_{y}}\tilde{\sigma}_{y}\tau_{x})+\Delta_{2}^{B_{u}}(\sin{k_{x}}\tilde{\sigma}_{y}\tau_{x}+\sin{k_{y}}\tilde{\sigma}_{x}\tau_{x})$ where $(\Delta_{1}^{B_{u}},\Delta_{2}^{B_{u}})=(\Delta,i\Delta)$.
	\\
	\\
	OAFM -- (ii) -- $A_{u}$\\
	\indent For the present case ($\min\lambda_{+}(\textbf{k})<|\mu|<\max\lambda_{-}(\textbf{k})$), there exist two Fermi surfaces, so we have to consider contribution from the both Fermi surfaces.
	\begin{align}\label{angleminimum}
	\frac{\partial F}{\partial \theta_{1}^{A_{u}}}&=-\frac{1}{\beta}\sum_{N}\sum_{\textbf{k}}\left[\frac{\partial\xi_{1}^{2}(\textbf{k})/\partial\theta_{1}^{A_{u}}}{\omega_{N}^{2}+\xi_{1}^{2}(\textbf{k})}+\frac{\partial\xi_{2}^{2}(\textbf{k})/\partial\theta_{1`}^{A_{u}}}{\omega_{N}^{2}+\xi_{2}^{2}(\textbf{k})}\right]\nonumber\\
	&=-\frac{1}{\beta}\sum_{N}\sum_{\textbf{k}}-\left[1-\frac{(m-\epsilon_{sp})^{2}}{|\lambda_{-}(\textbf{k})\mu|}\right]\frac{\sin^{2}{k_{x}}+\sin^{2}{k_{y}}}{|\lambda_{-}(\textbf{k})-\mu|\left[\omega_{N}^{2}+\xi_{1}^{2}(\textbf{k})\right]}|\Delta_{1}^{A_{u}}||\Delta_{2}^{A_{u}}|\cos{(\theta_{1}^{A_{u}}-\theta_{2}^{A_{u}})}\nonumber\\
	&\qquad+\left[1-\frac{(m+\epsilon_{sp})^{2}}{|\lambda_{+}(\textbf{k})\mu|}\right]\frac{\sin^{2}{k_{x}}+\sin^{2}{k_{y}}}{|\lambda_{+}(\textbf{k})-\mu|\left[\omega_{N}^{2}+\xi_{1}^{2}(\textbf{k})\right]}|\Delta_{1}^{A_{u}}||\Delta_{2}^{A_{u}}|\cos{(\theta_{1}^{A_{u}}-\theta_{2}^{A_{u}})}=0,\nonumber\\
	\frac{\partial F}{\partial \theta_{2}^{A_{u}}}&=-\frac{1}{\beta}\sum_{N}\sum_{\textbf{k}}\left[1-\frac{(m-\epsilon_{sp})^{2}}{|\lambda_{-}(\textbf{k})\mu|}\right]\frac{\sin^{2}{k_{x}}+\sin^{2}{k_{y}}}{|\lambda_{-}(\textbf{k})-\mu|\left[\omega_{N}^{2}+\xi_{1}^{2}(\textbf{k})\right]}|\Delta_{1}^{A_{u}}||\Delta_{2}^{A_{u}}|\cos{(\theta_{1}^{A_{u}}-\theta_{2}^{A_{u}})}\nonumber\\
	&\qquad-\left[1-\frac{(m+\epsilon_{sp})^{2}}{|\lambda_{+}(\textbf{k})\mu|}\right]\frac{\sin^{2}{k_{x}}+\sin^{2}{k_{y}}}{|\lambda_{+}(\textbf{k})-\mu|\left[\omega_{N}^{2}+\xi_{1}^{2}(\textbf{k})\right]}|\Delta_{1}^{A_{u}}||\Delta_{2}^{A_{u}}|\cos{(\theta_{1}^{A_{u}}-\theta_{2}^{A_{u}})}=0,
	\end{align}
	where we applied an approximation that near the transition temperature the superconducting order parameters are very small compared to other parameters. The two equations in Eq. (\ref{angleminimum}) imply that the free energy is minized when one of the two order parameters vanishes (i.e. $|\Delta_{1}^{A_{u}}|=0$ or $|\Delta_{2}^{A_{u}}|=0$), or $\cos{(\theta_{1}^{A_{u}}-\theta_{2}^{A_{u}})}=0$ (i.e. $|\Delta_{1}^{A_{u}}|=i|\Delta_{2}^{A_{u}}|$ or $|\Delta_{1}^{A_{u}}|=-i|\Delta_{2}^{A_{u}}|$). However, if $\cos{(\theta_{1}^{A_{u}}-\theta_{2}^{A_{u}})}=0$, we find that the system of linearized gap equations
	\begin{align}\label{}
	\frac{\partial F}{\partial |\Delta_{1}^{A_{u}}|}\Bigg\rvert_{\theta_{2}^{A_{u}}=\theta_{1}^{A_{u}}+\pi/2}&=\frac{2}{V}|\Delta_{1}^{A_{u}}|-\frac{I(\beta)}{\Omega}K_{1}(|\Delta_{1}^{A_{u}}|+|\Delta_{2}^{A_{u}}|)-\frac{I(\beta)}{\Omega}K_{2}(|\Delta_{1}^{A_{u}}|-|\Delta_{2}^{A_{u}}|)=0,\nonumber\\
	\frac{\partial F}{\partial |\Delta_{2}^{A_{u}}|}\Bigg\rvert_{\theta_{2}^{A_{u}}=\theta_{1}^{A_{u}}+\pi/2}&=\frac{2}{V}|\Delta_{2}^{A_{u}}|-\frac{I(\beta)}{\Omega}K_{1}(|\Delta_{1}^{A_{u}}|+|\Delta_{2}^{A_{u}}|)+\frac{I(\beta)}{\Omega}K_{2}(|\Delta_{1}^{A_{u}}|-|\Delta_{2}^{A_{u}}|)=0,
	\end{align}
	does not have an appropriate solution $I(\beta)$, where
	\begin{align}\label{k2substitution}
	K_{2}=\int_{\textrm{FS}_{2}} dk_{F2}^{\parallel}D_{2}(\textbf{k}_{F2})\left[1-\frac{(m+\epsilon_{sp})^{2}}{|\lambda_{+}(\textbf{k}_{F2})\mu|}\right]\frac{\sin^{2}{k_{F2,x}}+\sin^{2}{k_{F2,y}}}{|\lambda_{+}(\textbf{k}_{F2})-\mu|},
	\end{align}
	and $K_{1}\neq K_{2}$. On the other hand, if $|\Delta_{1}^{A_{u}}|=0$ or $|\Delta_{2}^{A_{u}}|=0$, we have
	\begin{align}\label{ii1}
	\frac{\partial F}{\partial |\Delta_{1}^{A_{u}}|}\Bigg\rvert_{|\Delta_{2}^{A_{u}}|=0}&=\frac{2}{V}|\Delta_{1}^{A_{u}}|-\frac{I(\beta)}{\Omega}K_{1}|\Delta_{1}^{A_{u}}|-\frac{I(\beta)}{\Omega}K_{2}|\Delta_{1}^{A_{u}}|=0,
	\end{align}
	or
	\begin{align}\label{ii2}
	\frac{\partial F}{\partial |\Delta_{2}^{A_{u}}|}\Bigg\rvert_{|\Delta_{1}^{A_{u}}|=0}&=\frac{2}{V}|\Delta_{2}^{A_{u}}|-\frac{I(\beta)}{\Omega}K_{1}|\Delta_{2}^{A_{u}}|-\frac{I(\beta)}{\Omega}K_{2}|\Delta_{2}^{A_{u}}|=0.
	\end{align}
	Since Eq. (\ref{ii1}) and Eq. (\ref{ii2}) have the same solution ($I(\beta_{c})=2\Omega/V(K_{1}+K_{2})$), the pairing interaction of the leading instability is given by $\Delta^{A_{u}}(\textbf{k})=\Delta_{1}^{A_{u}}(\sin{k_{x}}\tilde{\sigma}_{x}\tau_{x}+\sin{k_{y}}\tilde{\sigma}_{y}\tau_{x})+\Delta_{2}^{A_{u}}(\sin{k_{x}}\tilde{\sigma}_{y}\tau_{x}-\sin{k_{y}}\tilde{\sigma}_{x}\tau_{x})$ where $(\Delta_{1}^{A_{u}},\Delta_{2}^{A_{u}})=(\Delta,0)$ or $(0,\Delta)$.
	\\
	\\
	OAFM -- (ii) -- $B_{u}$\\
	\indent The pairing interaction of the leading instability is given by $\Delta^{B_{u}}(\textbf{k})=\Delta_{1}^{B_{u}}(\sin{k_{x}}\tilde{\sigma}_{x}\tau_{x}-\sin{k_{y}}\tilde{\sigma}_{y}\tau_{x})+\Delta_{2}^{B_{u}}(\sin{k_{x}}\tilde{\sigma}_{y}\tau_{x}+\sin{k_{y}}\tilde{\sigma}_{x}\tau_{x})$ where $(\Delta_{1}^{B_{u}},\Delta_{2}^{B_{u}})=(\Delta,0)$ or $(0,\Delta)$.
	\\
	\\
	OAFM -- (iii) -- $A_{u}$\\
	\indent The pairing interaction of the leading instability is given by $\Delta^{A_{u}}(\textbf{k})=\Delta_{1}^{A_{u}}(\sin{k_{x}}\tilde{\sigma}_{x}\tau_{x}+\sin{k_{y}}\tilde{\sigma}_{y}\tau_{x})+\Delta_{2}^{A_{u}}(\sin{k_{x}}\tilde{\sigma}_{y}\tau_{x}-\sin{k_{y}}\tilde{\sigma}_{x}\tau_{x})$ where $(\Delta_{1}^{A_{u}},\Delta_{2}^{A_{u}})=(\Delta,-i\Delta)$.
	\\
	\\
	OAFM -- (iii) -- $B_{u}$\\
	\indent The pairing interaction of the leading instability is given by $\Delta^{B_{u}}(\textbf{k})=\Delta_{1}^{B_{u}}(\sin{k_{x}}\tilde{\sigma}_{x}\tau_{x}-\sin{k_{y}}\tilde{\sigma}_{y}\tau_{x})+\Delta_{2}^{B_{u}}(\sin{k_{x}}\tilde{\sigma}_{y}\tau_{x}+\sin{k_{y}}\tilde{\sigma}_{x}\tau_{x})$ where $(\Delta_{1}^{B_{u}},\Delta_{2}^{B_{u}})=(\Delta,-i\Delta)$.
	\\
	\\
	IAFM -- (i) -- $A_{u}$\\
	\indent For the pairing channels that belong to $A_{u}$ representation, the relevant interaction is given by
	\begin{align}
	H_{\textrm{int}}^{A_{u}}&=-V\sum_{\textbf{k}\textbf{k}''}\sin{k_{y}}\sin{k_{y}''}(c_{\textbf{k},A\uparrow}^{\dagger}c_{-\textbf{k},B\uparrow}^{\dagger}c_{-\textbf{k}'',B\uparrow}c_{\textbf{k}'',A\uparrow}+c_{\textbf{k},B\uparrow}^{\dagger}c_{-\textbf{k},A\uparrow}^{\dagger}c_{-\textbf{k}'',A\uparrow}c_{\textbf{k}'',B\uparrow}\nonumber\\
	&\qquad\qquad\qquad\qquad\qquad+c_{\textbf{k},A\downarrow}^{\dagger}c_{-\textbf{k},B\downarrow}^{\dagger}c_{-\textbf{k}'',B\downarrow}c_{\textbf{k}'',A\downarrow}+c_{\textbf{k},B\downarrow}^{\dagger}c_{-\textbf{k},A\downarrow}^{\dagger}c_{-\textbf{k}'',A\downarrow}c_{\textbf{k}'',B\downarrow}\nonumber\\
	&\qquad\qquad\qquad\qquad\qquad+c_{\textbf{k},A\uparrow}^{\dagger}c_{-\textbf{k},B\downarrow}^{\dagger}c_{-\textbf{k}'',B\downarrow}c_{\textbf{k}'',A\uparrow}+c_{\textbf{k},B\uparrow}^{\dagger}c_{-\textbf{k},A\downarrow}^{\dagger}c_{-\textbf{k}'',A\downarrow}c_{\textbf{k}'',B\uparrow}\nonumber\\
	&\qquad\qquad\qquad\qquad\qquad+c_{\textbf{k},A\downarrow}^{\dagger}c_{-\textbf{k},B\uparrow}^{\dagger}c_{-\textbf{k}'',B\uparrow}c_{\textbf{k}'',A\downarrow}+c_{\textbf{k},B\downarrow}^{\dagger}c_{-\textbf{k},A\uparrow}^{\dagger}c_{-\textbf{k}'',A\uparrow}c_{\textbf{k}'',B\downarrow})
	\end{align}
	This time, we define $\Delta_{1}^{A_{u}}\equiv-V\langle \textbf{c}_{\textbf{k}}^{T}(-\sin{k_{y}}\tilde{\sigma}_{y}\tau_{x})\textbf{c}_{\textbf{k}}\rangle$ and $\Delta_{2}^{A_{u}}\equiv-V\langle \textbf{c}_{\textbf{k}}^{T}(\sin{k_{y}}\tilde{\sigma}_{z}\tau_{x})\textbf{c}_{\textbf{k}}\rangle$. Then we obtain
	\begin{align}
	F(\Delta_{1}^{A_{u}},\Delta_{2}^{A_{u}})=\frac{1}{V}|\Delta_{1}^{A_{u}}|^{2}+\frac{1}{V}|\Delta_{2}^{A_{u}}|^{2}-\sum_{N}\sum_{\textbf{k}n}\ln{\left[\omega_{N}^{2}+\xi_{n}^{2}(\textbf{k})\right]},
	\end{align}
	where
	\begin{align}
	\xi_{\textbf{k},1}^{2}&=\lambda_{-}^{2}(\textbf{k})+\mu^{2}+\eta_{+}\sin^{2}{k_{y}}-2\sqrt{\lambda_{-}^{2}(\textbf{k})\mu^{2}+\eta_{+}\sin^{2}{k_{y}}(m-\epsilon_{sp})^{2}},\nonumber\\
	\xi_{\textbf{k},2}^{2}&=\lambda_{+}^{2}(\textbf{k})+\mu^{2}+\eta_{-}\sin^{2}{k_{y}}-2\sqrt{\lambda_{+}^{2}(\textbf{k})\mu^{2}+\eta_{-}\sin^{2}{k_{y}}(m+\epsilon_{sp})^{2}},\nonumber\\
	\xi_{\textbf{k},3}^{2}&=\lambda_{-}^{2}(\textbf{k})+\mu^{2}+\eta_{+}\sin^{2}{k_{y}}+2\sqrt{\lambda_{-}^{2}(\textbf{k})\mu^{2}+\eta_{+}\sin^{2}{k_{y}}(m-\epsilon_{sp})^{2}},\nonumber\\
	\xi_{\textbf{k},4}^{2}&=\lambda_{+}^{2}(\textbf{k})+\mu^{2}+\eta_{-}\sin^{2}{k_{y}}+2\sqrt{\lambda_{+}^{2}(\textbf{k})\mu^{2}+\eta_{-}\sin^{2}{k_{y}}(m+\epsilon_{sp})^{2}}.
	\end{align}
	The pairing interaction of the leading instability is given by $\Delta^{A_{u}}(\textbf{k})=\Delta_{1}^{A_{u}}\sin{k_{y}}\tilde{\sigma}_{y}\tau_{x}+\Delta_{2}^{A_{u}}\sin{k_{y}}\tilde{\sigma}_{z}\tau_{x}$ where $(\Delta_{1}^{A_{u}},\Delta_{2}^{A_{u}})=(\Delta,i\Delta)$.
	\\
	\\
	IAFM -- (i) -- $B_{u}$\\
	\indent We define $\Delta_{1}^{B_{u}}\equiv-V\langle \textbf{c}_{\textbf{k}}^{T}(-\sin{k_{x}}\tilde{\sigma}_{y}\tau_{x})\textbf{c}_{\textbf{k}}\rangle$ and $\Delta_{2}^{B_{u}}\equiv-V\langle \textbf{c}_{\textbf{k}}^{T}(\sin{k_{x}}\tilde{\sigma}_{z}\tau_{x})\textbf{c}_{\textbf{k}}\rangle$. The pairing interaction of the leading instability is given by $\Delta^{B_{u}}(\textbf{k})=\Delta_{1}^{B_{u}}\sin{k_{x}}\tilde{\sigma}_{y}\tau_{x}+\Delta_{2}^{B_{u}}\sin{k_{x}}\tilde{\sigma}_{z}\tau_{x}$ where $(\Delta_{1}^{B_{u}},\Delta_{2}^{B_{u}})=(\Delta,i\Delta)$.
	\\
	\\
	IAFM -- (ii) -- $A_{u}$\\
	\indent The pairing interaction of the leading instability is given by $\Delta^{A_{u}}(\textbf{k})=\Delta_{1}^{A_{u}}\sin{k_{y}}\tilde{\sigma}_{y}\tau_{x}+\Delta_{2}^{A_{u}}\sin{k_{y}}\tilde{\sigma}_{z}\tau_{x}$ where $(\Delta_{1}^{A_{u}},\Delta_{2}^{A_{u}})=(\Delta,0)$ or $(0,\Delta)$. or $(0,\Delta)$.
	\\
	\\
	IAFM -- (ii) -- $B_{u}$\\
	\indent The pairing interaction of the leading instability is given by $\Delta^{B_{u}}(\textbf{k})=\Delta_{1}^{B_{u}}\sin{k_{x}}\tilde{\sigma}_{y}\tau_{x}+\Delta_{2}^{B_{u}}\sin{k_{x}}\tilde{\sigma}_{z}\tau_{x}$ where $(\Delta_{1}^{B_{u}},\Delta_{2}^{B_{u}})=(\Delta,0)$ or $(0,\Delta)$.
	\\
	\\
	IAFM -- (iii) -- $A_{u}$\\
	\indent The pairing interaction of the leading instability is given by $\Delta^{A_{u}}(\textbf{k})=\Delta_{1}^{A_{u}}\sin{k_{y}}\tilde{\sigma}_{y}\tau_{x}+\Delta_{2}^{A_{u}}\sin{k_{y}}\tilde{\sigma}_{z}\tau_{x}$ where $(\Delta_{1}^{A_{u}},\Delta_{2}^{A_{u}})=(\Delta,-i\Delta)$.
	\\
	\\
	IAFM -- (iii) -- $B_{u}$\\
	\indent The pairing interaction of the leading instability is given by $\Delta^{B_{u}}(\textbf{k})=\Delta_{1}^{B_{u}}\sin{k_{x}}\tilde{\sigma}_{y}\tau_{x}+\Delta_{2}^{B_{u}}\sin{k_{x}}\tilde{\sigma}_{z}\tau_{x}$ where $(\Delta_{1}^{B_{u}},\Delta_{2}^{B_{u}})=(\Delta,-i\Delta)$.
	\\
	\\
	\noindent
	\textbf{Net magnetic moment}\\
	Due to $\tilde{\Theta}$ symmetry breaking, the system has a small net magnetic moment. Here we show the net magnetic moment of the occupied states of our model, varying the chemical potential.
	\begin{figure}[h!]
		\begin{center}
			\includegraphics[scale=0.3]{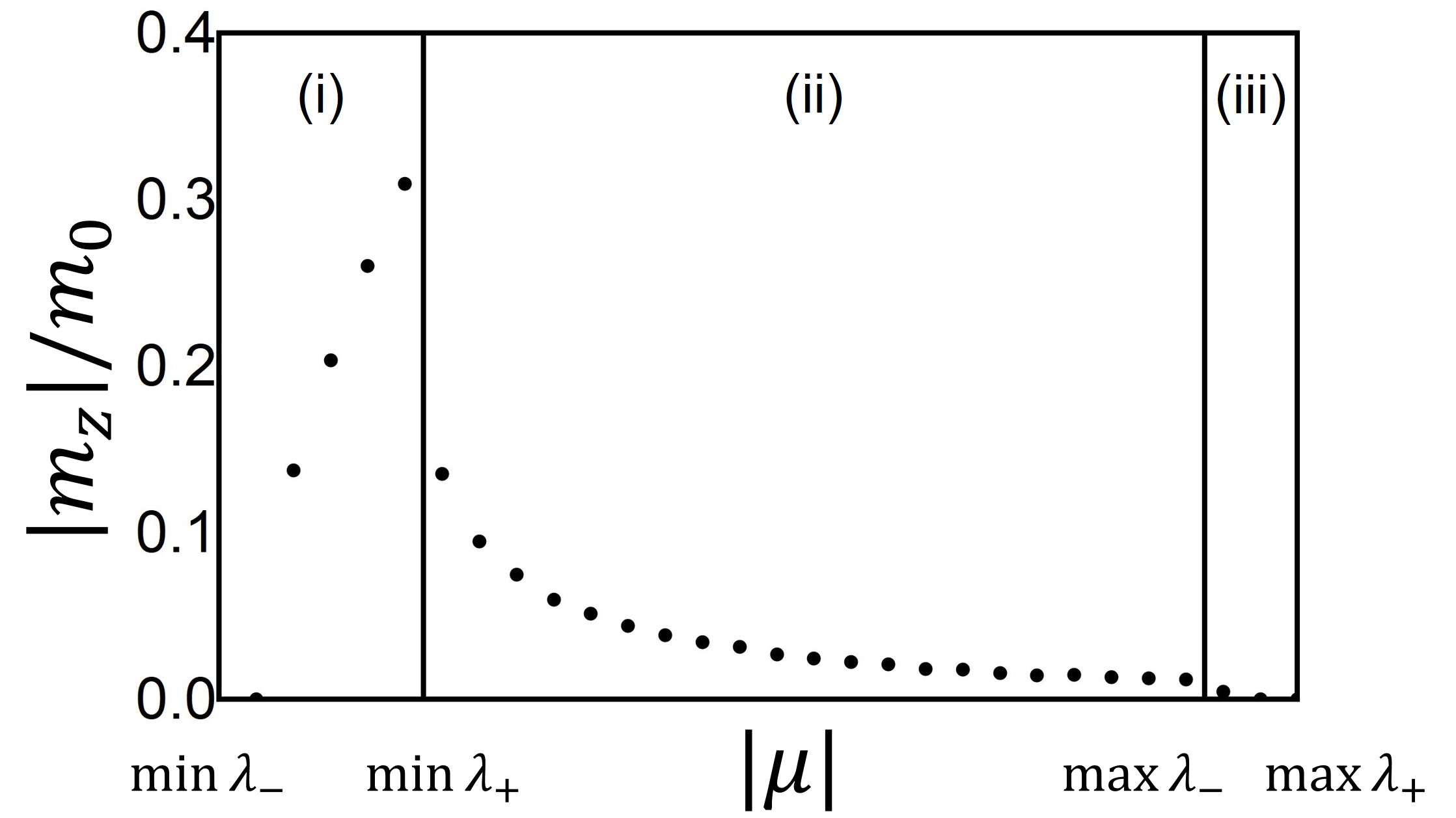}
			\caption{The variation in the magnitude of the net magnetic moment with respect to the change in the chemical potential.}
		\end{center}
	\end{figure}
	\\
	\noindent
	\textbf{Calculation of topological invariants}\\
	\indent In this section, we provide the result of calculation on the topological properties of the stable superconducting states not covered in the main text.
	\\
	\\
	O-AFM\\
	\indent Since the spin-up sector and spin-down sector of $\mathcal{H}_{\textrm{BdG}}^{\Gamma}(\textbf{k})$ are totally decoupled, $\mathcal{H}_{\textrm{BdG}}^{\Gamma}(\textbf{k})$ can be reduced into two blocks as
	\begin{align}
	\mathcal{H}_{\textrm{BdG}}^{\Gamma}(\textbf{k})=
	\begin{pmatrix}
	\mathcal{H}_{\textrm{BdG}}^{\Gamma,\uparrow\uparrow}(\textbf{k})	& 0 \\
	0	& \mathcal{H}_{\textrm{BdG}}^{\Gamma,\downarrow\downarrow}(\textbf{k})
	\end{pmatrix}.
	\end{align}
	The Chern numbers that the occupied bands of $\mathcal{H}_{\textrm{BdG}}^{\Gamma,\uparrow\uparrow}(\textbf{k})$ and $\mathcal{H}_{\textrm{BdG}}^{\Gamma,\downarrow\downarrow}(\textbf{k})$ carry can determined by the Wilson loop calculation~\cite{alexandradinata2014wilson}.
	The Wilson loop operator, defined as a path-ordered exponential of the Berry connection, is given by
	\begin{equation}
	W_{(k_{X}+\sqrt{2}\pi,k_{Y})\leftarrow(k_{X},k_{Y})}=\lim_{N\rightarrow\infty}F_{N-1}F_{N-2}\cdots F_{1}F_{0},
	\end{equation}
	where $[F_{i}]_{mn}=\left\langle u_{m}(\frac{\sqrt{2}\pi(i+1)}{N},k_{Y})\middle| u_{n}(\frac{\sqrt{2}\pi i}{N},k_{Y})\right\rangle$. $k_{X}$ and $k_{Y}$ are taken to be parallel to the reciprocal lattice vectors $\textbf{G}_{1}=\sqrt{2}(\pi,\pi)$ and $\textbf{G}_{2}=\sqrt{2}(-\pi,\pi)$, respectively. In Fig. \ref{NewSFig1}, we show the winding in the eigenvalue spectrum of $W$ which indicate the non-zero Chern number, together with the corresponding edge mode obtained from the finite-size system calculation for each case. We note that even when the next-nearest neighbor (NNN) interaction terms are included, the new pairing channels do not change the Chern numbers. Thus, the results with and without the NNN interaction are qualitatively the same.
	\begin{figure}[h!]
		\begin{center}
			\includegraphics[scale=0.4]{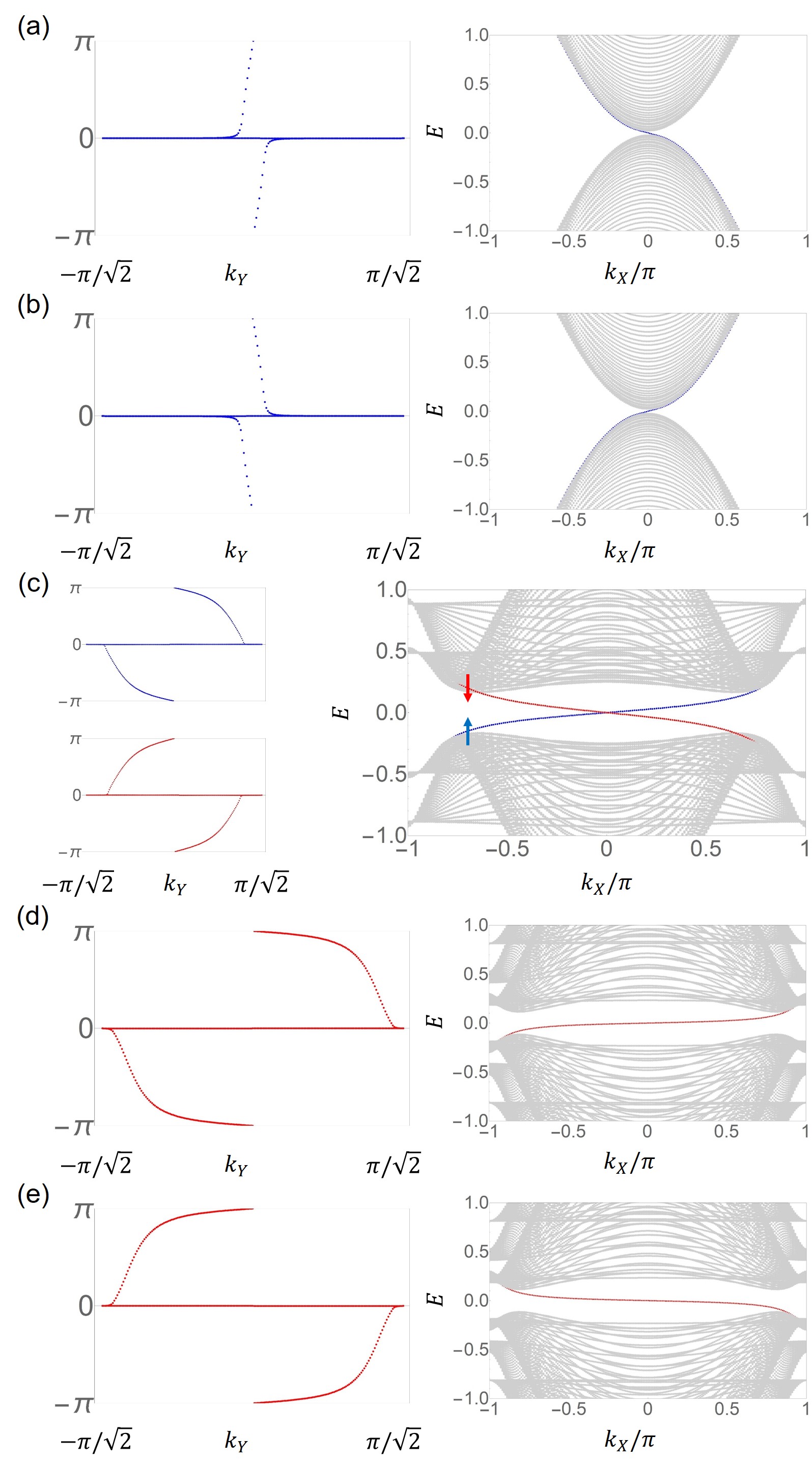}
			\caption{The Wilson loop spectrum of the topologically non-trivial occupied bands of $\mathcal{H}_{\textrm{BdG}}^{\Gamma,\sigma\sigma}(\textbf{k})$s (left panel), and the corresponding edge modes localized (right panel) on one of the edges of an OAFM system that is periodic along $k_{X}$ direction, but has a finite length along $k_{Y}$ direction. (a) OAFM -- (i) -- $A_{u}$, (b) OAFM -- (i) -- $B_{u}$, (c) OAFM -- (ii) -- $B_{u}$, (d) OAFM -- (iii) -- $A_{u}$, and (e) OAFM -- (iii) -- $B_{u}$. The blue (red) color indicates the spin-up (down) bands.}\label{NewSFig1}
		\end{center}
	\end{figure}
	\\
	\\
	I-AFM\\
	\indent For a system that supports both particle-hole symmetry and inversion symmetry, we consider a combination of the two symmetries $CP=U_{CP}K$ where $K$ is complex conjugation and $U_{CP}$ is a unitary matrix. If $(CP)^{2}=1$, it is always possible to choose a basis in which $U_{CP}=\mathbbm{1}$. In such a basis, the Hamiltonian of our system $H_{\textrm{BdG}}(\textbf{k})$ satisfies the following relation; $(CP)H_{\textrm{BdG}}(\textbf{k})(CP)^{1}=U_{CP}H_{\textrm{BdG}}^{*}(\textbf{k})U_{CP}^{-1}=H_{\textrm{BdG}}^{*}(\textbf{k})=-H_{\textrm{BdG}}(\textbf{k})$. Thus, in this basis, $H_{\textrm{BdG}}(\textbf{k})$ is purely imaginary. When $H_{\textrm{BdG}}(\textbf{k})$ is a $2^{n}\times2^{n}$ matrix, it can be written as $H_{\textrm{BdG}}(\textbf{k})=\sum_{i_{1},i_{2},\cdots,i_{n}}f_{i_{1},i_{2},\cdots,i_{n}}(\textbf{k})\sigma_{i_{1}}^{1}\otimes\sigma_{i_{2}}^{2}\otimes\cdots\otimes\sigma_{i_{n}}^{n}$, where $\sigma^{j}$s are the Pauli matrices ($i_{j}=0,x,y,z$). To make $H_{\textrm{BdG}}(\textbf{k})$ imaginary, odd numbers of $i_{j}$ must be $y$. It means that we can always find a basis in which $H_{\textrm{BdG}}(\textbf{k})$ is skew-symmetric. In such a basis, we can define a $\mathbb{Z}_{2}$ invariant $n_{\mathbb{Z}_{2}}^{\nu,\pm}$ for spin-up and -down sectors respectively as follows
	\begin{equation}
	n_{\mathbb{Z}_{2}}^{\nu,\pm}=\sgn[\pf H_{\textrm{BdG}}(k_{y}=0)]\sgn[\pf H_{\textrm{BdG}}(k_{y}=\pi)].
	\end{equation}
	$n_{\mathbb{Z}_{2}}^{\nu,\pm}$ is defined on two mirror reflection invariant lines $k_{x}=\nu=0,\pi$. It counts the change in the number of occupied states modulo 2 with mirror eigenvalue $+1$ ($-1$) at two points $k_{y}=0$ and $k_{y}=\pi$ on the mirror reflection invariant lines $k_{x}=0,\pi$. This $Z_{2}$ invariant is well-defined regardless of bulk gapped or gapless. It does not protect the band crossings in the bulk energy spectrum, but determines the existence of zero-energy state at the time-reversal invariant momenta (TRIM) points in the edge momentum space. When the bulk is gapped out by introduction of random perturbation that respects the symmetries, it can be seen as a mirror Chern superconductor like the O-AFM case.
	The solution for the tight-binding Hamiltonian of the finite-size IAFM system in the same geometry as for the OAFM case is displayed in Fig. \ref{NewSFig2}.
	\begin{figure}[h!]
		\begin{center}  
			\includegraphics[scale=0.194]{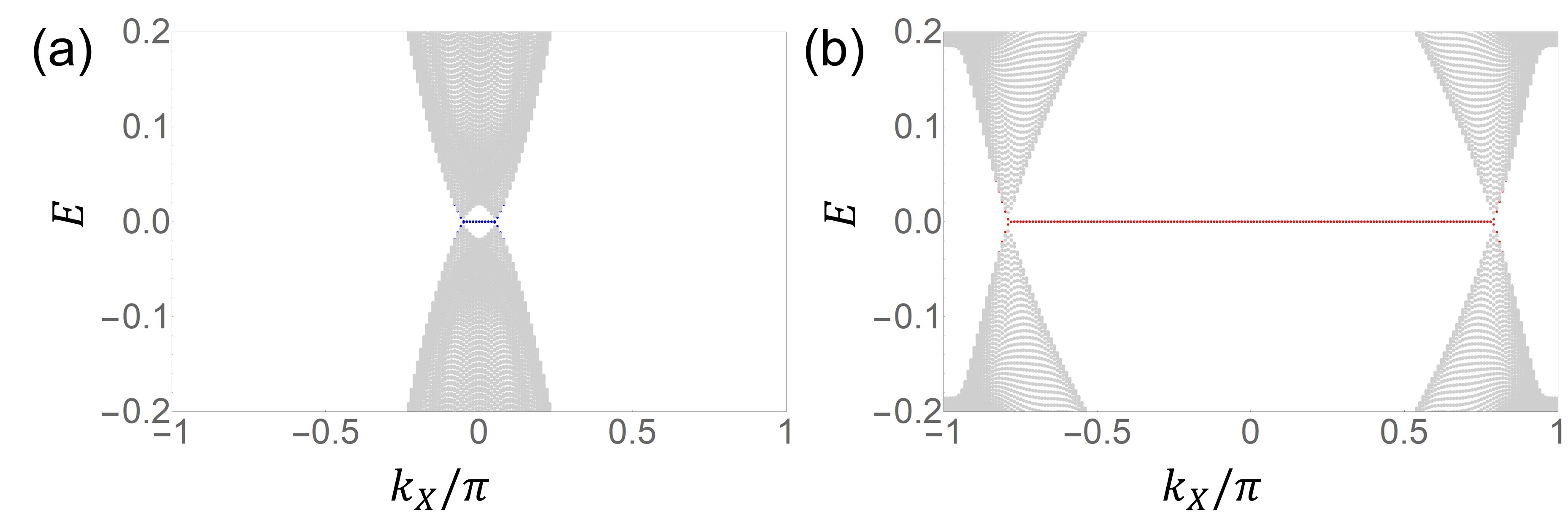}
			\caption{The edge modes localized (right panel) on the edges of an IAFM system that is periodic along $k_{X}$ direction, but has a finite length along $k_{Y}$ direction. (a) IAFM -- (i), and (b) IAFM --(iii).}\label{NewSFig2}
		\end{center}
	\end{figure}
	\\
	\\
	\\
	\noindent
	\textbf{$\mathbb{Z}_{2}$ indices for inversion symmetry protected second-order topological superconductors in two-dimension}\\
	\indent The second-order topological superconductors protected by inversion symmetry are characterized by a $\mathbb{Z}_{2}$ index $\nu_{2^{n}}^{\textrm{BdG}}$ defined by
	\begin{align}\label{index}
	v_{2^{n}}^{\textrm{BdG}}\equiv\sum_{\textbf{K}\in\textrm{TRIM}}\left[\frac{n_{-}^{\textrm{BdG};o}(\textbf{K})}{2^{n}}\right]_{\textrm{floor}},
	\end{align}
	where $[M+a]_{\textrm{floor}}=M$ for an integer $M$ and a real number $a\in[0,1]$~\cite{ahn2019higher}. $n_{-}^{\textrm{BdG};o}(\textbf{K})$ denotes the number of occupied bands with inversion eigenvalue $-1$ at the time-reversal invariant (TRIM) points $\textbf{K}=\Gamma,K_{1},K_{2},K_{3}$. $2^{n}$ in the subscript of $\nu$ index and the denominator of the RHS in Eq. \ref{index} denote the number of bands that meet the Fermi level in the normal state. It is nothing but the number of bands inverted during the band inversion process starting from the topologically trivial limit (i.e. the atomic limit)~\cite{skurativska2020atomic}.\\
	\indent We now verify that our models, both O-AFM and I-AFM, correspond to the non-trivial case according to Eq. \ref{index} when there are two Fermi surfaces in the normal states (i.e. Case (ii)). In the Nambu basis, inversion symmetry operator $P_{\textbf{k}}$ is given by
	\begin{equation}
	\begin{pmatrix}
	1	& 	& 	& \\
	& e^{-ik_{x}}	& 	& \\
	& 	& 1	& \\
	& 	& 	& e^{-ik_{x}}
	\end{pmatrix}\otimes\mathbbm{1}_{2}.
	\end{equation}
	After we transform the basis to that in which $H_{\textrm{BdG}}(\textbf{k})$ is periodic in the momentum space (i.e. $H_{\textrm{BdG}}(\textbf{k})=H_{\textrm{BdG}}(\textbf{k}+\textbf{G})$), we calculate the inversion symmetry eigenvalues of quasiparticle bands at following TRIM points: $\Gamma(0,0)$ $X'(\pi/2,-\pi/2)$, $Y'(\pi/2,\pi/2)$, $M'(\pi,0)$. In Fig. \ref{SFig3}, inversion symmetry eigenvalues of the normal state bands (solid lines) and their particle-hole partners (dashed bands) at TRIM points are shown.
	\begin{figure}[h!]
		\begin{center}
			\includegraphics[scale=0.28]{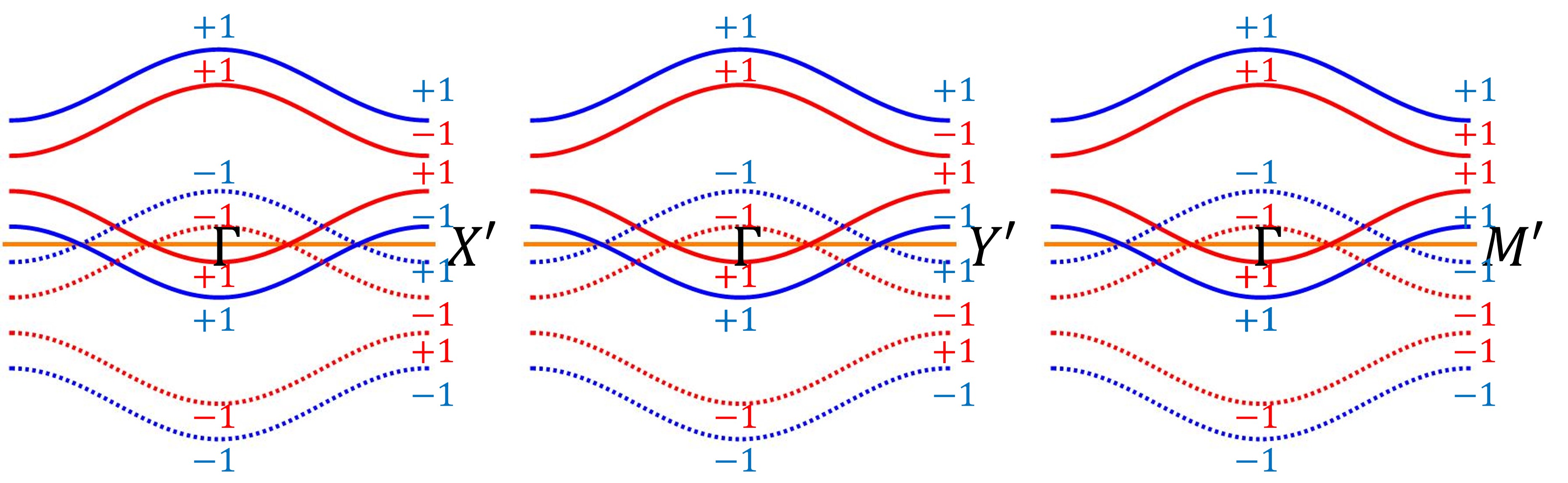}
			\caption{The inversion symmetry eigenvalues of the bands of $H_{\textrm{BdG}}^{\Gamma}$ at TRIM points.}\label{SFig3}
		\end{center}
	\end{figure}
	From Eq. \ref{index}, we obtain that $v_{2}^{\textrm{BdG}}=1+1+1+2\equiv1~(\textrm{mod}~2)$, since $n_{-}^{\textrm{BdG};o}(\Gamma)=n_{-}^{\textrm{BdG};o}(X')=n_{-}^{\textrm{BdG};o}(Y')=2$ and $n_{-}^{\textrm{BdG};o}(M')=4$.
	Indeed, we confirm that a pair of zero-energy mode localized at the corners arise when both the bulk and the edge energy spectrum is gapped in both O-AFM and I-AFM cases (Fig. \ref{SFig4}).
	\begin{figure}[h!]
		\begin{center}
			\includegraphics[scale=0.28]{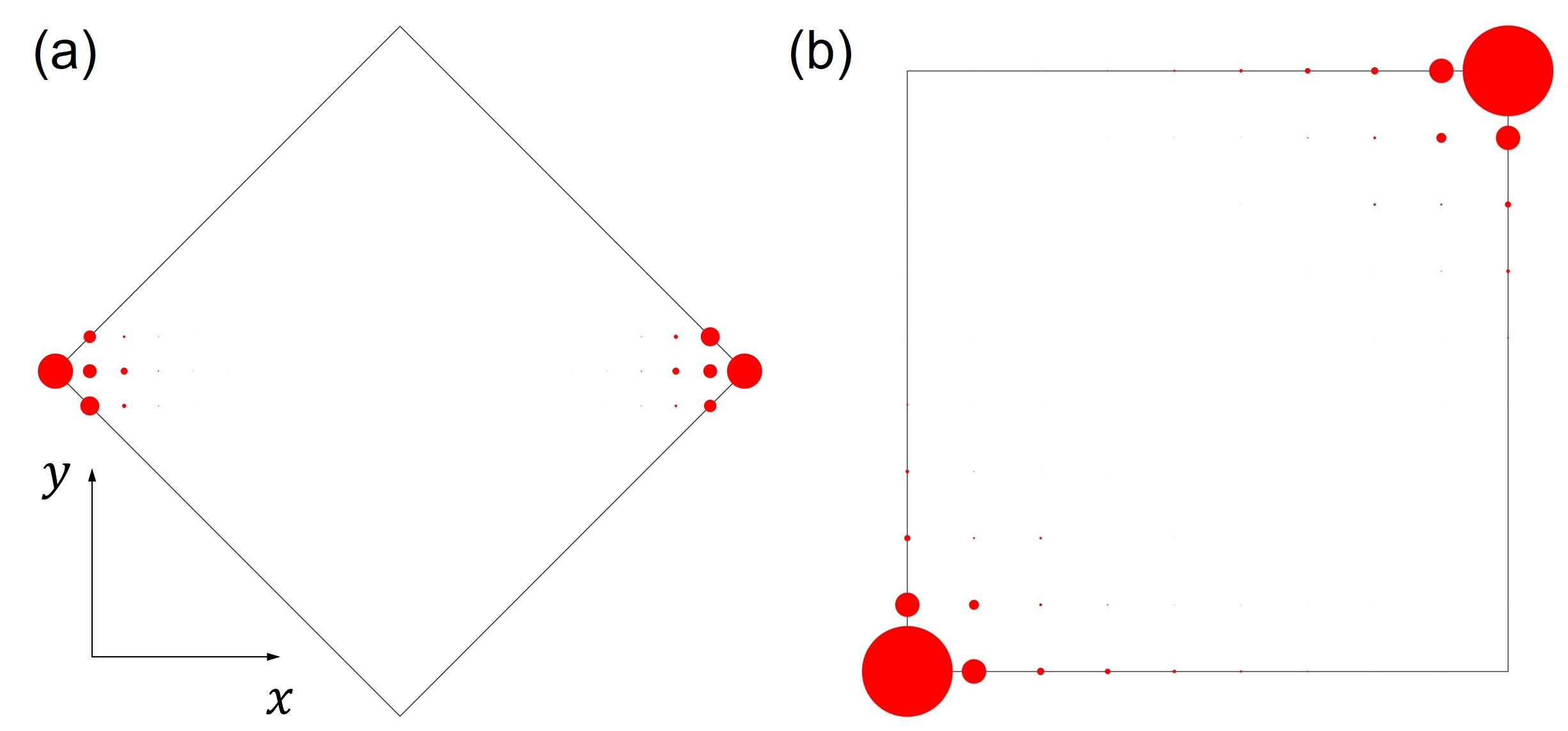}
			\caption{The charge distribution of zero-energy in-gap states localized at the corners of the finite-size systems in the square geometry: (a) O-AFM case with $11\times11$ unit cells, (b) I-AFM case with $21\times21$ unit cells. The radii of the red dots dictate the absolute value of the charge in the unit cells located at their positions}\label{SFig4}
		\end{center}
	\end{figure}
	
\end{bibunit}


\begin{thebibliography}{73}%
	\makeatletter
	\providecommand \@ifxundefined [1]{%
		\@ifx{#1\undefined}
	}%
	\providecommand \@ifnum [1]{%
		\ifnum #1\expandafter \@firstoftwo
		\else \expandafter \@secondoftwo
		\fi
	}%
	\providecommand \@ifx [1]{%
		\ifx #1\expandafter \@firstoftwo
		\else \expandafter \@secondoftwo
		\fi
	}%
	\providecommand \natexlab [1]{#1}%
	\providecommand \enquote  [1]{``#1''}%
	\providecommand \bibnamefont  [1]{#1}%
	\providecommand \bibfnamefont [1]{#1}%
	\providecommand \citenamefont [1]{#1}%
	\providecommand \href@noop [0]{\@secondoftwo}%
	\providecommand \href [0]{\begingroup \@sanitize@url \@href}%
	\providecommand \@href[1]{\@@startlink{#1}\@@href}%
	\providecommand \@@href[1]{\endgroup#1\@@endlink}%
	\providecommand \@sanitize@url [0]{\catcode `\\12\catcode `\$12\catcode
		`\&12\catcode `\#12\catcode `\^12\catcode `\_12\catcode `\%12\relax}%
	\providecommand \@@startlink[1]{}%
	\providecommand \@@endlink[0]{}%
	\providecommand \url  [0]{\begingroup\@sanitize@url \@url }%
	\providecommand \@url [1]{\endgroup\@href {#1}{\urlprefix }}%
	\providecommand \urlprefix  [0]{URL }%
	\providecommand \Eprint [0]{\href }%
	\providecommand \doibase [0]{http://dx.doi.org/}%
	\providecommand \selectlanguage [0]{\@gobble}%
	\providecommand \bibinfo  [0]{\@secondoftwo}%
	\providecommand \bibfield  [0]{\@secondoftwo}%
	\providecommand \translation [1]{[#1]}%
	\providecommand \BibitemOpen [0]{}%
	\providecommand \bibitemStop [0]{}%
	\providecommand \bibitemNoStop [0]{.\EOS\space}%
	\providecommand \EOS [0]{\spacefactor3000\relax}%
	\providecommand \BibitemShut  [1]{\csname bibitem#1\endcsname}%
	\let\auto@bib@innerbib\@empty
	\bibitem [{\citenamefont {Taillefer}(2010)}]{taillefer2010scattering}%
	\BibitemOpen
	\bibfield  {author} {\bibinfo {author} {\bibfnamefont {L.}~\bibnamefont
			{Taillefer}},\ }\href@noop {} {\bibfield  {journal} {\bibinfo  {journal}
			{Annu. Rev. Condens. Matter Phys.}\ }\textbf {\bibinfo {volume} {1}},\
		\bibinfo {pages} {51} (\bibinfo {year} {2010})}\BibitemShut {NoStop}%
	\bibitem [{\citenamefont {Nagaosa}(1997)}]{nagaosa1997superconductivity}%
	\BibitemOpen
	\bibfield  {author} {\bibinfo {author} {\bibfnamefont {N.}~\bibnamefont
			{Nagaosa}},\ }\href@noop {} {\bibfield  {journal} {\bibinfo  {journal}
			{Science}\ }\textbf {\bibinfo {volume} {275}},\ \bibinfo {pages} {1078}
		(\bibinfo {year} {1997})}\BibitemShut {NoStop}%
	\bibitem [{\citenamefont {Sachdev}(2012)}]{sachdev2012entangling}%
	\BibitemOpen
	\bibfield  {author} {\bibinfo {author} {\bibfnamefont {S.}~\bibnamefont
			{Sachdev}},\ }\href@noop {} {\bibfield  {journal} {\bibinfo  {journal}
			{Science}\ }\textbf {\bibinfo {volume} {336}},\ \bibinfo {pages} {1510}
		(\bibinfo {year} {2012})}\BibitemShut {NoStop}%
	\bibitem [{\citenamefont {Kuwabara}\ and\ \citenamefont
		{Ogata}(2000)}]{kuwabara2000spin}%
	\BibitemOpen
	\bibfield  {author} {\bibinfo {author} {\bibfnamefont {T.}~\bibnamefont
			{Kuwabara}}\ and\ \bibinfo {author} {\bibfnamefont {M.}~\bibnamefont
			{Ogata}},\ }\href@noop {} {\bibfield  {journal} {\bibinfo  {journal} {Phys.
				Rev. Lett.}\ }\textbf {\bibinfo {volume} {85}},\ \bibinfo {pages} {4586}
		(\bibinfo {year} {2000})}\BibitemShut {NoStop}%
	\bibitem [{\citenamefont {Meng}\ \emph {et~al.}(2014)\citenamefont {Meng},
		\citenamefont {Kim},\ and\ \citenamefont {Kee}}]{meng2014odd}%
	\BibitemOpen
	\bibfield  {author} {\bibinfo {author} {\bibfnamefont {Z.~Y.}\ \bibnamefont
			{Meng}}, \bibinfo {author} {\bibfnamefont {Y.~B.}\ \bibnamefont {Kim}}, \
		and\ \bibinfo {author} {\bibfnamefont {H.-Y.}\ \bibnamefont {Kee}},\
	}\href@noop {} {\bibfield  {journal} {\bibinfo  {journal} {Phys. Rev. Lett.}\
		}\textbf {\bibinfo {volume} {113}},\ \bibinfo {pages} {177003} (\bibinfo
		{year} {2014})}\BibitemShut {NoStop}%
	\bibitem [{\citenamefont {Sumita}\ \emph {et~al.}(2017)\citenamefont {Sumita},
		\citenamefont {Nomoto},\ and\ \citenamefont {Yanase}}]{sumita2017multipole}%
	\BibitemOpen
	\bibfield  {author} {\bibinfo {author} {\bibfnamefont {S.}~\bibnamefont
			{Sumita}}, \bibinfo {author} {\bibfnamefont {T.}~\bibnamefont {Nomoto}}, \
		and\ \bibinfo {author} {\bibfnamefont {Y.}~\bibnamefont {Yanase}},\
	}\href@noop {} {\bibfield  {journal} {\bibinfo  {journal} {Phys. Rev. Lett.}\
		}\textbf {\bibinfo {volume} {119}},\ \bibinfo {pages} {027001} (\bibinfo
		{year} {2017})}\BibitemShut {NoStop}%
	\bibitem [{\citenamefont {Almeida}\ \emph {et~al.}(2017)\citenamefont
		{Almeida}, \citenamefont {Fernandes},\ and\ \citenamefont
		{Miranda}}]{almeida2017induced}%
	\BibitemOpen
	\bibfield  {author} {\bibinfo {author} {\bibfnamefont {D.~E.}\ \bibnamefont
			{Almeida}}, \bibinfo {author} {\bibfnamefont {R.~M.}\ \bibnamefont
			{Fernandes}}, \ and\ \bibinfo {author} {\bibfnamefont {E.}~\bibnamefont
			{Miranda}},\ }\href@noop {} {\bibfield  {journal} {\bibinfo  {journal} {Phys.
				Rev. B}\ }\textbf {\bibinfo {volume} {96}},\ \bibinfo {pages} {014514}
		(\bibinfo {year} {2017})}\BibitemShut {NoStop}%
	\bibitem [{\citenamefont {Setty}\ \emph {et~al.}(2019)\citenamefont {Setty},
		\citenamefont {Bhattacharyya}, \citenamefont {Kreisel},\ and\ \citenamefont
		{Hirschfeld}}]{setty2019ultranodal}%
	\BibitemOpen
	\bibfield  {author} {\bibinfo {author} {\bibfnamefont {C.}~\bibnamefont
			{Setty}}, \bibinfo {author} {\bibfnamefont {S.}~\bibnamefont
			{Bhattacharyya}}, \bibinfo {author} {\bibfnamefont {A.}~\bibnamefont
			{Kreisel}}, \ and\ \bibinfo {author} {\bibfnamefont {P.}~\bibnamefont
			{Hirschfeld}},\ }\href@noop {} {\bibfield  {journal} {\bibinfo  {journal}
			{Nat. Commun.}\ }\textbf {\bibinfo {volume} {11}},\ \bibinfo {pages}
		{523} (\bibinfo {year} {2019})}\BibitemShut
	{NoStop}%
	\bibitem [{\citenamefont {Fay}\ and\ \citenamefont
		{Appel}(1980)}]{fay1980coexistence}%
	\BibitemOpen
	\bibfield  {author} {\bibinfo {author} {\bibfnamefont {D.}~\bibnamefont
			{Fay}}\ and\ \bibinfo {author} {\bibfnamefont {J.}~\bibnamefont {Appel}},\
	}\href@noop {} {\bibfield  {journal} {\bibinfo  {journal} {Phys. Rev. B}\
		}\textbf {\bibinfo {volume} {22}},\ \bibinfo {pages} {3173} (\bibinfo {year}
		{1980})}\BibitemShut {NoStop}%
	\bibitem [{\citenamefont {Tada}\ \emph {et~al.}(2013)\citenamefont {Tada},
		\citenamefont {Fujimoto}, \citenamefont {Kawakami}, \citenamefont {Hattori},
		\citenamefont {Ihara}, \citenamefont {Ishida}, \citenamefont {Deguchi},
		\citenamefont {Sato},\ and\ \citenamefont {Satoh}}]{tada2013spin}%
	\BibitemOpen
	\bibfield  {author} {\bibinfo {author} {\bibfnamefont {Y.}~\bibnamefont
			{Tada}}, \bibinfo {author} {\bibfnamefont {S.}~\bibnamefont {Fujimoto}},
		\bibinfo {author} {\bibfnamefont {N.}~\bibnamefont {Kawakami}}, \bibinfo
		{author} {\bibfnamefont {T.}~\bibnamefont {Hattori}}, \bibinfo {author}
		{\bibfnamefont {Y.}~\bibnamefont {Ihara}}, \bibinfo {author} {\bibfnamefont
			{K.}~\bibnamefont {Ishida}}, \bibinfo {author} {\bibfnamefont
			{K.}~\bibnamefont {Deguchi}}, \bibinfo {author} {\bibfnamefont
			{N.}~\bibnamefont {Sato}}, \ and\ \bibinfo {author} {\bibfnamefont
			{I.}~\bibnamefont {Satoh}},\ }in\ \href@noop {} {\emph {\bibinfo {booktitle}
			{Journal of Physics: Conference Series}}},\ Vol.\ \bibinfo {volume} {449}\
	(\bibinfo {organization} {IOP Publishing},\ \bibinfo {year} {2013})\ p.\
	\bibinfo {pages} {012029}\BibitemShut {NoStop}%
	\bibitem [{\citenamefont {Ishizuka}\ and\ \citenamefont
		{Yanase}(2018)}]{ishizuka2018odd}%
	\BibitemOpen
	\bibfield  {author} {\bibinfo {author} {\bibfnamefont {J.}~\bibnamefont
			{Ishizuka}}\ and\ \bibinfo {author} {\bibfnamefont {Y.}~\bibnamefont
			{Yanase}},\ }\href@noop {} {\bibfield  {journal} {\bibinfo  {journal} {Phys.
				Rev. B}\ }\textbf {\bibinfo {volume} {98}},\ \bibinfo {pages} {224510}
		(\bibinfo {year} {2018})}\BibitemShut {NoStop}%
	\bibitem [{\citenamefont {Machida}\ \emph {et~al.}(1980)\citenamefont
		{Machida}, \citenamefont {Nokura},\ and\ \citenamefont
		{Matsubara}}]{machida1980theory}%
	\BibitemOpen
	\bibfield  {author} {\bibinfo {author} {\bibfnamefont {K.}~\bibnamefont
			{Machida}}, \bibinfo {author} {\bibfnamefont {K.}~\bibnamefont {Nokura}}, \
		and\ \bibinfo {author} {\bibfnamefont {T.}~\bibnamefont {Matsubara}},\
	}\href@noop {} {\bibfield  {journal} {\bibinfo  {journal} {Phys. Rev. B}\
		}\textbf {\bibinfo {volume} {22}},\ \bibinfo {pages} {2307} (\bibinfo {year}
		{1980})}\BibitemShut {NoStop}%
	\bibitem [{\citenamefont {Fujimoto}(2006)}]{fujimoto2006emergent}%
	\BibitemOpen
	\bibfield  {author} {\bibinfo {author} {\bibfnamefont {S.}~\bibnamefont
			{Fujimoto}},\ }\href@noop {} {\bibfield  {journal} {\bibinfo  {journal}
			{Journal of the Physical Society of Japan}\ }\textbf {\bibinfo {volume}
			{75}},\ \bibinfo {pages} {083704} (\bibinfo {year} {2006})}\BibitemShut
	{NoStop}%
	\bibitem [{\citenamefont {Milovanovi{\'c}}\ and\ \citenamefont
		{Predin}(2012)}]{milovanovic2012coexistence}%
	\BibitemOpen
	\bibfield  {author} {\bibinfo {author} {\bibfnamefont {M.}~\bibnamefont
			{Milovanovi{\'c}}}\ and\ \bibinfo {author} {\bibfnamefont {S.}~\bibnamefont
			{Predin}},\ }\href@noop {} {\bibfield  {journal} {\bibinfo  {journal} {Phys.
				Rev. B}\ }\textbf {\bibinfo {volume} {86}},\ \bibinfo {pages} {195113}
		(\bibinfo {year} {2012})}\BibitemShut {NoStop}%
	\bibitem [{\citenamefont {Qi}\ \emph {et~al.}(2017)\citenamefont {Qi},
		\citenamefont {Fu}, \citenamefont {Sun},\ and\ \citenamefont
		{Gu}}]{qi2017coexistence}%
	\BibitemOpen
	\bibfield  {author} {\bibinfo {author} {\bibfnamefont {Y.}~\bibnamefont
			{Qi}}, \bibinfo {author} {\bibfnamefont {L.}~\bibnamefont {Fu}}, \bibinfo
		{author} {\bibfnamefont {K.}~\bibnamefont {Sun}}, \ and\ \bibinfo {author}
		{\bibfnamefont {Z.}~\bibnamefont {Gu}},\ }\href@noop {} {\bibfield  {journal}
		{\bibinfo  {journal} {arXiv preprint arXiv:1709.08232}\ } (\bibinfo {year}
		{2017})}\BibitemShut {NoStop}%
	\bibitem [{\citenamefont {Powell}\ \emph {et~al.}(2003)\citenamefont {Powell},
		\citenamefont {Annett},\ and\ \citenamefont {Gy{\"o}rffy}}]{powell2003gap}%
	\BibitemOpen
	\bibfield  {author} {\bibinfo {author} {\bibfnamefont {B.}~\bibnamefont
			{Powell}}, \bibinfo {author} {\bibfnamefont {J.~F.}\ \bibnamefont {Annett}},
		\ and\ \bibinfo {author} {\bibfnamefont {B.}~\bibnamefont {Gy{\"o}rffy}},\
	}\href@noop {} {\bibfield  {journal} {\bibinfo  {journal} {Journal of Physics
				A: Mathematical and General}\ }\textbf {\bibinfo {volume} {36}},\ \bibinfo
		{pages} {9289} (\bibinfo {year} {2003})}\BibitemShut {NoStop}%
	\bibitem [{\citenamefont {Cheung}\ and\ \citenamefont
		{Raghu}(2016)}]{cheung2016topological}%
	\BibitemOpen
	\bibfield  {author} {\bibinfo {author} {\bibfnamefont {A.~K.}\ \bibnamefont
			{Cheung}}\ and\ \bibinfo {author} {\bibfnamefont {S.}~\bibnamefont {Raghu}},\
	}\href@noop {} {\bibfield  {journal} {\bibinfo  {journal} {Phys. Rev. B}\
		}\textbf {\bibinfo {volume} {93}},\ \bibinfo {pages} {134516} (\bibinfo
		{year} {2016})}\BibitemShut {NoStop}%
	\bibitem [{\citenamefont {K{\k{a}}dzielawa-Major}\ \emph
		{et~al.}(2018)\citenamefont {K{\k{a}}dzielawa-Major}, \citenamefont
		{Fidrysiak}, \citenamefont {Kubiczek},\ and\ \citenamefont
		{Spa{\l}ek}}]{kkadzielawa2018spin}%
	\BibitemOpen
	\bibfield  {author} {\bibinfo {author} {\bibfnamefont {E.}~\bibnamefont
			{K{\k{a}}dzielawa-Major}}, \bibinfo {author} {\bibfnamefont {M.}~\bibnamefont
			{Fidrysiak}}, \bibinfo {author} {\bibfnamefont {P.}~\bibnamefont {Kubiczek}},
		\ and\ \bibinfo {author} {\bibfnamefont {J.}~\bibnamefont {Spa{\l}ek}},\
	}\href@noop {} {\bibfield  {journal} {\bibinfo  {journal} {Phys. Rev. B}\
		}\textbf {\bibinfo {volume} {97}},\ \bibinfo {pages} {224519} (\bibinfo
		{year} {2018})}\BibitemShut {NoStop}%
	\bibitem [{\citenamefont {Feyerherm}\ \emph {et~al.}(1994)\citenamefont
		{Feyerherm}, \citenamefont {Amato}, \citenamefont {Gygax}, \citenamefont
		{Schenck}, \citenamefont {Geibel}, \citenamefont {Steglich}, \citenamefont
		{Sato},\ and\ \citenamefont {Komatsubara}}]{feyerherm1994coexistence}%
	\BibitemOpen
	\bibfield  {author} {\bibinfo {author} {\bibfnamefont {R.}~\bibnamefont
			{Feyerherm}}, \bibinfo {author} {\bibfnamefont {A.}~\bibnamefont {Amato}},
		\bibinfo {author} {\bibfnamefont {F.}~\bibnamefont {Gygax}}, \bibinfo
		{author} {\bibfnamefont {A.}~\bibnamefont {Schenck}}, \bibinfo {author}
		{\bibfnamefont {C.}~\bibnamefont {Geibel}}, \bibinfo {author} {\bibfnamefont
			{F.}~\bibnamefont {Steglich}}, \bibinfo {author} {\bibfnamefont
			{N.}~\bibnamefont {Sato}}, \ and\ \bibinfo {author} {\bibfnamefont
			{T.}~\bibnamefont {Komatsubara}},\ }\href@noop {} {\bibfield  {journal}
		{\bibinfo  {journal} {Phys. Rev. Lett.}\ }\textbf {\bibinfo {volume} {73}},\
		\bibinfo {pages} {1849} (\bibinfo {year} {1994})}\BibitemShut {NoStop}%
	\bibitem [{\citenamefont {Pagliuso}\ \emph {et~al.}(2001)\citenamefont
		{Pagliuso}, \citenamefont {Petrovic}, \citenamefont {Movshovich},
		\citenamefont {Hall}, \citenamefont {Hundley}, \citenamefont {Sarrao},
		\citenamefont {Thompson},\ and\ \citenamefont
		{Fisk}}]{pagliuso2001coexistence}%
	\BibitemOpen
	\bibfield  {author} {\bibinfo {author} {\bibfnamefont {P.}~\bibnamefont
			{Pagliuso}}, \bibinfo {author} {\bibfnamefont {C.}~\bibnamefont {Petrovic}},
		\bibinfo {author} {\bibfnamefont {R.}~\bibnamefont {Movshovich}}, \bibinfo
		{author} {\bibfnamefont {D.}~\bibnamefont {Hall}}, \bibinfo {author}
		{\bibfnamefont {M.}~\bibnamefont {Hundley}}, \bibinfo {author} {\bibfnamefont
			{J.}~\bibnamefont {Sarrao}}, \bibinfo {author} {\bibfnamefont
			{J.}~\bibnamefont {Thompson}}, \ and\ \bibinfo {author} {\bibfnamefont
			{Z.}~\bibnamefont {Fisk}},\ }\href@noop {} {\bibfield  {journal} {\bibinfo
			{journal} {Phys. Rev. B}\ }\textbf {\bibinfo {volume} {64}},\ \bibinfo
		{pages} {100503} (\bibinfo {year} {2001})}\BibitemShut {NoStop}%
	\bibitem [{\citenamefont {Knebel}\ \emph {et~al.}(2006)\citenamefont {Knebel},
		\citenamefont {Aoki}, \citenamefont {Braithwaite}, \citenamefont {Salce},\
		and\ \citenamefont {Flouquet}}]{knebel2006coexistence}%
	\BibitemOpen
	\bibfield  {author} {\bibinfo {author} {\bibfnamefont {G.}~\bibnamefont
			{Knebel}}, \bibinfo {author} {\bibfnamefont {D.}~\bibnamefont {Aoki}},
		\bibinfo {author} {\bibfnamefont {D.}~\bibnamefont {Braithwaite}}, \bibinfo
		{author} {\bibfnamefont {B.}~\bibnamefont {Salce}}, \ and\ \bibinfo {author}
		{\bibfnamefont {J.}~\bibnamefont {Flouquet}},\ }\href@noop {} {\bibfield
		{journal} {\bibinfo  {journal} {Phys. Rev. B}\ }\textbf {\bibinfo {volume}
			{74}},\ \bibinfo {pages} {020501} (\bibinfo {year} {2006})}\BibitemShut
	{NoStop}%
	\bibitem [{\citenamefont {Saxena}\ \emph {et~al.}(2000)\citenamefont {Saxena},
		\citenamefont {Agarwal}, \citenamefont {Ahilan}, \citenamefont {Grosche},
		\citenamefont {Haselwimmer}, \citenamefont {Steiner}, \citenamefont {Pugh},
		\citenamefont {Walker}, \citenamefont {Julian}, \citenamefont {Monthoux}
		\emph {et~al.}}]{saxena2000superconductivity}%
	\BibitemOpen
	\bibfield  {author} {\bibinfo {author} {\bibfnamefont {S.}~\bibnamefont
			{Saxena}}, \bibinfo {author} {\bibfnamefont {P.}~\bibnamefont {Agarwal}},
		\bibinfo {author} {\bibfnamefont {K.}~\bibnamefont {Ahilan}}, \bibinfo
		{author} {\bibfnamefont {F.}~\bibnamefont {Grosche}}, \bibinfo {author}
		{\bibfnamefont {R.}~\bibnamefont {Haselwimmer}}, \bibinfo {author}
		{\bibfnamefont {M.}~\bibnamefont {Steiner}}, \bibinfo {author} {\bibfnamefont
			{E.}~\bibnamefont {Pugh}}, \bibinfo {author} {\bibfnamefont {I.}~\bibnamefont
			{Walker}}, \bibinfo {author} {\bibfnamefont {S.}~\bibnamefont {Julian}},
		\bibinfo {author} {\bibfnamefont {P.}~\bibnamefont {Monthoux}},  \emph
		{et~al.},\ }\href@noop {} {\bibfield  {journal} {\bibinfo  {journal}
			{Nature}\ }\textbf {\bibinfo {volume} {406}},\ \bibinfo {pages} {587}
		(\bibinfo {year} {2000})}\BibitemShut {NoStop}%
	\bibitem [{\citenamefont {Aoki}\ \emph {et~al.}(2001)\citenamefont {Aoki},
		\citenamefont {Huxley}, \citenamefont {Ressouche}, \citenamefont
		{Braithwaite}, \citenamefont {Flouquet}, \citenamefont {Brison},
		\citenamefont {Lhotel},\ and\ \citenamefont {Paulsen}}]{aoki2001coexistence}%
	\BibitemOpen
	\bibfield  {author} {\bibinfo {author} {\bibfnamefont {D.}~\bibnamefont
			{Aoki}}, \bibinfo {author} {\bibfnamefont {A.}~\bibnamefont {Huxley}},
		\bibinfo {author} {\bibfnamefont {E.}~\bibnamefont {Ressouche}}, \bibinfo
		{author} {\bibfnamefont {D.}~\bibnamefont {Braithwaite}}, \bibinfo {author}
		{\bibfnamefont {J.}~\bibnamefont {Flouquet}}, \bibinfo {author}
		{\bibfnamefont {J.-P.}\ \bibnamefont {Brison}}, \bibinfo {author}
		{\bibfnamefont {E.}~\bibnamefont {Lhotel}}, \ and\ \bibinfo {author}
		{\bibfnamefont {C.}~\bibnamefont {Paulsen}},\ }\href@noop {} {\bibfield
		{journal} {\bibinfo  {journal} {Nature}\ }\textbf {\bibinfo {volume} {413}},\
		\bibinfo {pages} {613} (\bibinfo {year} {2001})}\BibitemShut {NoStop}%
	\bibitem [{\citenamefont {Huy}\ \emph {et~al.}(2007)\citenamefont {Huy},
		\citenamefont {Gasparini}, \citenamefont {De~Nijs}, \citenamefont {Huang},
		\citenamefont {Klaasse}, \citenamefont {Gortenmulder}, \citenamefont
		{de~Visser}, \citenamefont {Hamann}, \citenamefont {G{\"o}rlach},\ and\
		\citenamefont {L{\"o}hneysen}}]{huy2007superconductivity}%
	\BibitemOpen
	\bibfield  {author} {\bibinfo {author} {\bibfnamefont {N.}~\bibnamefont
			{Huy}}, \bibinfo {author} {\bibfnamefont {A.}~\bibnamefont {Gasparini}},
		\bibinfo {author} {\bibfnamefont {D.}~\bibnamefont {De~Nijs}}, \bibinfo
		{author} {\bibfnamefont {Y.}~\bibnamefont {Huang}}, \bibinfo {author}
		{\bibfnamefont {J.}~\bibnamefont {Klaasse}}, \bibinfo {author} {\bibfnamefont
			{T.}~\bibnamefont {Gortenmulder}}, \bibinfo {author} {\bibfnamefont
			{A.}~\bibnamefont {de~Visser}}, \bibinfo {author} {\bibfnamefont
			{A.}~\bibnamefont {Hamann}}, \bibinfo {author} {\bibfnamefont
			{T.}~\bibnamefont {G{\"o}rlach}}, \ and\ \bibinfo {author} {\bibfnamefont
			{H.~v.}\ \bibnamefont {L{\"o}hneysen}},\ }\href@noop {} {\bibfield  {journal}
		{\bibinfo  {journal} {Phys. Rev. Lett.}\ }\textbf {\bibinfo {volume} {99}},\
		\bibinfo {pages} {067006} (\bibinfo {year} {2007})}\BibitemShut {NoStop}%
	\bibitem [{\citenamefont {Linder}\ \emph {et~al.}(2008)\citenamefont {Linder},
		\citenamefont {Sperstad}, \citenamefont {Nevidomskyy}, \citenamefont
		{Cuoco},\ and\ \citenamefont {Sudb{\o}}}]{linder2008coexistence}%
	\BibitemOpen
	\bibfield  {author} {\bibinfo {author} {\bibfnamefont {J.}~\bibnamefont
			{Linder}}, \bibinfo {author} {\bibfnamefont {I.~B.}\ \bibnamefont
			{Sperstad}}, \bibinfo {author} {\bibfnamefont {A.~H.}\ \bibnamefont
			{Nevidomskyy}}, \bibinfo {author} {\bibfnamefont {M.}~\bibnamefont {Cuoco}},
		\ and\ \bibinfo {author} {\bibfnamefont {A.}~\bibnamefont {Sudb{\o}}},\
	}\href@noop {} {\bibfield  {journal} {\bibinfo  {journal} {Phys. Rev. B}\
		}\textbf {\bibinfo {volume} {77}},\ \bibinfo {pages} {184511} (\bibinfo
		{year} {2008})}\BibitemShut {NoStop}%
	\bibitem [{\citenamefont {Gasparini}\ \emph {et~al.}(2010)\citenamefont
		{Gasparini}, \citenamefont {Huang}, \citenamefont {Huy}, \citenamefont
		{Klaasse}, \citenamefont {Naka}, \citenamefont {Slooten},\ and\ \citenamefont
		{De~Visser}}]{gasparini2010superconducting}%
	\BibitemOpen
	\bibfield  {author} {\bibinfo {author} {\bibfnamefont {A.}~\bibnamefont
			{Gasparini}}, \bibinfo {author} {\bibfnamefont {Y.}~\bibnamefont {Huang}},
		\bibinfo {author} {\bibfnamefont {N.}~\bibnamefont {Huy}}, \bibinfo {author}
		{\bibfnamefont {J.}~\bibnamefont {Klaasse}}, \bibinfo {author} {\bibfnamefont
			{T.}~\bibnamefont {Naka}}, \bibinfo {author} {\bibfnamefont {E.}~\bibnamefont
			{Slooten}}, \ and\ \bibinfo {author} {\bibfnamefont {A.}~\bibnamefont
			{De~Visser}},\ }\href@noop {} {\bibfield  {journal} {\bibinfo  {journal}
			{Journal of Low Temperature Physics}\ }\textbf {\bibinfo {volume} {161}},\
		\bibinfo {pages} {134} (\bibinfo {year} {2010})}\BibitemShut {NoStop}%
	\bibitem [{\citenamefont {Wu}\ \emph {et~al.}(2017)\citenamefont {Wu},
		\citenamefont {Bastien}, \citenamefont {Taupin}, \citenamefont {Paulsen},
		\citenamefont {Howald}, \citenamefont {Aoki},\ and\ \citenamefont
		{Brison}}]{wu2017pairing}%
	\BibitemOpen
	\bibfield  {author} {\bibinfo {author} {\bibfnamefont {B.}~\bibnamefont
			{Wu}}, \bibinfo {author} {\bibfnamefont {G.}~\bibnamefont {Bastien}},
		\bibinfo {author} {\bibfnamefont {M.}~\bibnamefont {Taupin}}, \bibinfo
		{author} {\bibfnamefont {C.}~\bibnamefont {Paulsen}}, \bibinfo {author}
		{\bibfnamefont {L.}~\bibnamefont {Howald}}, \bibinfo {author} {\bibfnamefont
			{D.}~\bibnamefont {Aoki}}, \ and\ \bibinfo {author} {\bibfnamefont {J.-P.}\
			\bibnamefont {Brison}},\ }\href@noop {} {\bibfield  {journal} {\bibinfo
			{journal} {Nat. Commun.}\ }\textbf {\bibinfo {volume} {8}},\ \bibinfo {pages}
		{14480} (\bibinfo {year} {2017})}\BibitemShut {NoStop}%
	\bibitem [{\citenamefont {Ran}\ \emph {et~al.}(2019)\citenamefont {Ran},
		\citenamefont {Eckberg}, \citenamefont {Ding}, \citenamefont {Furukawa},
		\citenamefont {Metz}, \citenamefont {Saha}, \citenamefont {Liu},
		\citenamefont {Zic}, \citenamefont {Kim}, \citenamefont {Paglione} \emph
		{et~al.}}]{ran2019nearly}%
	\BibitemOpen
	\bibfield  {author} {\bibinfo {author} {\bibfnamefont {S.}~\bibnamefont
			{Ran}}, \bibinfo {author} {\bibfnamefont {C.}~\bibnamefont {Eckberg}},
		\bibinfo {author} {\bibfnamefont {Q.-P.}\ \bibnamefont {Ding}}, \bibinfo
		{author} {\bibfnamefont {Y.}~\bibnamefont {Furukawa}}, \bibinfo {author}
		{\bibfnamefont {T.}~\bibnamefont {Metz}}, \bibinfo {author} {\bibfnamefont
			{S.~R.}\ \bibnamefont {Saha}}, \bibinfo {author} {\bibfnamefont {I.-L.}\
			\bibnamefont {Liu}}, \bibinfo {author} {\bibfnamefont {M.}~\bibnamefont
			{Zic}}, \bibinfo {author} {\bibfnamefont {H.}~\bibnamefont {Kim}}, \bibinfo
		{author} {\bibfnamefont {J.}~\bibnamefont {Paglione}},  \emph {et~al.},\
	}\href@noop {} {\bibfield  {journal} {\bibinfo  {journal} {Science}\ }\textbf
		{\bibinfo {volume} {365}},\ \bibinfo {pages} {684} (\bibinfo {year}
		{2019})}\BibitemShut {NoStop}%
	\bibitem [{\citenamefont {Sundar}\ \emph {et~al.}(2019)\citenamefont {Sundar},
		\citenamefont {Gheidi}, \citenamefont {Akintola}, \citenamefont
		{C{\^o}t{\'e}}, \citenamefont {Dunsiger}, \citenamefont {Ran}, \citenamefont
		{Butch}, \citenamefont {Saha}, \citenamefont {Paglione},\ and\ \citenamefont
		{Sonier}}]{sundar2019coexistence}%
	\BibitemOpen
	\bibfield  {author} {\bibinfo {author} {\bibfnamefont {S.}~\bibnamefont
			{Sundar}}, \bibinfo {author} {\bibfnamefont {S.}~\bibnamefont {Gheidi}},
		\bibinfo {author} {\bibfnamefont {K.}~\bibnamefont {Akintola}}, \bibinfo
		{author} {\bibfnamefont {A.}~\bibnamefont {C{\^o}t{\'e}}}, \bibinfo {author}
		{\bibfnamefont {S.}~\bibnamefont {Dunsiger}}, \bibinfo {author}
		{\bibfnamefont {S.}~\bibnamefont {Ran}}, \bibinfo {author} {\bibfnamefont
			{N.}~\bibnamefont {Butch}}, \bibinfo {author} {\bibfnamefont
			{S.}~\bibnamefont {Saha}}, \bibinfo {author} {\bibfnamefont {J.}~\bibnamefont
			{Paglione}}, \ and\ \bibinfo {author} {\bibfnamefont {J.}~\bibnamefont
			{Sonier}},\ }\href@noop {} {\bibfield  {journal} {\bibinfo  {journal} {Phys.
				Rev. B}\ }\textbf {\bibinfo {volume} {100}},\ \bibinfo {pages} {140502}
		(\bibinfo {year} {2019})}\BibitemShut {NoStop}%
	\bibitem [{\citenamefont {Ikeda}\ \emph {et~al.}(2018)\citenamefont {Ikeda},
		\citenamefont {Tsuchiya}, \citenamefont {Zhang}, \citenamefont {Kishimoto},
		\citenamefont {Kikegawa}, \citenamefont {Yoda}, \citenamefont {Nakamura},
		\citenamefont {Machida}, \citenamefont {Glasbrenner},\ and\ \citenamefont
		{Kobayashi}}]{ikeda2018new}%
	\BibitemOpen
	\bibfield  {author} {\bibinfo {author} {\bibfnamefont {S.}~\bibnamefont
			{Ikeda}}, \bibinfo {author} {\bibfnamefont {Y.}~\bibnamefont {Tsuchiya}},
		\bibinfo {author} {\bibfnamefont {X.-W.}\ \bibnamefont {Zhang}}, \bibinfo
		{author} {\bibfnamefont {S.}~\bibnamefont {Kishimoto}}, \bibinfo {author}
		{\bibfnamefont {T.}~\bibnamefont {Kikegawa}}, \bibinfo {author}
		{\bibfnamefont {Y.}~\bibnamefont {Yoda}}, \bibinfo {author} {\bibfnamefont
			{H.}~\bibnamefont {Nakamura}}, \bibinfo {author} {\bibfnamefont
			{M.}~\bibnamefont {Machida}}, \bibinfo {author} {\bibfnamefont {J.~K.}\
			\bibnamefont {Glasbrenner}}, \ and\ \bibinfo {author} {\bibfnamefont
			{H.}~\bibnamefont {Kobayashi}},\ }\href {\doibase 10.1103/PhysRevB.98.100502}
	{\bibfield  {journal} {\bibinfo  {journal} {Phys. Rev. B}\ }\textbf {\bibinfo
			{volume} {98}},\ \bibinfo {pages} {100502} (\bibinfo {year}
		{2018})}\BibitemShut {NoStop}%
	\bibitem [{\citenamefont {Lu}\ \emph {et~al.}(2015)\citenamefont {Lu},
		\citenamefont {Wang}, \citenamefont {Wu}, \citenamefont {Wu}, \citenamefont
		{Zhao}, \citenamefont {Zeng}, \citenamefont {Luo}, \citenamefont {Wu},
		\citenamefont {Bao}, \citenamefont {Zhang} \emph
		{et~al.}}]{lu2015coexistence}%
	\BibitemOpen
	\bibfield  {author} {\bibinfo {author} {\bibfnamefont {X.}~\bibnamefont
			{Lu}}, \bibinfo {author} {\bibfnamefont {N.}~\bibnamefont {Wang}}, \bibinfo
		{author} {\bibfnamefont {H.}~\bibnamefont {Wu}}, \bibinfo {author}
		{\bibfnamefont {Y.}~\bibnamefont {Wu}}, \bibinfo {author} {\bibfnamefont
			{D.}~\bibnamefont {Zhao}}, \bibinfo {author} {\bibfnamefont {X.}~\bibnamefont
			{Zeng}}, \bibinfo {author} {\bibfnamefont {X.}~\bibnamefont {Luo}}, \bibinfo
		{author} {\bibfnamefont {T.}~\bibnamefont {Wu}}, \bibinfo {author}
		{\bibfnamefont {W.}~\bibnamefont {Bao}}, \bibinfo {author} {\bibfnamefont
			{G.}~\bibnamefont {Zhang}},  \emph {et~al.},\ }\href@noop {} {\bibfield
		{journal} {\bibinfo  {journal} {Nature Mat.}\ }\textbf {\bibinfo {volume}
			{14}},\ \bibinfo {pages} {325} (\bibinfo {year} {2015})}\BibitemShut
	{NoStop}%
	\bibitem [{\citenamefont {Pratt}\ \emph {et~al.}(2009)\citenamefont {Pratt},
		\citenamefont {Tian}, \citenamefont {Kreyssig}, \citenamefont {Zarestky},
		\citenamefont {Nandi}, \citenamefont {Ni}, \citenamefont {Bud’ko},
		\citenamefont {Canfield}, \citenamefont {Goldman},\ and\ \citenamefont
		{McQueeney}}]{pratt2009coexistence}%
	\BibitemOpen
	\bibfield  {author} {\bibinfo {author} {\bibfnamefont {D.}~\bibnamefont
			{Pratt}}, \bibinfo {author} {\bibfnamefont {W.}~\bibnamefont {Tian}},
		\bibinfo {author} {\bibfnamefont {A.}~\bibnamefont {Kreyssig}}, \bibinfo
		{author} {\bibfnamefont {J.}~\bibnamefont {Zarestky}}, \bibinfo {author}
		{\bibfnamefont {S.}~\bibnamefont {Nandi}}, \bibinfo {author} {\bibfnamefont
			{N.}~\bibnamefont {Ni}}, \bibinfo {author} {\bibfnamefont {S.}~\bibnamefont
			{Bud’ko}}, \bibinfo {author} {\bibfnamefont {P.}~\bibnamefont {Canfield}},
		\bibinfo {author} {\bibfnamefont {A.}~\bibnamefont {Goldman}}, \ and\
		\bibinfo {author} {\bibfnamefont {R.}~\bibnamefont {McQueeney}},\ }\href@noop
	{} {\bibfield  {journal} {\bibinfo  {journal} {Phys. Rev. Lett.}\ }\textbf
		{\bibinfo {volume} {103}},\ \bibinfo {pages} {087001} (\bibinfo {year}
		{2009})}\BibitemShut {NoStop}%
	\bibitem [{\citenamefont {Liu}\ \emph {et~al.}(2019)\citenamefont {Liu},
		\citenamefont {Hao}, \citenamefont {Khalaf}, \citenamefont {Lee},
		\citenamefont {Watanabe}, \citenamefont {Taniguchi}, \citenamefont
		{Vishwanath},\ and\ \citenamefont {Kim}}]{liu2019spin}%
	\BibitemOpen
	\bibfield  {author} {\bibinfo {author} {\bibfnamefont {X.}~\bibnamefont
			{Liu}}, \bibinfo {author} {\bibfnamefont {Z.}~\bibnamefont {Hao}}, \bibinfo
		{author} {\bibfnamefont {E.}~\bibnamefont {Khalaf}}, \bibinfo {author}
		{\bibfnamefont {J.~Y.}\ \bibnamefont {Lee}}, \bibinfo {author} {\bibfnamefont
			{K.}~\bibnamefont {Watanabe}}, \bibinfo {author} {\bibfnamefont
			{T.}~\bibnamefont {Taniguchi}}, \bibinfo {author} {\bibfnamefont
			{A.}~\bibnamefont {Vishwanath}}, \ and\ \bibinfo {author} {\bibfnamefont
			{P.}~\bibnamefont {Kim}},\ }\href@noop {} {\bibfield  {journal} {\bibinfo
			{journal} {arXiv preprint arXiv:1903.08130}\ } (\bibinfo {year}
		{2019})}\BibitemShut {NoStop}%
	\bibitem [{\citenamefont {Lee}\ \emph {et~al.}(2019)\citenamefont {Lee},
		\citenamefont {Khalaf}, \citenamefont {Liu}, \citenamefont {Liu},
		\citenamefont {Hao}, \citenamefont {Kim},\ and\ \citenamefont
		{Vishwanath}}]{lee2019theory}%
	\BibitemOpen
	\bibfield  {author} {\bibinfo {author} {\bibfnamefont {J.~Y.}\ \bibnamefont
			{Lee}}, \bibinfo {author} {\bibfnamefont {E.}~\bibnamefont {Khalaf}},
		\bibinfo {author} {\bibfnamefont {S.}~\bibnamefont {Liu}}, \bibinfo {author}
		{\bibfnamefont {X.}~\bibnamefont {Liu}}, \bibinfo {author} {\bibfnamefont
			{Z.}~\bibnamefont {Hao}}, \bibinfo {author} {\bibfnamefont {P.}~\bibnamefont
			{Kim}}, \ and\ \bibinfo {author} {\bibfnamefont {A.}~\bibnamefont
			{Vishwanath}},\ }\href@noop {} {\bibfield  {journal} {\bibinfo  {journal}
			{Nat. Commun.}\ }\textbf {\bibinfo {volume} {10}},\ \bibinfo {pages}
		{5333} (\bibinfo {year} {2019})}\BibitemShut
	{NoStop}%
	\bibitem [{\citenamefont {Wu}\ and\ \citenamefont
		{Das~Sarma}(2019)}]{wu2019ferromagnetism}%
	\BibitemOpen
	\bibfield  {author} {\bibinfo {author} {\bibfnamefont {F.}~\bibnamefont
			{Wu}}\ and\ \bibinfo {author} {\bibfnamefont {S.}~\bibnamefont {Das~Sarma}},\
	}\href@noop {} {\bibfield  {journal} {\bibinfo  {journal} {arXiv preprint
				arXiv:1906.07302}\ } (\bibinfo {year} {2019})}\BibitemShut {NoStop}%
	\bibitem [{\citenamefont {Sigrist}(2009)}]{sigrist2009introduction}%
	\BibitemOpen
	\bibfield  {author} {\bibinfo {author} {\bibfnamefont {M.}~\bibnamefont
			{Sigrist}},\ }in\ \href@noop {} {\emph {\bibinfo {booktitle} {AIP Conference
				Proceedings}}},\ Vol.\ \bibinfo {volume} {1162}\ (\bibinfo {organization}
	{AIP},\ \bibinfo {year} {2009})\ pp.\ \bibinfo {pages} {55--96}\BibitemShut
	{NoStop}%
	\bibitem [{\citenamefont {Ramazashvili}(2008)}]{ramazashvili2008kramers}%
	\BibitemOpen
	\bibfield  {author} {\bibinfo {author} {\bibfnamefont {R.}~\bibnamefont
			{Ramazashvili}},\ }\href@noop {} {\bibfield  {journal} {\bibinfo  {journal}
			{Phys. Rev. Lett.}\ }\textbf {\bibinfo {volume} {101}},\ \bibinfo {pages}
		{137202} (\bibinfo {year} {2008})}\BibitemShut {NoStop}%
	\bibitem [{\citenamefont {Ramazashvili}(2009)}]{ramazashvili2009kramers}%
	\BibitemOpen
	\bibfield  {author} {\bibinfo {author} {\bibfnamefont {R.}~\bibnamefont
			{Ramazashvili}},\ }\href@noop {} {\bibfield  {journal} {\bibinfo  {journal}
			{Phys. Rev. B}\ }\textbf {\bibinfo {volume} {79}},\ \bibinfo {pages} {184432}
		(\bibinfo {year} {2009})}\BibitemShut {NoStop}%
	\bibitem [{\citenamefont {{\v{S}}mejkal}\ \emph {et~al.}(2019)\citenamefont
		{{\v{S}}mejkal}, \citenamefont {Gonz{\'a}lez-Hern{\'a}ndez}, \citenamefont
		{Jungwirth},\ and\ \citenamefont {Sinova}}]{vsmejkal2019crystal}%
	\BibitemOpen
	\bibfield  {author} {\bibinfo {author} {\bibfnamefont {L.}~\bibnamefont
			{{\v{S}}mejkal}}, \bibinfo {author} {\bibfnamefont {R.}~\bibnamefont
			{Gonz{\'a}lez-Hern{\'a}ndez}}, \bibinfo {author} {\bibfnamefont
			{T.}~\bibnamefont {Jungwirth}}, \ and\ \bibinfo {author} {\bibfnamefont
			{J.}~\bibnamefont {Sinova}},\ }\href@noop {} {\bibfield  {journal} {\bibinfo
			{journal} {arXiv preprint arXiv:1901.00445}\ } (\bibinfo {year}
		{2019})}\BibitemShut {NoStop}%
	\bibitem [{Note1()}]{Note1}%
	\BibitemOpen
	\bibinfo {note} {The combination of inversion symmetry and the eTRS is an
		antiunitary symmetry whose square is $-1$. Since the combined symmetry is
		local in the momentum space, it protects the Kramers degeneracy at any
		$k$-point.}\BibitemShut {Stop}%
	\bibitem [{\citenamefont {Fischer}\ \emph {et~al.}(2011)\citenamefont
		{Fischer}, \citenamefont {Loder},\ and\ \citenamefont
		{Sigrist}}]{fischer2011superconductivity}%
	\BibitemOpen
	\bibfield  {author} {\bibinfo {author} {\bibfnamefont {M.~H.}\ \bibnamefont
			{Fischer}}, \bibinfo {author} {\bibfnamefont {F.}~\bibnamefont {Loder}}, \
		and\ \bibinfo {author} {\bibfnamefont {M.}~\bibnamefont {Sigrist}},\
	}\href@noop {} {\bibfield  {journal} {\bibinfo  {journal} {Phys. Rev. B}\
		}\textbf {\bibinfo {volume} {84}},\ \bibinfo {pages} {184533} (\bibinfo
		{year} {2011})}\BibitemShut {NoStop}%
	\bibitem [{\citenamefont {Goryo}\ \emph {et~al.}(2012)\citenamefont {Goryo},
		\citenamefont {Fischer},\ and\ \citenamefont {Sigrist}}]{goryo2012possible}%
	\BibitemOpen
	\bibfield  {author} {\bibinfo {author} {\bibfnamefont {J.}~\bibnamefont
			{Goryo}}, \bibinfo {author} {\bibfnamefont {M.~H.}\ \bibnamefont {Fischer}},
		\ and\ \bibinfo {author} {\bibfnamefont {M.}~\bibnamefont {Sigrist}},\
	}\href@noop {} {\bibfield  {journal} {\bibinfo  {journal} {Phys. Rev. B}\
		}\textbf {\bibinfo {volume} {86}},\ \bibinfo {pages} {100507} (\bibinfo
		{year} {2012})}\BibitemShut {NoStop}%
	\bibitem [{Note2()}]{Note2}%
	\BibitemOpen
	\bibinfo {note} {More rigorously, $\protect \mathaccentV {tilde}07E{\protect
			\mathit {\Theta }}_{\protect \textbf {k}}=t_{1/2}\protect \mathit {\Theta
		}=ie^{i\protect \textbf {k}\cdot (\protect \textbf {a}/2)}\sigma _{y}\tau
		_{x}K$ and $\protect \mathaccentV {tilde}07E{\protect \mathit {\Theta
		}}_{\protect \textbf {k}}\protect \mathcal {H}(\protect \textbf {k})\protect
		\mathaccentV {tilde}07E{\protect \mathit {\Theta }}_{\protect \textbf
			{k}}^{-1}=\protect \mathcal {H}(-\protect \textbf {k})$, where $t_{1/2}$ and
		$\protect \textbf {a}$ denote the half translation operator and the lattice
		vector, respectively. But the extra phase factor comming from the translation
		by half the lattice vector is not important in this context, so we omit it
		and write $\protect \mathaccentV {tilde}07E{\protect \mathit {\Theta }}$ as
		$i\sigma _{y}\tau _{x}K$ from now on.}\BibitemShut {Stop}%
	\bibitem [{\citenamefont {Sigrist}\ and\ \citenamefont
		{Ueda}(1991)}]{sigrist1991phenomenological}%
	\BibitemOpen
	\bibfield  {author} {\bibinfo {author} {\bibfnamefont {M.}~\bibnamefont
			{Sigrist}}\ and\ \bibinfo {author} {\bibfnamefont {K.}~\bibnamefont {Ueda}},\
	}\href@noop {} {\bibfield  {journal} {\bibinfo  {journal} {Rev. Mod. Phys.}\
		}\textbf {\bibinfo {volume} {63}},\ \bibinfo {pages} {239} (\bibinfo {year}
		{1991})}\BibitemShut {NoStop}%
	\bibitem [{\citenamefont {Fu}\ and\ \citenamefont {Berg}(2010)}]{fu2010odd}%
	\BibitemOpen
	\bibfield  {author} {\bibinfo {author} {\bibfnamefont {L.}~\bibnamefont
			{Fu}}\ and\ \bibinfo {author} {\bibfnamefont {E.}~\bibnamefont {Berg}},\
	}\href@noop {} {\bibfield  {journal} {\bibinfo  {journal} {Physical review
				letters}\ }\textbf {\bibinfo {volume} {105}},\ \bibinfo {pages} {097001}
		(\bibinfo {year} {2010})}\BibitemShut {NoStop}%
	\bibitem [{\citenamefont {Sato}(2010)}]{sato2010topological}%
	\BibitemOpen
	\bibfield  {author} {\bibinfo {author} {\bibfnamefont {M.}~\bibnamefont
			{Sato}},\ }\href@noop {} {\bibfield  {journal} {\bibinfo  {journal} {Phys.
				Rev. B}\ }\textbf {\bibinfo {volume} {81}},\ \bibinfo {pages} {220504}
		(\bibinfo {year} {2010})}\BibitemShut {NoStop}%
	\bibitem [{\citenamefont {Nakosai}\ \emph {et~al.}(2012)\citenamefont
		{Nakosai}, \citenamefont {Tanaka},\ and\ \citenamefont
		{Nagaosa}}]{nakosai2012topological}%
	\BibitemOpen
	\bibfield  {author} {\bibinfo {author} {\bibfnamefont {S.}~\bibnamefont
			{Nakosai}}, \bibinfo {author} {\bibfnamefont {Y.}~\bibnamefont {Tanaka}}, \
		and\ \bibinfo {author} {\bibfnamefont {N.}~\bibnamefont {Nagaosa}},\
	}\href@noop {} {\bibfield  {journal} {\bibinfo  {journal} {Phys. Rev. Lett.}\
		}\textbf {\bibinfo {volume} {108}},\ \bibinfo {pages} {147003} (\bibinfo
		{year} {2012})}\BibitemShut {NoStop}%
	\bibitem [{\citenamefont {Altland}\ and\ \citenamefont
		{Zirnbauer}(1997)}]{altland1997nonstandard}%
	\BibitemOpen
	\bibfield  {author} {\bibinfo {author} {\bibfnamefont {A.}~\bibnamefont
			{Altland}}\ and\ \bibinfo {author} {\bibfnamefont {M.~R.}\ \bibnamefont
			{Zirnbauer}},\ }\href@noop {} {\bibfield  {journal} {\bibinfo  {journal}
			{Phys. Rev. B}\ }\textbf {\bibinfo {volume} {55}},\ \bibinfo {pages} {1142}
		(\bibinfo {year} {1997})}\BibitemShut {NoStop}%
	\bibitem [{\citenamefont {Kitaev}(2009)}]{kitaev2009periodic}%
	\BibitemOpen
	\bibfield  {author} {\bibinfo {author} {\bibfnamefont {A.}~\bibnamefont
			{Kitaev}},\ }in\ \href@noop {} {\emph {\bibinfo {booktitle} {AIP conference
				proceedings}}},\ Vol.\ \bibinfo {volume} {1134}\ (\bibinfo {organization}
	{AIP},\ \bibinfo {year} {2009})\ pp.\ \bibinfo {pages} {22--30}\BibitemShut
	{NoStop}%
	\bibitem [{\citenamefont {Chiu}\ \emph {et~al.}(2016)\citenamefont {Chiu},
		\citenamefont {Teo}, \citenamefont {Schnyder},\ and\ \citenamefont
		{Ryu}}]{chiu2016classification}%
	\BibitemOpen
	\bibfield  {author} {\bibinfo {author} {\bibfnamefont {C.-K.}\ \bibnamefont
			{Chiu}}, \bibinfo {author} {\bibfnamefont {J.~C.}\ \bibnamefont {Teo}},
		\bibinfo {author} {\bibfnamefont {A.~P.}\ \bibnamefont {Schnyder}}, \ and\
		\bibinfo {author} {\bibfnamefont {S.}~\bibnamefont {Ryu}},\ }\href@noop {}
	{\bibfield  {journal} {\bibinfo  {journal} {Rev. Mod. Phys.}\ }\textbf
		{\bibinfo {volume} {88}},\ \bibinfo {pages} {035005} (\bibinfo {year}
		{2016})}\BibitemShut {NoStop}%
	\bibitem [{\citenamefont {Schnyder}\ \emph {et~al.}(2016)\citenamefont {Schnyder},
		\citenamefont {Ryu}, \citenamefont {Furusaki},\ and\ \citenamefont
		{Ludwig}}]{schnyder2008classification}%
	\BibitemOpen
	\bibfield  {author} {\bibinfo {author} {\bibfnamefont {A. P.}\ \bibnamefont
			{Schnyder}}, \bibinfo {author} {\bibfnamefont {S.}\ \bibnamefont {Ryu}},
		\bibinfo {author} {\bibfnamefont {A.}\ \bibnamefont {Furusaki}}, \ and\
		\bibinfo {author} {\bibfnamefont {A.}~\bibnamefont {Ludwig}},\ }\href@noop {}
	{\bibfield  {journal} {\bibinfo  {journal} {Phys. Rev. B}\ }\textbf
		{\bibinfo {volume} {78}},\ \bibinfo {pages} {195125} (\bibinfo {year}
		{2008})}\BibitemShut {NoStop}%
	\bibitem [{\citenamefont {Sato}\ and\ \citenamefont
		{Ando}(2017)}]{sato2017topological}%
	\BibitemOpen
	\bibfield  {author} {\bibinfo {author} {\bibfnamefont {M.}~\bibnamefont
			{Sato}}\ and\ \bibinfo {author} {\bibfnamefont {Y.}~\bibnamefont {Ando}},\
	}\href@noop {} {\bibfield  {journal} {\bibinfo  {journal} {Reports on
				Progress in Physics}\ }\textbf {\bibinfo {volume} {80}},\ \bibinfo {pages}
		{076501} (\bibinfo {year} {2017})}\BibitemShut {NoStop}%
	\bibitem [{\citenamefont {Fidkowski}\ \emph {et~al.}(2011)\citenamefont
		{Fidkowski}, \citenamefont {Jackson},\ and\ \citenamefont
		{Klich}}]{fidkowski2011model}%
	\BibitemOpen
	\bibfield  {author} {\bibinfo {author} {\bibfnamefont {L.}~\bibnamefont
			{Fidkowski}}, \bibinfo {author} {\bibfnamefont {T.}~\bibnamefont {Jackson}},
		\ and\ \bibinfo {author} {\bibfnamefont {I.}~\bibnamefont {Klich}},\
	}\href@noop {} {\bibfield  {journal} {\bibinfo  {journal} {Phys. Rev. Lett.}\
		}\textbf {\bibinfo {volume} {107}},\ \bibinfo {pages} {036601} (\bibinfo
		{year} {2011})}\BibitemShut {NoStop}%
	\bibitem [{\citenamefont {Alexandradinata}\ \emph {et~al.}(2014)\citenamefont
		{Alexandradinata}, \citenamefont {Dai},\ and\ \citenamefont
		{Bernevig}}]{alexandradinata2014wilson}%
	\BibitemOpen
	\bibfield  {author} {\bibinfo {author} {\bibfnamefont {A.}~\bibnamefont
			{Alexandradinata}}, \bibinfo {author} {\bibfnamefont {X.}~\bibnamefont
			{Dai}}, \ and\ \bibinfo {author} {\bibfnamefont {B.~A.}\ \bibnamefont
			{Bernevig}},\ }\href@noop {} {\bibfield  {journal} {\bibinfo  {journal}
			{Phys. Rev. B}\ }\textbf {\bibinfo {volume} {89}},\ \bibinfo {pages} {155114}
		(\bibinfo {year} {2014})}\BibitemShut {NoStop}%
	\bibitem [{\citenamefont {Laughlin}(1981)}]{laughlin1981quantized}%
	\BibitemOpen
	\bibfield  {author} {\bibinfo {author} {\bibfnamefont {R.~B.}\ \bibnamefont
			{Laughlin}},\ }\href@noop {} {\bibfield  {journal} {\bibinfo  {journal}
			{Phys. Rev. B}\ }\textbf {\bibinfo {volume} {23}},\ \bibinfo {pages} {5632}
		(\bibinfo {year} {1981})}\BibitemShut {NoStop}%
	\bibitem [{\citenamefont {Halperin}(1982)}]{halperin1982quantized}%
	\BibitemOpen
	\bibfield  {author} {\bibinfo {author} {\bibfnamefont {B.~I.}\ \bibnamefont
			{Halperin}},\ }\href@noop {} {\bibfield  {journal} {\bibinfo  {journal}
			{Phys. Rev. B}\ }\textbf {\bibinfo {volume} {25}},\ \bibinfo {pages} {2185}
		(\bibinfo {year} {1982})}\BibitemShut {NoStop}%
	\bibitem [{\citenamefont {Hatsugai}(1993{\natexlab{a}})}]{hatsugai1993chern}%
	\BibitemOpen
	\bibfield  {author} {\bibinfo {author} {\bibfnamefont {Y.}~\bibnamefont
			{Hatsugai}},\ }\href@noop {} {\bibfield  {journal} {\bibinfo  {journal}
			{Phys. Rev. Lett.}\ }\textbf {\bibinfo {volume} {71}},\ \bibinfo {pages}
		{3697} (\bibinfo {year} {1993}{\natexlab{a}})}\BibitemShut {NoStop}%
	\bibitem [{\citenamefont {Hatsugai}(1993{\natexlab{b}})}]{hatsugai1993edge}%
	\BibitemOpen
	\bibfield  {author} {\bibinfo {author} {\bibfnamefont {Y.}~\bibnamefont
			{Hatsugai}},\ }\href@noop {} {\bibfield  {journal} {\bibinfo  {journal}
			{Phys. Rev. B}\ }\textbf {\bibinfo {volume} {48}},\ \bibinfo {pages} {11851}
		(\bibinfo {year} {1993}{\natexlab{b}})}\BibitemShut {NoStop}%
	\bibitem [{\citenamefont {Qi}\ \emph {et~al.}(2006)\citenamefont {Qi},
		\citenamefont {Wu},\ and\ \citenamefont {Zhang}}]{qi2006general}%
	\BibitemOpen
	\bibfield  {author} {\bibinfo {author} {\bibfnamefont {X.-L.}\ \bibnamefont
			{Qi}}, \bibinfo {author} {\bibfnamefont {Y.-S.}\ \bibnamefont {Wu}}, \ and\
		\bibinfo {author} {\bibfnamefont {S.-C.}\ \bibnamefont {Zhang}},\ }\href@noop
	{} {\bibfield  {journal} {\bibinfo  {journal} {Phys. Rev. B}\ }\textbf
		{\bibinfo {volume} {74}},\ \bibinfo {pages} {045125} (\bibinfo {year}
		{2006})}\BibitemShut {NoStop}%
	\bibitem [{\citenamefont {Fukui}\ \emph {et~al.}(2012)\citenamefont {Fukui},
		\citenamefont {Shiozaki}, \citenamefont {Fujiwara},\ and\ \citenamefont
		{Fujimoto}}]{fukui2012bulk}%
	\BibitemOpen
	\bibfield  {author} {\bibinfo {author} {\bibfnamefont {T.}~\bibnamefont
			{Fukui}}, \bibinfo {author} {\bibfnamefont {K.}~\bibnamefont {Shiozaki}},
		\bibinfo {author} {\bibfnamefont {T.}~\bibnamefont {Fujiwara}}, \ and\
		\bibinfo {author} {\bibfnamefont {S.}~\bibnamefont {Fujimoto}},\ }\href@noop
	{} {\bibfield  {journal} {\bibinfo  {journal} {Journal of the Physical
				Society of Japan}\ }\textbf {\bibinfo {volume} {81}},\ \bibinfo {pages}
		{114602} (\bibinfo {year} {2012})}\BibitemShut {NoStop}%
	\bibitem [{\citenamefont {Imai}\ \emph {et~al.}(2017)\citenamefont {Imai},
		\citenamefont {Wakabayashi},\ and\ \citenamefont
		{Sigrist}}]{imai2017thermal}%
	\BibitemOpen
	\bibfield  {author} {\bibinfo {author} {\bibfnamefont {Y.}~\bibnamefont
			{Imai}}, \bibinfo {author} {\bibfnamefont {K.}~\bibnamefont {Wakabayashi}}, \
		and\ \bibinfo {author} {\bibfnamefont {M.}~\bibnamefont {Sigrist}},\
	}\href@noop {} {\bibfield  {journal} {\bibinfo  {journal} {Phys. Rev. B}\
		}\textbf {\bibinfo {volume} {95}},\ \bibinfo {pages} {024516} (\bibinfo
		{year} {2017})}\BibitemShut {NoStop}%
	\bibitem [{Note3()}]{Note3}%
	\BibitemOpen
	\bibinfo {note} {See Supplementary Material at [].}\BibitemShut {Stop}%
	\bibitem [{\citenamefont {Khalaf}(2018)}]{khalaf2018higher}%
	\BibitemOpen
	\bibfield  {author} {\bibinfo {author} {\bibfnamefont {E.}~\bibnamefont
			{Khalaf}},\ }\href@noop {} {\bibfield  {journal} {\bibinfo  {journal} {Phys.
				Rev. B}\ }\textbf {\bibinfo {volume} {97}},\ \bibinfo {pages} {205136}
		(\bibinfo {year} {2018})}\BibitemShut {NoStop}%
	\bibitem [{\citenamefont {Ahn}\ and\ \citenamefont
		{Yang}(2020)}]{ahn2020higher}%
	\BibitemOpen
	\bibfield  {author} {\bibinfo {author} {\bibfnamefont {J.}~\bibnamefont
			{Ahn}}\ and\ \bibinfo {author} {\bibfnamefont {B.-J.}\ \bibnamefont {Yang}},\
	}\href@noop {} {\bibfield  {journal} {\bibinfo  {journal} {Phys. Rev.
				Research}\ }\textbf {\bibinfo {volume} {2}},\ \bibinfo {pages} {012060}
		(\bibinfo {year} {2020})}\BibitemShut {NoStop}%
	\bibitem [{\citenamefont {Hwang}\ and\ \citenamefont
		{Ahn} and\ \citenamefont{Yang}(2019)}]{hwang2019fragile}%
	\BibitemOpen
	\bibfield  {author} {\bibinfo {author} {\bibfnamefont {Y.}~\bibnamefont
			{Hwang}}\ and\ \bibinfo {author} {\bibfnamefont {J.}~\bibnamefont
		{Ahn}}\ and\ \bibinfo {author} {\bibfnamefont {B.-J.}\ \bibnamefont {Yang}},\
	}\href@noop {} {\bibfield  {journal} {\bibinfo  {journal} {Phys. Rev.
				B}\ }\textbf {\bibinfo {volume} {100}},\ \bibinfo {pages} {205126}
		(\bibinfo {year} {2019})}\BibitemShut {NoStop}%
	\bibitem [{\citenamefont {Chiu}\ and\ \citenamefont
		{Schnyder}(2014)}]{chiu2014classification}%
	\BibitemOpen
	\bibfield  {author} {\bibinfo {author} {\bibfnamefont {C.-K.}\ \bibnamefont
			{Chiu}}\ and\ \bibinfo {author} {\bibfnamefont {A.~P.}\ \bibnamefont
			{Schnyder}},\ }\href@noop {} {\bibfield  {journal} {\bibinfo  {journal}
			{Phys. Rev. B}\ }\textbf {\bibinfo {volume} {90}},\
			\bibinfo {pages} {205136} (\bibinfo {year} {2014})}\BibitemShut {NoStop}%
	\bibitem [{\citenamefont {Schnyder}\ and\ \citenamefont
		{Brydon}(2015)}]{schnyder2015topological}%
	\BibitemOpen
	\bibfield  {author} {\bibinfo {author} {\bibfnamefont {A.~P.}\ \bibnamefont
			{Schnyder}}\ and\ \bibinfo {author} {\bibfnamefont {P.~M.}\ \bibnamefont
			{Brydon}},\ }\href@noop {} {\bibfield  {journal} {\bibinfo  {journal}
			{Journal of Physics: Condensed Matter}\ }\textbf {\bibinfo {volume} {27}},\
		\bibinfo {pages} {243201} (\bibinfo {year} {2015})}\BibitemShut {NoStop}%
	\bibitem [{\citenamefont {Sumita}\ and\ \citenamefont
		{Yanase}(2018)}]{sumita2018unconventional}%
	\BibitemOpen
	\bibfield  {author} {\bibinfo {author} {\bibfnamefont {S.}~\bibnamefont
			{Sumita}}\ and\ \bibinfo {author} {\bibfnamefont {Y.}~\bibnamefont
			{Yanase}},\ }\href@noop {} {\bibfield  {journal} {\bibinfo  {journal} {Phys.
				Rev. B}\ }\textbf {\bibinfo {volume} {97}},\ \bibinfo {pages} {134512}
		(\bibinfo {year} {2018})}\BibitemShut {NoStop}%
	\bibitem [{\citenamefont {Saha-Dasgupta}(2020)}]{Saha_Dasgupta_2020}%
	\BibitemOpen
	\bibfield  {author} {\bibinfo {author} {\bibfnamefont {T.}~\bibnamefont
			{Saha-Dasgupta}},\ }\href@noop {} {\bibfield  {journal} {\bibinfo  {journal} {Mater. Res. Express}\ }\textbf {\bibinfo {volume} {7}},\ \bibinfo {pages} {014003}
		(\bibinfo {year} {2020})}\BibitemShut {NoStop}%
	\bibitem [{\citenamefont {Kobayashi}\ and\ \citenamefont {Kimura}\ \citenamefont {Sawada}\ and\ \citenamefont {Terakura}\ and\ \citenamefont {Tokura}(1998)}]{kobayashi1998room}%
	\BibitemOpen
	\bibfield  {author} {\bibinfo {author} {\bibfnamefont {K-I.}~\bibnamefont
			{Kobayashi}, \bibfnamefont {T.}~\bibnamefont
			{Kimura}, \bibfnamefont {H.}~\bibnamefont
			{Sawada}, \bibfnamefont {K.}~\bibnamefont
			{Terakura}, and\ \bibfnamefont {Y.}~\bibnamefont
			{Tokura}},\ }\href@noop {} {\bibfield  {journal} {\bibinfo  {journal} {Nature}\ }\textbf {\bibinfo {volume} {395}},\ \bibinfo {pages} {677--680}
		(\bibinfo {year} {1998})}\BibitemShut {NoStop}%
	\bibitem [{\citenamefont {Sanyal}\ and\ \citenamefont {Das}\ and\ \citenamefont {Saha-Dasgupta}(2020)}]{sanyal2009evidence}%
	\BibitemOpen
	\bibfield  {author} {\bibinfo {author} {\bibfnamefont {P.}~\bibnamefont
			{Sanyal}, \bibfnamefont {H.}~\bibnamefont
		{Das}, and\ \bibfnamefont {T.}~\bibnamefont
			{Saha-Dasgupta}},\ }\href@noop {} {\bibfield  {journal} {\bibinfo  {journal} {Phys.
				Rev. B}\ }\textbf {\bibinfo {volume} {80}},\ \bibinfo {pages} {224412}
		(\bibinfo {year} {2009})}\BibitemShut {NoStop}%
	\bibitem [{\citenamefont {Connor}\ and\ \citenamefont
		{Leppert} and\ \citenamefont {Smith} and\ \citenamefont	{Neaton} and\ \citenamefont	{Karunadasa}(2018)}]{connor2018layered}%
	\BibitemOpen
	\bibfield  {author} {\bibinfo {author} {\bibfnamefont {B. A.}~\bibnamefont
			{Connor}}\, \bibinfo {author} {\bibfnamefont {L.}~\bibnamefont
			{Leppert}}\, \bibinfo {author} {\bibfnamefont {M. D.}~\bibnamefont
			{Smith}}\, \bibinfo {author} {\bibfnamefont {J. B.}~\bibnamefont
			{Neaton}}\, and\ \bibinfo {author} {\bibfnamefont {H. I.}~\bibnamefont
			{Karunadasa}},\ }\href@noop {} {\bibfield  {journal} {\bibinfo  {journal} J. Am. Chem. Soc.}\ \textbf {\bibinfo {volume} {140}},
		\ \bibinfo {pages} {5235--5240}
		(\bibinfo {year} {2018})} \BibitemShut {NoStop}%
\end{thebibliography}

\begin{thebibliography}{6}
		\bibitem{sigrist1991phenomenological} 
		M. Sigrist, and K. Ueda,
		Rev. Mod. Phys. \textbf{63}, 239 (1991).
		
		\bibitem{altland2010condensed} 
		A. Altland, and B. D. Simons,
		\textit{Condensed matter field theory},
		(Cambridge university press, Cambridge, 2010)
		
		\bibitem{alexandradinata2014wilson} 
		A. Alexandradinata, X. Dai, and B. A. Bernevig,
		Phys. Rev. B \textbf{89}, 155114 (2014).
		
		\bibitem{bernevig2013topological} 
		A. B. Bernevig,  and T. L. Hughes,
		\textit{Topological insulators and topological superconductors},
		(Princeton university press, Princeton, 2013).
		
		\bibitem{ahn2019higher}
		J. Ahn and B.-J. Yang,
		Phys. Rev. Research \textbf{2}, 012060 (2020).
		
		\bibitem{skurativska2020atomic}
		A. Skurativska, T. Neupert, and M. H. Fischer,
		Phys. Rev. Research \textbf{2}, 013064 (2020).
	\end{thebibliography}
\end{document}